\documentclass[fleqn,usenatbib]{mnras}

\usepackage{newtxtext,newtxmath}
\usepackage[T1]{fontenc}

\DeclareRobustCommand{\VAN}[3]{#2}
\let\VANthebibliography\thebibliography
\def\thebibliography{\DeclareRobustCommand{\VAN}[3]{##3}\VANthebibliography}

\usepackage{graphicx}	% Including figure files
\usepackage{amsmath}	% Advanced maths commands
\usepackage{hyperref} % For clickable links and \texorpdfstring

\title[The \textit{Fermi}-TRAPUM UHF Survey]{Pulsar Discoveries from the TRAPUM UHF Survey of \textit{Fermi}-LAT Sources}

\author[T. Thongmeearkom, et al.]{T. Thongmeearkom$^{1,2}$\thanks{
\href{mailto:tinn.thongmeearkom@manchester.ac.uk}{tinn.thongmeearkom@manchester.ac.uk} / 
\href{mailto:tinn@narit.or.th}{tinn@narit.or.th} (TT);
\protect\newline
\href{mailto:colin.clark@aei.mpg.de}{colin.clark@aei.mpg.de} (CJC);
\href{mailto:rene.breton@manchester.ac.uk}{rene.breton@manchester.ac.uk} (RPB)
},
C. J. Clark$^{3,4}$\footnotemark[1],
R. P. Breton$^{1}$\footnotemark[1],
M. Burgay$^{5}$,
L. Nieder$^{3,4}$,
O. G. Dodge$^{3,4,1}$,
\newauthor B. McGloughlin$^{3,4}$,
E. D. Barr$^{6}$,
S. Buchner$^{7}$,
B. W. Stappers$^{1}$,
J. Berteaud$^{8,9}$,
E. C. Ferrara$^{9,10,8}$,
\newauthor P. C. C. Freire$^{6}$,
L. Levin$^{1}$,
S. M. Ransom$^{11}$,
L. Vleeschower$^{12,1}$,
S. Belmonte Díaz$^{1}$,
F. Calore$^{13}$,
\newauthor I. Cognard$^{14,15}$,
V. S. Dhillon$^{16,17}$,
J.-M. Grie{\ss}meier$^{14,15}$,
R. Karuppusamy$^{6}$,
M. R. Kennedy$^{18,1}$,
\newauthor M. Kramer$^{6,1}$,
P. V. Padmanabh$^{3,4}$,
M. A. Papa$^{3,4}$,
A. Phosrisom$^{2,1}$,
 and B. Steltner$^{3,4}$
\\\\\\
$^{1}$Jodrell Bank Centre for Astrophysics, Department of Physics and Astronomy, The University of Manchester, Manchester M13 9PL, United Kingdom\\
$^{2}$ National Astronomical Research Institute of Thailand, Don Kaeo, Mae Rim, Chiang Mai 50180, Thailand\\
$^{3}$ Max Planck Institute for Gravitational Physics (Albert Einstein Institute), D-30167 Hannover, Germany\\
$^{4}$ Leibniz Universit\"{a}t Hannover, D-30167 Hannover, Germany\\
$^{5}$ INAF -- Osservatorio Astronomico di Cagliari, Via della Scienza 5, I-09047 Selargius (CA), Italy\\
$^{6}$ Max-Planck-Institut f\"{u}r Radioastronomie, Auf dem H\"{u}gel 69, D-53121 Bonn, Germany\\
$^{7}$ South African Radio Astronomy Observatory, 2 Fir Street, Black River Park, Observatory 7925, South Africa\\
$^{8}$NASA Goddard Space Flight Center, Code 662, Greenbelt, MD 20771, USA\\
$^{9}$University of Maryland, Department of Astronomy, College Park, MD 20742, USA\\
$^{10}$Center for Research and Exploration in Space Science \& Technology II (CRESST II), NASA/GSFC, Greenbelt, MD 20771, USA\\
$^{11}$National Radio Astronomy Observatory, 520 Edgemont Rd., Charlottesville, VA 22903, USA\\
$^{12}$Center for Gravitation, Cosmology, and Astrophysics, Department of Physics, University of Wisconsin-Milwaukee, P.O. Box 413, Milwaukee, WI 53201, USA\\
$^{13}$LAPTh, CNRS, USMB, F-74940 Annecy, France\\
$^{14}$LPC2E, OSUC, Univ Orleans, CNRS, CNES, Observatoire de Paris, F-45071 Orleans, France\\
$^{15}$ORN, Observatoire de Paris, Universit\'{e} PSL, Univ Orl\'{e}ans, CNRS, 18330 Nan\c{c}ay, France\\
$^{16}$Astrophysics Research Cluster, School of Mathematical \& Physical Sciences, University of Sheffield, Sheffield S3 7RH, United Kingdom\\
$^{17}$Instituto de Astrof\'{i}sica de Canarias, E-38205 La Laguna, Tenerife, Spain\\
$^{18}$ School of Physics, Kane Building, University College Cork, Cork, T12 K8AF, Ireland}

\date{Accepted 2026 February 23. Received 2026 February 23; in original form 2026 January 12}

\pubyear{2026}

\begin{document}
\label{firstpage}
\pagerange{\pageref{firstpage}--\pageref{lastpage}}
\maketitle

% Abstract of the paper
\begin{abstract}
The \textit{Fermi} Large Area Telescope (LAT) provides advantages for radio pulsar searches by enabling efficient target selection. We can confidently point radio telescopes to the positions of \textit{Fermi} unidentified gamma-ray sources that have a high probability of hosting a pulsar. As part of Transients and Pulsars with MeerKAT (TRAPUM), we conducted a survey of \textit{Fermi}-LAT sources using the Ultra High Frequency (UHF; 544--1088\,MHz) receiver of the MeerKAT radio telescope. We observed 79 sources that were identified as pulsar-like candidates using a random forest technique from the \textit{Fermi}-LAT Fourth Source Catalogue. We observed each target for 10 minutes at two separate epochs. As a result, we discovered nine new millisecond pulsars (MSPs) and six slow pulsars. Based on the radio discoveries, we also searched for gamma-ray pulsations, confirming that seven of the newly discovered MSPs are associated with \textit{Fermi}-LAT sources, and performed joint radio and gamma-ray pulsar timing. Companion mass estimates and evidence of radio eclipses indicate that among the nine MSPs there are three black widows and three redbacks. Lastly, we compared the discovered pulsars in the MeerKAT UHF survey against the previous \textit{Fermi} sources TRAPUM survey at \textit{L} band, concluding the superiority of UHF observations in sensitivity to fainter pulsars and in detection rate than \textit{L} band for finding new gamma-ray MSPs.
\end{abstract}

% Select between one and six entries from the list of approved keywords.
% Don't make up new ones.

%%%%%%%%%%%%%%%%%%%%%%%%%%%%%%%%%%%%%%%%%%%%%%%%%%

%%%%%%%%%%%%%%%%% BODY OF PAPER %%%%%%%%%%%%%%%%%%
%\clearpage
\section{Introduction}
\label{S:intro}

The \textit{Fermi Gamma-ray Space Telescope}, launched in 2008, carries the Large Area Telescope \citep[LAT;][]{Atwood2009+LAT} as its main instrument. The \textit{Fermi}-LAT has been used to observe high-energy gamma-ray phenomena involving non-thermal processes. Over the years, it has accumulated gamma-ray data resulting in the latest catalogue, the \textit{Fermi}-LAT Fourth Source Catalogue \citep{4FGL,Abdollahi2022+4FGLDR3} Data Release 4 \citep[4FGL DR4;][]{4FGLDR4}, which contains $\sim$7,000 gamma-ray sources, $\sim$300 of which are pulsars \citep{Smith2023+3PC}, and $\sim$2,600 are unidentified sources\footnote{\url{https://fermi.gsfc.nasa.gov/ssc/data/access/lat/14yr_catalog/}}. Gamma-ray pulsars (the subject of this study) have unique gamma-ray emission characteristics, with highly curved spectra and low flux variability \citep{SazParkinson2016+ML}, which distinguish them from other gamma-ray objects like active galactic nuclei (AGNs) \citep{Ackermann2012+classification}.

The \textit{Fermi}-LAT unassociated sources that are localised with precisions of a few arcminutes and that have pulsar-like spectra provide optimal targets for pulsar searches. Consequently, many radio surveys have been searching for radio pulsations using these pulsar-like unassociated gamma-ray sources as targets under the \textit{Fermi} Pulsar Search Consortium \citep[PSC;][]{Ray2012+PSC}. This strategy has proven effective in numerous surveys using radio telescopes worldwide \citep[e.g.,][]{Kerr2012+Parkes,Cromartie2016+Arecibo,Wang2021+FASTMSP,Bangale2024,Kerr2025}. By targeting \textit{Fermi} gamma-ray sources, this search technique is particularly sensitive for detecting millisecond pulsars (MSPs), as they are often gamma-ray bright because of their high spin-down power. MSPs are especially important astrophysical laboratories because of their stable rotation, which enables precise timing studies. Moreover, many MSPs reside in diverse binary environments, which provide valuable opportunities to study binary evolution. 

In particular, compact binary MSPs known as ``spider pulsars'' feature an MSP with an irradiated, low-mass companion in a sub-day orbit: black widows (BWs) have an extremely low-mass, semi-degenerate companion ($\ll 0.1\,M_{\odot}$), while redbacks (RBs) have a low-mass, non-degenerate companion ($0.2 - 0.4\,{\rm M}_\odot$) \citep{Roberts2013}. Some spiders appear to be intermediate between the conventional black widow and redback categories, bridging the two types \citep[e.g., PSR~J1242$-$4712;][]{Ghosh2024+J1242}. Over the years, many spider systems have been discovered, helping in the development of binary evolution models \citep[e.g.,][]{Chen2013,Benvenuto2014}. As of this writing, there are 82 confirmed spider pulsars\footnote{\url{https://astro.phys.ntnu.no/SpiderCAT/}} \citep{SpiderCAT}. Spider pulsars are challenging to detect as they often exhibit long radio eclipses due to intra-binary material and obstruction by the companion, which varies depending on the system structure and inclination \citep{Polzin2018+J1810,Polzin2019+J2051}. \citet{Shang2024+J1816} studied the radio eclipse of PSR~J1816+4510, finding that the pulse profile broadens in the eclipse region due to increased scattering. Similarly, observations of PSR~J1653$-$0158 \citep{Nieder2020+J1653} suggest that some spider binaries are continuously obstructed by material, underscoring the value of observing across multiple wavelengths (e.g., radio and gamma rays). Examples of extreme radio-eclipsing spider systems include PSRs~J1048+2339 \citep{Deneva2016+J1048}, J0212+5321 \citep{Perez2023+J0212}, and J0838$-$2827 \citep{Thongmeearkom2024+RBs}.

This paper is part of the Transients and Pulsars with MeerKAT \citep[TRAPUM;][]{Stappers2018+TRAPUM} survey. The TRAPUM collaboration comprises several groups with varied search strategies, targeting different regions and sources across the sky to maximise the discovery of new pulsars. These groups include the Globular Clusters Working Group \citep{Ridolfi2021+GC,Vleeschower2024+M62}, the Nearby Galaxies Working Group \citep{Carli2024+SMC,Prayag2024+LMC}, the Supernova Remnants Working Group \citep{Turner2024+SNR}, and the \textit{Fermi} Sources Working Group \citep{Clark2023+Lband,Thongmeearkom2024+RBs}. So far, 307 new pulsars\footnote{\url{https://trapum.org/discoveries/}} have been discovered by TRAPUM, and by the MPIfR-MeerKAT Galactic Plane Survey \citep[MMGPS;][]{Padmanabh2023+MGPS}, which uses the same processing infrastructure. The \textit{Fermi} Sources Working Group has contributed 52 of these discoveries. Of these, 47 are from the shallow (10-minute) surveys (\citealp{Clark2023+Lband}; this paper; Thongmeearkom et al., in preparation), while five pulsars are from the targeted deep (1-h) surveys (\citealp{Thongmeearkom2024+RBs, Belmonte2025+J1544}; Belmonte Díaz et al., in preparation).

Following the TRAPUM shallow survey for \textit{Fermi}-LAT unassociated sources at \textit{L} band \citep{Clark2023+Lband}, this paper presents the first shallow pulsar survey of \textit{Fermi}-LAT sources using the MeerKAT radio telescope at Ultra High Frequency (UHF; 544--1088 MHz). We report results from 10-minute observations of 79 unidentified \textit{Fermi} gamma-ray sources (67 of which were previously observed with MeerKAT at \textit{L} band, while 13 were newly added). We also included follow-up observations with other radio telescopes for detection confirmation and pulsar timing. Additionally, we compared the discoveries at \textit{L} band \citep{Clark2023+Lband} with those at UHF (this study) to inform the forthcoming surveys with MeerKAT (the expanded survey; Thongmeearkom et al., in preparation), as well as other pulsar surveys in general. The outline of this report is as follows: Section \ref{S:survey_UHF} describes the survey strategy, Section \ref{S:result_UHF} presents the discoveries, localisation, and timing, Section \ref{S:discussion_UHF} discusses the characteristics of the discovered pulsars and compares the \textit{L}-band and UHF surveys, and Section \ref{S:conclusion_UHF} provides a summary and future work plans.

\section{Survey Properties}
\label{S:survey_UHF}
\subsection{MeerKAT and TRAPUM Processing Infrastructure}
The MeerKAT radio telescope is an interferometer array of 64 radio dishes, each with a diameter of 13.5 m, located in the Karoo, a desert region in the Northern Cape, South Africa. This region experiences very low levels of radio frequency interference (RFI). MeerKAT is equipped with three types of receivers: the Ultra High Frequency (UHF) receiver (544--1088\,MHz), the \textit{L}-band receiver (856--1712\,MHz), and the \textit{S}-band receiver (1750--3499\,MHz). At the time of this survey, only the \textit{L}-band and UHF receivers were available. Full technical details can be found in \citet{Booth2009+MeerKAT} and \citet{Jonas2016+MeerKAT}. In this paper, we present discoveries from the shallow survey conducted with the UHF receiver.

In this survey, we planned to use all the antennas to create coherent beams with high sensitivity and high resolution to cover our targets. However, not all antennas were available depending on the observing date. As a result, 56 or 60 antennas were utilised during the observations. These numbers respect the additional constraint of being divisible by four, as required by the beam-forming algorithm. This process is performed by the Filterbanking BeamFormer User Supplied Equipment (FBFUSE), a 32-node, GPU-based software beamformer developed by the Max Planck Institute for Radio Astronomy \citep[e.g.,][]{Barr2018+FBFUSE,Chen2021+Beamformer}. We used FBFUSE to form multiple coherent beams to cover the localisation regions for sources from the \textit{Fermi} catalogue. For each coherent beam, the channelised data were recorded onto a distributed file system accessible from the Accelerated Pulsar Search User Supplied Equipment (APSUSE). The number of beams was limited by the data rate that APSUSE could record \citep[see Section 2.1 of][for more details]{Clark2023+Lband}. For this UHF survey, the 4096-channel MeerKAT F-engine was used to record all spectra for all targets. Therefore, the native time resolution (4096/544 MHz = 7.529 $\mu$s) was down sampled by a factor of 16 to provide 120\,$\mu$s time resolution. Using this setup, APSUSE could record up to 480 coherent beams to cover the \textit{Fermi}-LAT localised region of each source, though we reduced this number to 276 for most observations to ensure stability on the recording file-system. An example of beam tiling can be found in Figure 1 of \citet{Clark2023+Lband}; note that the UHF tiling has fewer beams, but these are wider than in the \textit{L}-band tiling due to the longer wavelength.

\subsection{Observing Strategy}

As reported by \citet{Bailes2020+MeerTIME}, the full MeerKAT array offers a high gain ($G = 2.8\,\rm K\,Jy^{-1}$) and a low system temperature\footnote{\url{https://skaafrica.atlassian.net/wiki/spaces/ESDKB/pages/277315585/MeerKAT+specifications}} ($T_{\rm sys} = 25\,\rm K$). Using the radiometer equation \citep{PSRHandbook} and sky temperature ($T_{\rm sky}$) estimates from the 408 MHz all-sky map \citep{Haslam1982,Remazeilles2015}, we calculated the minimum detectable flux density ($S_{\rm min}$) for each target, 
\begin{equation}
S_{\rm min} = \frac{ {\rm S/N} \, (T_{\rm sys} + T_{\rm sky}) }{ \beta \, G \sqrt{n_{\rm pol} t_{\rm obs} {\rm BW}} } \sqrt{\frac{W}{P-W}}  \,,
\label{E:radiometerpulsar_UHF}
\end{equation}
where S/N is the signal-to-noise ratio, $\beta$ is the correction factor due to digitisation, $n_{\rm pol}$ is the number of polarisations, BW is the bandwidth, $t_{\rm obs}$ is the observing time, $P$ is the pulsar's rotation period, and $W$ is the pulse width. Based on these calculations \citep{Thongmeearkom2021}, we selected a 10-minute integration time, which provides a sensitivity comparable to that achieved by previous surveys with other telescopes \citep[e.g.,][]{Keith2011+Parkes,Ransom2011+GBT,Camilo2015+Parkes}. To further increase our chances of detecting faint or variable pulsars, each target was observed at two separate epochs. We expected that UHF observations would allow us to re-detect discoveries from the \textit{L}-band survey and to discover dimmer pulsars, since pulsars generally have steep spectra making them brighter at lower frequencies \citep{Jankowski2018}. However, searching at UHF may present challenges for some MSPs, particularly eclipsing systems, as eclipses at higher frequencies are usually shorter and less opaque than those at lower frequencies \citep{Polzin2020}. Large dispersion measure (DM) smearing in UHF is more likely to broaden the pulse of a very fast pulsar compared to the DM smearing in \textit{L} band (0.26 ms vs. 0.10 ms for DM = 100 pc cm$^{-3}$ within a channel). In addition, a pulsar's pulse profile tends to be broader at lower frequencies \citep[e.g.,][]{Jankowski2018,Johnston2020+PTA}. Despite these limitations, UHF remains a promising frequency range to explore, and in Section \ref{S:compare} we present a detailed comparison of \textit{L} band and UHF based on the results of our survey.

\subsection{Search Pipeline}
\label{S:search_pipeline}
We used the \texttt{Peasoup}\footnote{\url{https://github.com/ewanbarr/peasoup}} GPU-accelerated pulsar search code to search all coherent tied-array and incoherent beams. This programme utilises a Fast Fourier Transform (FFT)-based acceleration search using time-domain resampling with incoherent harmonic summing \citep{Morello2019+HTRUrepro,Barr2020+peasoup}. We selected a maximum DM of 1000 pc cm$^{-3}$ with a variable step size of 0.03--0.2 pc cm$^{-3}$, following the \texttt{DDplan.py} routine from \texttt{PRESTO}\footnote{\url{https://github.com/scottransom/presto}} \citep{Ransom2011+PRESTO}. This range covers the maximum predicted DM along the line of sight for each target, as predicted by the YMW16 electron density model \citep{YMW16}. We then chose a maximum acceleration of $|a| = 50 \rm\,m\,s^{-2}$. This acceleration range covers the maximum acceleration expected from redback systems (\citealt{Strader2019+RBs}, see calculation in \citealt{Thongmeearkom2024+RBs}). After the search, the resulting candidate were folded using the \texttt{PulsarX} software\footnote{\url{https://github.com/ypmen/PulsarX}} \citep{Men2023+PulsarX} and then scored by the Pulsar Image-based Classification System (PICS) machine learning classifier \citep{Zhu2014+PICS}. The last step is to investigate the pulsar candidates using a viewing programme such as \texttt{PSR\_labeller}\footnote{\url{https://github.com/01tinn/PSR_labeller}} or \texttt{Candyjar}\footnote{\url{https://github.com/vivekvenkris/CandyJar}}.

\subsection{Target Selection}
The target list was created using \textit{Fermi}-LAT unidentified gamma-ray sources with pulsar-like gamma-ray spectra (high curvature significance and low variability) from the 4FGL catalogue \citep{4FGL}, which was the latest \textit{Fermi}-LAT catalogue at the time the UHF survey commenced. Machine-learning classification techniques are an effective way to distinguish pulsar candidates from the population of unassociated sources, with successful results from many studies \citep{Lee2012+GMM,SazParkinson2016+ML,Luo2020+ML,Finke2021+DNN}. However, the visual inspection method has also proven effective in identifying pulsar candidates \citep{Camilo2015+Parkes}. These selection criteria stem from the fact that gamma-ray pulsars have higher curvature significance and lower variability compared to other classes of sources, such as AGNs.

We applied the Random Forest technique of \citet{Clark2023+Lband} to each unassociated source in the 4FGL catalogue \citep{4FGL}, which estimates the probability that a given source is a pulsar. Using this pulsar probability, we created a list of ranked unassociated sources. We then removed sources with declinations greater than 20$\degr$, due to the observing limits imposed by MeerKAT’s location, and sources with Galactic latitudes within $\pm 10\degr$ of the plane, in order to avoid the Galactic plane since this region of the sky was already covered by the MMGPS project \citep{Padmanabh2023+MGPS}. Furthermore, we excluded sources whose 95\% confidence regions (localisation from \textit{Fermi}-LAT) had semi-major axes exceeding 7 arcminutes, as this is the maximum area that can be covered in a single pointing using hundreds of coherent beams (e.g., 276, 288, or 480) with 50\% sensitivity overlap. For the first UHF epoch, we removed 13 confirmed pulsars from the first epoch at \textit{L} band \citep{Clark2023+Lband} and replaced them with 13 new sources. For the second epoch, we removed eight candidates from the first epoch of UHF as they have been discovered with MeerKAT and other telescopes (e.g., PSR~J1947$-$1120 discovered with GBT; \citealp{Strader2025+J1947}). A list of UHF observations from the first epoch and the second epoch can be found in Tables \ref{T:source_list_1} and \ref{T:source_list_2}, respectively.

\section{Results}
\label{S:result_UHF}
\begin{figure*}
    \centering
	\includegraphics[width=\textwidth]{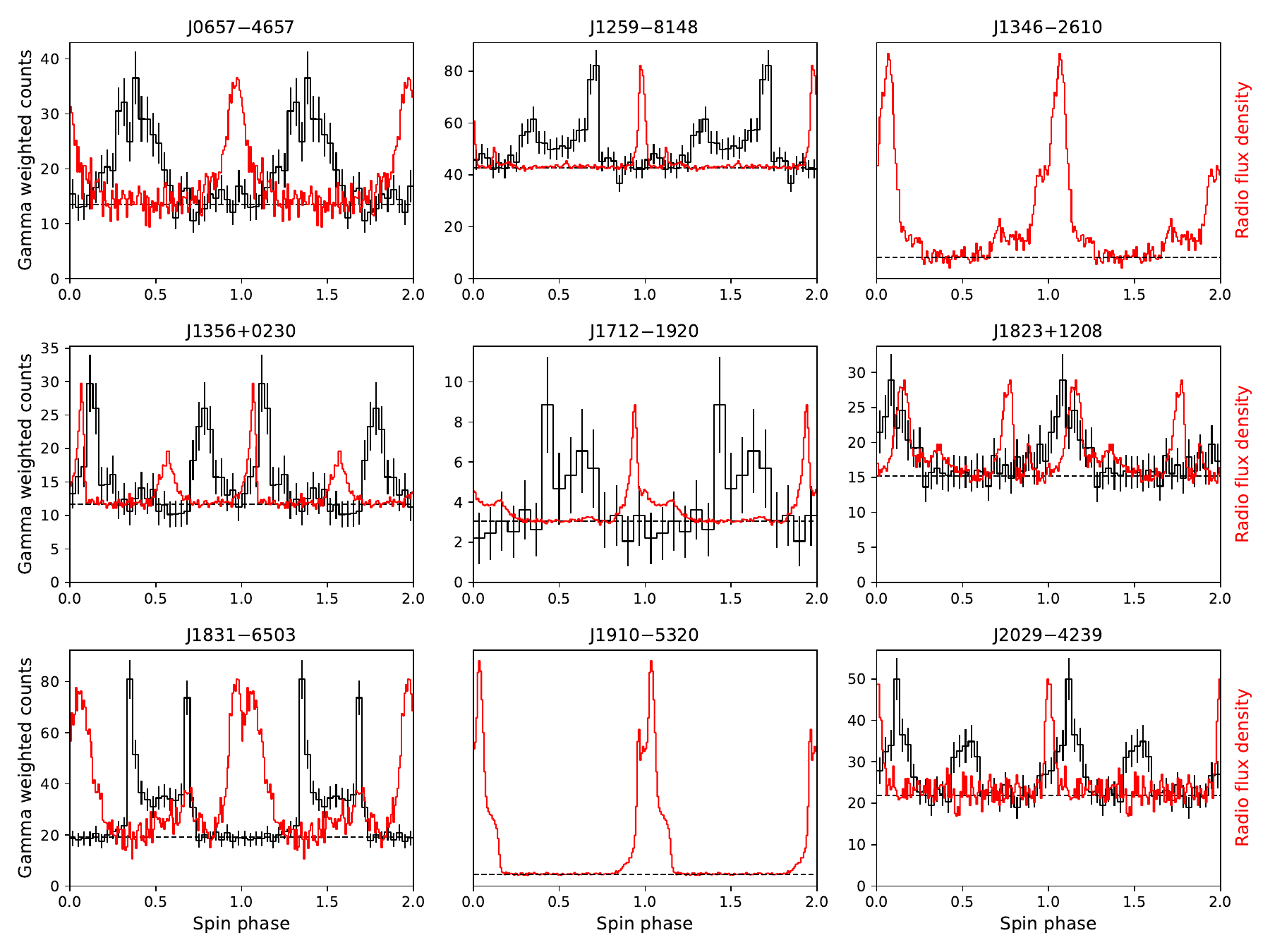}
    \caption{Radio (red) pulse profiles for all nine MSPs with phase-aligned gamma-ray (black) profiles for seven of them, showing two identical rotations for clarity. The dashed line represents the gamma-ray background level, estimated from photon weights as $b=\sum_i w_i(1-w_i)/n_{\rm bins}$. The radio profiles are shown in arbitrary units, scaling the amplitude to match the highest gamma-ray peak and moving the baseline flux level to the gamma-ray background level.}
    \label{F:9_MSPs}
\end{figure*}

\begin{figure*}
    \centering
	\includegraphics[width=\textwidth]{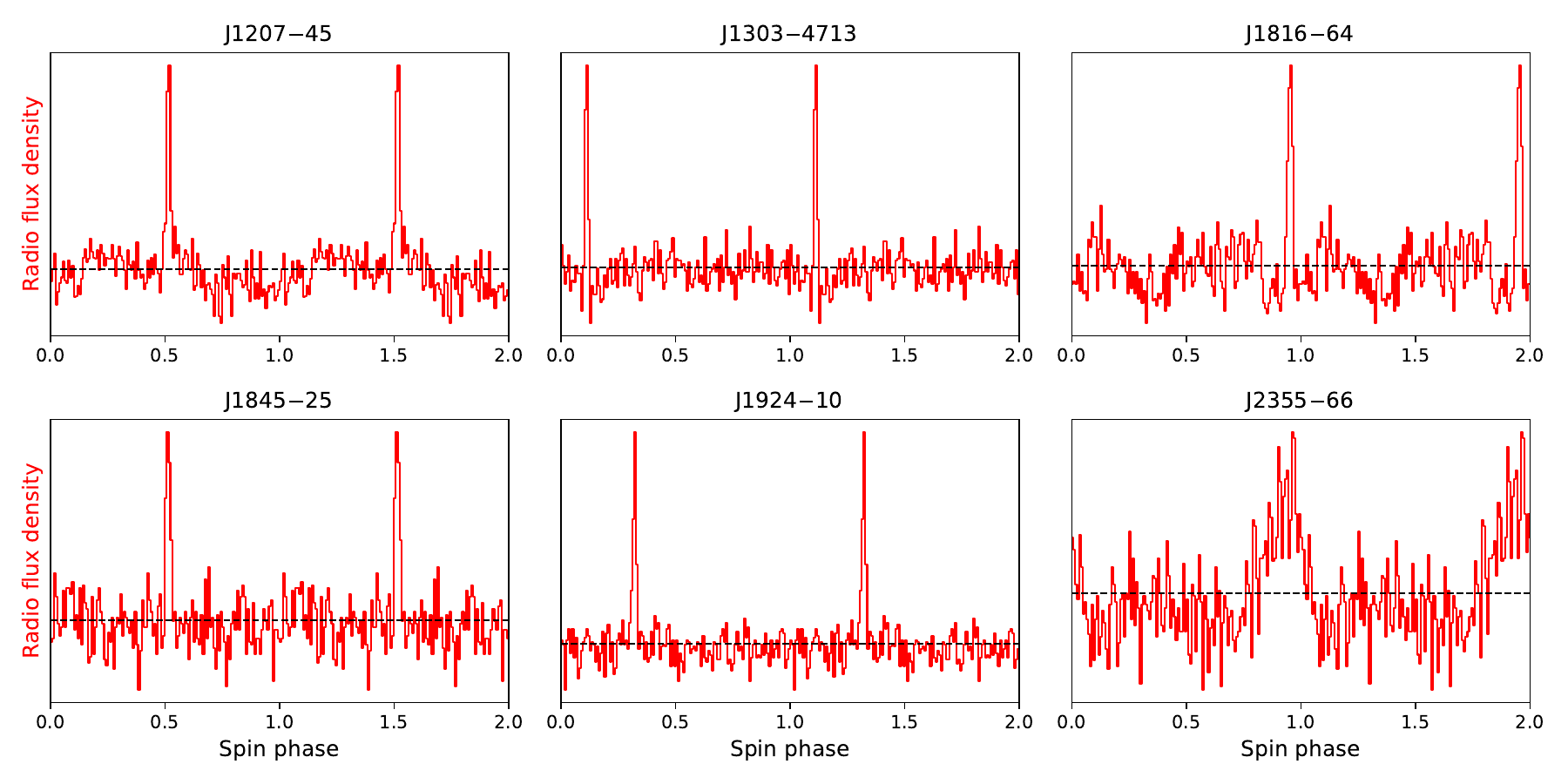}
    \caption{Radio pulse profiles for the six slow pulsars, showing two identical rotations. The profiles display the radio flux density in arbitrary units. The dashed line represents the baseline, estimated as the median of the off-pulse region.}
    \label{F:6_slow}
\end{figure*}

We discovered nine MSPs and six slow pulsars. The pulse profiles of the MSPs are shown in Figure~\ref{F:9_MSPs}, while the profiles of the slow pulsars are displayed in Figure~\ref{F:6_slow}.

Seven of these nine new MSPs are confirmed to be counterparts of the targeted \textit{Fermi}-LAT sources through the detection of gamma-ray pulsations (see Section \ref{S:gamma-ray pulsations}). These MSPs have spin periods ranging from 1.81 to 5.31\,ms and DMs from 10 to 126\,pc cm$^{-3}$. One of these MSPs, PSR~J2029$-$4239, is an isolated pulsar, while the remaining eight are binary systems. All were detected in coherent tied-array beams. At the time we conducted this survey, \citet{Au2023+J1910} identified that PSR~J1910$-$5320 is a redback candidate with an orbital period $P_{\rm b}$ of 4.28\,h, using Catalina Real-Time Transient Survey \citep[CRTS;][]{Djorgovski2011+CRTS} for optical light curves and the Southern Astrophysical Research (SOAR) telescope with the Goodman spectrograph \citep{Clemens2004+SOAR} for spectroscopy (see Section \ref{S:spider_pulsars} and \citealp{Dodge2024+J1910} for more details). Additionally, \citet{Tian2025+MeerTRAP} detected single pulses from an unknown Galactic source on two epochs in the incoherent beam from the same observations where we detected PSR~J0657$-$4657, under the name MTP0066.

In contrast, five of the six slow pulsars (see Section \ref{S:slow_pulsars}) were detected in the incoherent beam, which covers a much wider region (with FWHM of $1.9\degr$ for MeerKAT's 13.5-m antennas at the central frequency of 816\,MHz), while one pulsar (PSR~J1303$-$4713) was discovered in a coherent beam. This source was independently detected by \citet{Tian2025+MeerTRAP} as MTP0075.

\subsection{Localisation}
\label{S:localisation}

By using the S/N values from the detection in the coherent beam and adjacent beams, a precise sky position with two-sigma localisation ellipses of less than 10 arcseconds can be estimated. The localisation method is described in \citet{SeeKAT}, which is based on the concept of \citet{Obrocka2015+reloc}. We used the \texttt{SeeKAT} package\footnote{\url{https://github.com/BezuidenhoutMC/SeeKAT}} on each confirmed MSP to refine each pulsar position \citep[see Section 3.2 of][for an example]{Clark2023+Lband}. This improved positional accuracy helps us identify potential counterparts at other wavelengths and also helps in determining phase-connected timing solutions. 

\subsection{Radio Timing}
\label{S:timing_UHF}
Following the framework of \citet{Burgay2024+timing} at \textit{L} band, we initiated a follow-up timing campaign for the nine newly discovered UHF MSPs. For pulsars visible from the Northern Hemisphere, the campaign was conducted using the Nan\c cay Radio Telescope (NRT) and the Effelsberg telescope, while for those in the Southern Hemisphere, we used the Murriyang (Parkes) telescope equipped with the Ultra-Wideband Low receiver \citep[UWL;][]{Hobbs2020+UWL}. We decided on 1- to 2-h observations for follow-up sessions depending on the S/N of the discovered pulsars at MeerKAT. For fainter sources, we performed 5- or 10-minute MeerKAT observations. The details of the timing observations for each source can be found in Table~\ref{T:timing_uhf}. In order to obtain an initial orbital solution, we also performed multiple 5-minute MeerKAT observations with a pseudo-logarithmic cadence. In this campaign, we used the Pulsar Timing User Supplied Equipment \citep[PTUSE;][]{Bailes2020+MeerTIME}, which provides coherent de-dispersion.

Prior to obtaining a phase-connected timing solution, we obtained a preliminary timing solution for the eight binary MSPs by estimating the binary parameters, including the orbital period ($P_{\rm b}$), the epoch of ascending node ($T_{\rm asc}$), and the projected semi-major axis ($x$), from the observed modulation of barycentric spin periods using the standard circular-orbit relation. Subsequently, we used \texttt{PRESTO}\footnote{\url{https://github.com/scottransom/presto}}’s \texttt{fit\_circular\_orbit.py} \citep{Ransom2011+PRESTO} to fit a sinusoidal modulation to the observed barycentric spin periods from multiple observations, without requiring phase alignment across observations. We then used the software \texttt{tempo2}\footnote{\url{https://bitbucket.org/psrsoft/tempo2/src/master/}} \citep{tempo2} to obtain a phase-connected timing solution by flattening the residuals of the times-of-arrival (ToAs). We also added constant time offsets, referred to as ``JUMPs'' to account for time delays between distinct observatories and different instruments (e.g., APSUSE, PTUSE, and other telescopes). For some binary pulsars, we employed \texttt{Dracula}\footnote{\url{https://github.com/pfreire163/Dracula}} \citep{Freire2018+DRACULA} as an additional step for determining the global rotation count and finding phase-connected timing solutions.

\begin{table*}
  \centering
  \caption{Parameters of the radio timing campaign of the nine newly discovered MSPs. The columns report the pulsar name, the data span of the radio timing solution, the total duration of the observations used to obtain the timing solution, and the number of ToAs that have been extracted from the observation, where each ToA corresponds to an average pulse profile over a time segment containing many individual pulses. The telescopes, from left to right, are MeerKAT (MK), Parkes (PKS), Nan\c{c}ay (NRT), and Effelsberg (EFF). We do not include observations with no detections in the total counts of this table.}
  \label{T:timing_uhf}
  \begin{tabular}{c @{\hspace{1 truecm}} c @{\hspace{1 truecm}} cccc @{\hspace{1 truecm}} cccc @{\hspace{1 truecm}} cccc}
    \hline
    PSR & Data span & \multicolumn{4}{c}{N. obs} & \multicolumn{4}{c}{$t_{\rm obs}$} & \multicolumn{4}{c}{N. ToA} \\
        & (MJD)     & \multicolumn{4}{c}{}       & \multicolumn{4}{c}{(h)}  & \multicolumn{4}{c}{}       \\
        \hline
        &           & MK & PKS & NRT & EFF & MK & PKS & NRT & EFF & MK & PKS & NRT & EFF  \\
    \hline
J0657$-$4657 & 59767 -- 60093 & 8 & 16 & -- & -- & 1.4 & 22.0 & --  &  --  & 46  & 75  & -- & -- \\
J1259$-$8148 & 59536 -- 59846 & 26 &  12 & -- & -- &  2.5 & 15.5 & -- &  --  & 317  & 446  & -- & -- \\
J1346$-$2610 & 60084 -- 60185 & 13 & -- & -- & -- & 1.0 & -- & --  &  --  & 192 & --  & -- & -- \\
J1356+0230 & 59674 -- 59851 & 11 &  -- & 1 & -- & 2.3 &  -- & 0.9  & -- & 133 & --  & 5 & --  \\ % ntoa 244 adding simultaneous APSUSE's
J1712$-$1920 & 60084 -- 60258 & 11 & 21 & 5 & 7 & 1.1 & 26.6 & 3.8 &  10.7  & 389 & 205 & 44 & 74 \\ % 439 inclu APSUSE simult
J1823+1208 & 59674 -- 60125 & 27 & -- & 3 & -- & 6.0 & -- & 2.5  & -- & 179 & -- & 26 & -- \\
J1831$-$6503 & 60084 -- 60185 & 11 & -- & -- & -- & 2.8 & -- & -- &  --  & 113 & -- & -- & -- \\ % 481 inclu APSUSE's simult
J1910$-$5320 & 59759 -- 60028 & 15 & 10 & -- & -- & 1.1 & 7.3 & --  &  --  & 864 & 246 & -- & -- \\ % 383 including APSUSE's simult
J2029$-$4239 & 59496 -- 59837 & 4 & 6 & -- & -- & 0.4 & 3.3 & --  & -- & 44 & 52 & -- & -- \\ % 404 including APSUSE's simult
    \hline
  \end{tabular}
\end{table*}

\subsection{Spider Pulsars}
\label{S:spider_pulsars}
We found that PSRs~J1259$-$8148, J1346$-$2610, J1356+0230, J1712$-$1920, J1831$-$6503, and J1910$-$5320 are possible spider binaries since their spin periods \citep[$P \lesssim\,5$ ms;][]{Strader2019+RBs} and their orbital periods \citep[$P_{\rm b} \lesssim\,1$ d;][]{Swihart2022+J1408} lie in the range of spider pulsars. Afterwards, we became more confident in their classifications by measuring their companion masses from radio and gamma-ray timing (see Tables \ref{T:timing} and \ref{T:timing2} in this work, and Table\,1 of \citealt{Dodge2024+J1910}). In addition, the redbacks PSRs~J1346$-$2610, J1712$-$1920, and J1910$-$5320, along with the black widows PSRs~J1356+0230 and J1831$-$6503, all show evidence of radio eclipses, a characteristic signature of spider pulsars. Although PSR~J1259$-$8148 shows no sign of radio eclipses, its compact orbit and very low companion mass still suggests it is a black widow rather than a He~white dwarf system.

We can interrogate the companion to the neutron star in spider systems at optical wavelengths. Spider companions show characteristic sinusoidal optical variability, driven by two key mechanisms. Firstly, due to compact orbits and tidal locking, the hemisphere of the companion facing the pulsar is heavily irradiated by the pulsar wind. This irradiated face (the dayside) becomes hotter and thus brighter than the unirradiated face (the nightside), producing a single-peaked light curve \citep[e.g.,][]{Callanan1995+B1957,Breton2013}, with the peak occuring when the irradiated face is presented to the observer. Secondly, the gravitational distortion of the companion leads to ellipsoidal modulation, producing a double-peaked light curve. Both effects are present in most systems, but their relative strength determines the observed morphology. In black widows, strong irradiation often dominates and masks the weaker ellipsoidal component, while in redbacks, the larger and more distorted companions make the ellipsoidal modulation more prominent. However, some strongly irradiated redbacks still show single-peaked, asymmetric double-peaked, or irregular light curves. For example, PSR~J1910$-$5320 has been observed using ULTRCAM/NTT and SOAR/Goodman to obtain optical light curves and spectroscopy \citep{Dodge2024+J1910}. The study found that the optical light curves of PSR~J1910$-$5320 display a single-peaked light curve and found that this system has a companion mass of 0.28$\,M_\odot$ and a neutron star mass of 1.4$\,M_\odot$, confirming its classification as an irradiated redback.

To characterise the companion in the system, PSR~J1346$-$2610 was observed over two nights, the 18th and 19th of March 2023, with ULTRACAM \citep{Dhillon2007}. ULTRACAM, mounted on the European Southern Observatory's New Technology Telescope (NTT) at the La Silla observatory, offers high time-resolution multi-band simultaneous photometry, in our case providing $r_s$, $g_s$, and $u_s$ light curves. Here, the subscript $s$ denotes ``Super-SDSS'' filters which cover the same wavelength range as the traditional SDSS filters, but with a higher throughput \citep{Dhillon2021}. The details of the observations are shown in Table \ref{t:optical obs}. The data were reduced using the HiPERCAM pipeline \citep{Dhillon2021} following the typical process: de-biasing, flat-fielding and aperture selection. For the $r_s$ and $g_s$ filters, conversion from instrumental to calibrated flux required 15 nearby sources whose magnitudes are within the Pan-STARRS1 survey catalogue \citep{PS1,PS1DB} to be extracted alongside the target, with the same reference stars used for each night of data. As Pan-STARRS1 uses standard SDSS ($g^\prime$$r^\prime$$i^\prime$$z^\prime$) filters, as opposed to the HiPERCAM super-SDSS filters ($u_s$$g_s$$r_s$), 
we transformed the listed source magnitudes using the colour-transforms listed in \citet{Brown2022}. Pan-STARRS1 does not include a $u^\prime$ filter, thus this simple calibration could not be used for our $u_s$ data. Instead, we determined the instrumental zeropoint by observation of the standard stars GD 108 (on the first night) and GD 71 (on the second night). The reference aperture magnitudes could then be bootstrapped according to this zeropoint, and a consistent $u_s$ reduction between nights was obtained. 

\begin{figure}
  \centering
  	\includegraphics[width=.5\textwidth]{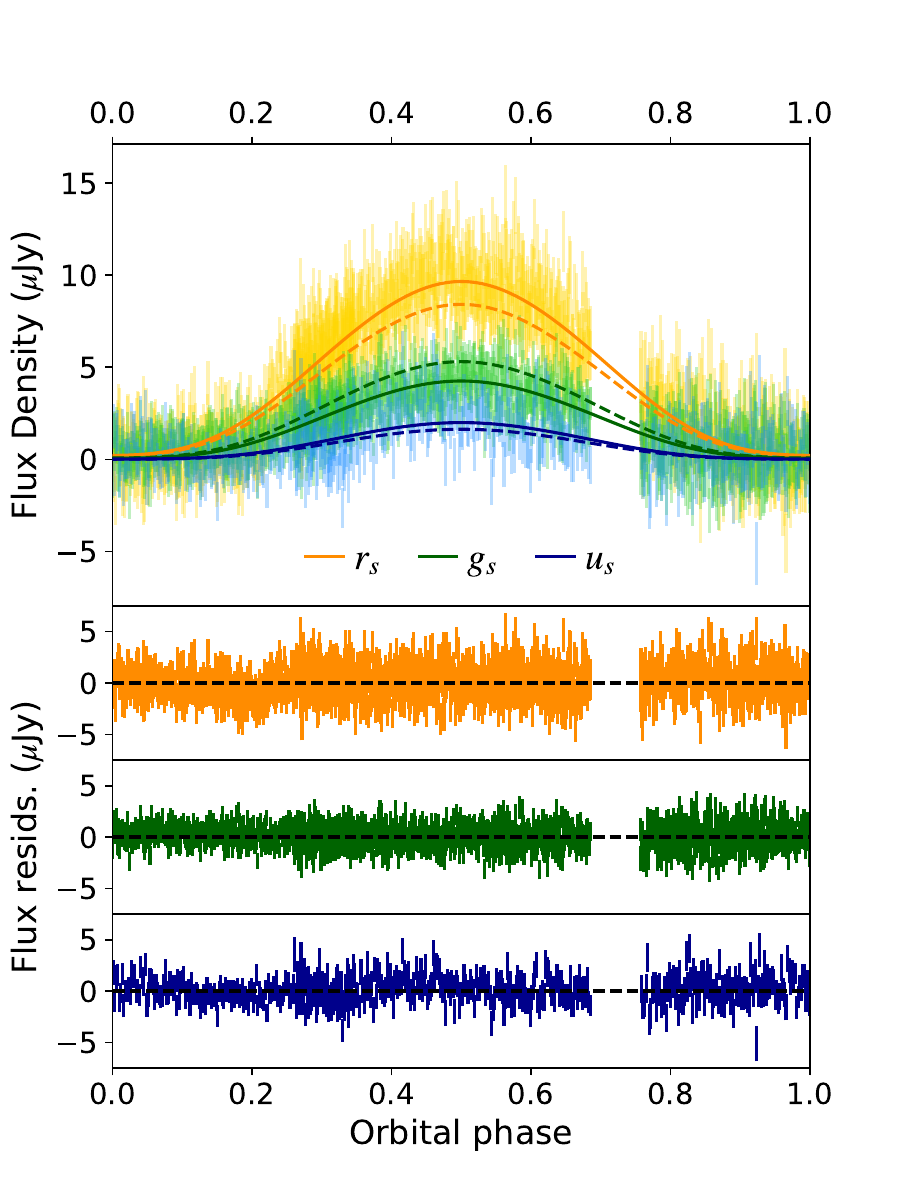}
        \caption{Optical light curves (points), \texttt{Icarus} model fits (solid lines) and resiudals (bottom panels) for PSR~J1346$-$2610 from ULTRACAM data.}
    \label{F:model_1346}
\end{figure}

We found a single-peaked light curve (Figure \ref{F:model_1346}). As described previously, this feature is more commonly seen in black widows, although some redbacks also exhibit such profiles. A short orbital period of $\sim$3.8 h supports this black widow classification, as they tend to have more compact orbits. However, the \texttt{Icarus} (see \citealp{Breton2012} for details) optical modelling results presented in Table \ref{T:j1346 modelling} strongly suggest a redback, with a companion mass of $0.23\,M_\odot$, albeit an unusually cold example. Whilst, given the lack of spectroscopy, the component masses cannot be precisely estimated, we can marginalise over the neutron star mass with a loose top-hat prior allowing for any neutron star mass in the range $1.2\,M_\odot < M_\text{NS} < 2.5\,M_\odot$. 

By averaging the surface element temperatures over the night and daysides, (phases 0.25 and 0.75, respectively), we can characterise the temperatures of each face. The nightside temperature is low ($T_N = 2300\,\rm K$) in comparison with other field redbacks \citep{Simpson2025}, consistent with the unusual faintness of the source. 
The distance determined from our modelling ($d\sim5.3$\,kpc) is also further than the expectation given the DM ($d\sim1.2$\,kpc). This discrepancy exceeds the nominal precision of electron density models, suggesting that the electron density along the line of sight has been overestimated. However, in this direction the DM quickly plateaus thus our modelled distance is not inconsistent.

\begin{table*}
	\caption{Timing solutions for eight newly discovered MSPs. Timing parameters are obtained from joint radio and gamma-ray timing, as described in Sections \ref{S:timing_UHF} and \ref{S:gamma-ray pulsations}. PSR~J1346$-$2610 has only a radio timing solution, marked with an asterisk (*), due to the absence of gamma-ray pulsations. The minimum companion mass ($M_{\rm c,min}$) was calculated from the mass function, assuming a neutron star mass of $M_\text{NS}$ = 1.4 $M_{\odot}$ and an inclination of $i = 90\degr$. We compute derived parameters from the intrinsic period derivative ($\dot{P}_{\rm int}$). For pulsars with measured proper motions, $\dot{P}_{\rm int}$ is obtained by correcting the observed period derivative for the Shklovskii effect \citep{Shklovskii1970} and Galactic acceleration \citep{Damour1991+Pdot}, assuming YMW16 distances \citep{YMW16}. For pulsars without measured proper motions, we apply only the correction for Galactic acceleration. $95\%$ upper limits on parameters are marked with a dagger ($\dagger$). Timing parameter (.par) files, following the standard \texttt{tempo} format \citep{tempo2}, are available as supplementary material accompanying this paper.}
	\label{T:timing}
	\centering
    \renewcommand{\arraystretch}{1.25}
    \begin{tabular}{lcccc}
		\hline
		  Parameter & PSR~J0657$-$4657 & PSR~J1259$-$8148 & PSR~J1346$-$2610* & PSR~J1356+0230\\
		\hline
        \multicolumn{5}{c}{Timing parameters}\\
        \hline
		  R.A., $\alpha$ (J2000) & $06^{\rm h}57^{\rm m}22\fs0545(3)$ & $12^{\rm h}59^{\rm m}31\fs5832(4)$ & $13^{\rm h}46^{\rm m}03\fs0270(6)$ & $13^{\rm h}56^{\rm m}37\fs1262(2)$ \\
        Decl., $\delta$ (J2000) & $-46\degr57\arcmin52\farcs611(5)$ & $-81\degr48\arcmin51\farcs3752(4)$ & $-26\degr10\arcmin18\farcs73(2)$ & $+02\degr30\arcmin30\farcs441(7)$ \\
        Proper motion in R.A., $\mu_{\alpha}\cos\delta$ (mas\,yr$^{-1}$) & 4.1 $\pm$ 3.1 & $-$5.9 $\pm$ 0.5 & -- & $-$17.3 $\pm$ 2.1 \\
        Proper motion in Decl., $\mu_{\delta}$ (mas\,yr$^{-1}$) & 2.4 $\pm$ 3.2 & 6.3 $\pm$ 0.3 & -- & $-$9.2 $\pm$ 5.2\\
        Dispersion measure, DM (pc $\rm cm^{-3}$) & 126.059(8) & 44.314(7) & 20.145(4) & 17.787(1)\\
        Spin frequency, $\nu$ (Hz) & 253.4542797024(2) & 479.4813415355(4) & 361.2127648726(5) & 353.7362124742(2)\\
        Spin-down rate, $\dot{\nu}$ (Hz $\rm s^{-1}$) & $-4.609(9)\times 10^{-16}$ & $-7.67(2)\times 10^{-16}$ & $-7.0(2)\times 10^{-16}$ & $-9.781(7)\times 10^{-16}$\\
        Orbital period, $P_{\rm b}$ (d) & 84.0524734(6) & 0.185680580(2) & 0.1577240538(2) & 0.141879279(2)\\
        Projected semi-major axis, $x$ (lt s) & 36.002523(8) & 0.084884(5) & 0.323356(6) & 0.022654(3)\\
        Epoch of ascending node, $T_{\rm asc}$ (MJD) & 59825.373197(4) & 59635.516072(2)  & 60084.9496878(4) & 59674.080619(2)\\
        1st Laplace-Lagrange parameter, $\epsilon_1$ & $8.55(6)\times 10^{-5}$ & --  & -- & -- \\
        2nd Laplace-Lagrange parameter, $\epsilon_2$ & $-2.38(4)\times 10^{-5}$ & --  & -- & -- \\
        \hline
        \multicolumn{5}{c}{Derived parameters}\\
        \hline
        Spin period, $P$ (ms) & 3.945484768196(3) & 2.0855868901959(2) & 2.768451442608(4) & 2.826965305603(1)\\
        Spin period derivative, $\dot{P}$ (s s$^{-1}$) & $7.17(1)\times10^{-21}$ & $3.3360(8)\times10^{-21}$ & $5.4(1)\times10^{-21}$ & $7.817(5)\times10^{-21}$\\
        Intrinsic period derivative, $\dot{P}_{\rm int}$ (s s$^{-1}$) & $7.31\times10^{-21}$ & $3.08\times10^{-21}$ & $5.76\times10^{-21}$ & $3.77\times10^{-21}$ \\
        Spin-down power, $\dot{E}$ (erg $\rm s^{-1}$) & $4.7\times10^{33}$ & $1.3\times10^{34}$  & $1.1\times10^{34}$ & $6.6\times10^{33}$\\
        Surface magnetic field strength, $B_{\rm S}$ (G) &  $1.7 \times 10^8$ & $8.1\times10^7$ & $1.3 \times 10^8$ &  $1.0 \times 10^8$\\
        Light-cylinder magnetic field strength, $B_{\rm LC}$ (G) & $2.6 \times 10^4$ & $8.2 \times 10^4$ & $5.6 \times 10^4$ & $4.3 \times 10^4$\\
        Minimum companion mass, $M_{\rm c,min}$ ($M_{\odot}$) & 0.27 & 0.03 & 0.15 & 0.01\\
        Eccentricity, $e$ & $8.88^{+0.03}_{-0.08} \times 10^{-5}$ & -- & -- & -- \\
        %Longitude of periastron, $\omega$ (deg) & -- & -- & -- & -- \\
        Proper motion, $\mu$ (mas\,yr$^{-1}$) & $<11^{\dagger}$ & 8.6 $\pm$ 0.4 & -- & 19.6 $\pm$ 3.1 \\
        Distance from YMW16 (kpc) & 0.5 & 1.4 & 1.2 & 1.8\\
        Distance from NE2001 (kpc) & $>50.0$ & 1.5 & 0.9 & 1.2\\
        System class & pulsar--white dwarf & black widow & redback & black widow\\
		\hline
	\end{tabular}
\end{table*}

\begin{table*}
	\caption{Continuation of Table~\ref{T:timing}.}
	\label{T:timing2}
	\centering
    \renewcommand{\arraystretch}{1.25}
   \begin{tabular}{lcccc}
		\hline
		   & PSR~J1712$-$1920 & PSR~J1823+1208 & PSR~J1831$-$6503 & PSR~J2029$-$4239\\
		\hline
        \multicolumn{5}{c}{Timing parameters}\\
        \hline
		  R.A., $\alpha$ (J2000) & $17^{\rm h}12^{\rm m}01\fs4376(1)$ & $18^{\rm h}23^{\rm m}18\fs3158(3)$ & $18^{\rm h}31^{\rm m}04\fs3241(3)$ & $20^{\rm h}29^{\rm m}35\fs2006(9)$ \\
        Decl., $\delta$ (J2000) & $-19\degr20\arcmin24\farcs52(2)$ & $+12\degr08\arcmin38\farcs715(9)$ & $-65\degr03\arcmin13\farcs984(4)$ & $-42\degr39\arcmin35\farcs99(2)$ \\
        Proper motion in R.A., $\mu_{\alpha}\cos\delta$ (mas\,yr$^{-1}$) & -- & 9.4 $\pm$ 4.1 & $-$5.5 $\pm$ 0.3 & 5.6 $\pm$ 2.8 \\
        Proper motion in Decl., $\mu_{\delta}$ (mas\,yr$^{-1}$) & -- & $-$5.7 $\pm$ 7.9 & $-$2.7 $\pm$ 0.6 & 2.8 $\pm$ 6.0\\
        Dispersion measure, DM (pc $\rm cm^{-3}$) & 49.4049(2) & 40.687(2) & 25.715(9) & 10.170(2)\\
        Spin frequency, $\nu$ (Hz) & 552.2592655671(2) & 192.08456416966(9) & 539.62314331721(4) & 188.1764568362(1) \\
        Spin-down rate, $\dot{\nu}$ (Hz $\rm s^{-1}$) & $-3.03(1)\times10^{-15}$ & $-3.287(6)\times10^{-16}$ & $-5.749(2)\times10^{-16}$ & $-3.339(5)\times10^{-16}$\\
        Orbital period, $P_{\rm b}$ (d) & 0.183439029(3) & 7.79893996(4) & 0.157913023(2) & --\\
        Projected semi-major axis, $x$ (lt s) & 0.497066(7) & 6.169776(4) & 0.063017(3) & --\\
        Epoch of ascending node, $T_{\rm asc}$ (MJD) & 60084.88545055(8) & 59676.197511(1) & 60084.926947(3) & --\\
        1st Laplace-Lagrange parameter, $\epsilon_1$ & $(-3.8 \pm17) \times 10^{-6}$ & $9(2)\times 10^{-6}$ & $3(1)\times 10^{-4}$ & -- \\
        2nd Laplace-Lagrange parameter, $\epsilon_2$ & $1.9(9)\times 10^{-5}$ & $2(1)\times 10^{-6}$  & $-1.5(9)\times 10^{-4}$ & -- \\
        OPV amplitude, $h_{\rm opv}$ & -- & -- & -6.7(2) & -- \\
        OPV cutoff frequency, $\nu_{\rm c,opv}$ (yr$^{-1}$) & -- & -- & -1.4(6) & -- \\
        OPV spectral index, $\gamma_{\rm opv}$ & -- & -- & 2(1) & -- \\
        \hline
        \multicolumn{5}{c}{Derived parameters}\\
        \hline
        Spin period, $P$ (ms) & 1.8107437255454(5) & 5.206040393317(2) & 1.8531451298637(1) & 5.314161063573(3)\\
        Spin period derivative, $\dot{P}$ (s s$^{-1}$)  & $9.92(3)\times10^{-21}$ & $8.91(1)\times10^{-21}$ & $1.9742(6)\times10^{-21}$ & $9.43(1)\times10^{-21}$ \\
        Intrinsic period derivative, $\dot{P}_{\rm int}$ (s s$^{-1}$) & $9.87\times10^{-21}$ & $6.57\times10^{-21}$ & $1.89\times10^{-21}$ & $6.95\times10^{-21}$\\
        Spin-down power, $\dot{E}$ (erg $\rm s^{-1}$) & $6.6\times10^{34}$ & $1.8\times10^{33}$ & $1.2\times10^{34}$ & $1.8\times10^{33}$\\
        Surface magnetic field strength, $B_{\rm S}$ (G) & $1.4\times 10^8$ & $1.9\times 10^8$ & $5.9\times 10^7$ & $1.9\times 10^8$ \\
        Light-cylinder magnetic field strength, $B_{\rm LC}$ (G) & $2.1\times 10^5$ & $1.2\times 10^4$ & $8.7\times 10^4$ & $1.2\times 10^4$ \\
        Minimum companion mass, $M_{\rm c,min}$ ($M_{\odot}$) & 0.22 & 0.22 & 0.03 & --\\
        Eccentricity, $e$ & $< 3.8 \times 10^{-5}$$^{\dagger}$ & $< 1.3 \times 10^{-5}$$^{\dagger}$ & $< 5.3 \times 10^{-4}$$^{\dagger}$ & -- \\
        %Longitude of periastron, $\omega$ (deg) & -- & -- & -- & -- \\
        Proper motion, $\mu$ (mas\,yr$^{-1}$) & -- & $<21^{\dagger}$ & 6.1 $\pm$ 0.4 & 18.5 $\pm$ 5.8 \\
        Distance from YMW16 (kpc) & 1.6 & 1.7 & 1.0 & 0.6\\
        Distance from NE2001 (kpc) & 1.4 & 1.9 & 0.9 & 0.5\\
        System class & redback & pulsar--white dwarf & black widow & isolated pulsar\\
		\hline
	\end{tabular}
\end{table*}

\begin{table*}
\caption{ULTRACAM observations of PSR~J1346$-$2610. The exposure time was 10\,s, with only 24 milliseconds of dead time between each frame. To increase signal-to-noise ratio, a single $u_s$ datapoint consists of 3 stacked exposures, resulting in the lower number of $u_s$ datapoints.}
\begin{tabular}{lccccc}
\hline
 Start time (UTC) & Length (h) & Orbital coverage & \multicolumn{3}{c}{n datapoints} \\
 & & & $r_s$ & $g_s$ & $u_s$ \\
 \hline
 2023/03/19 04:06:28 & 2.25 & 0.01 - 0.60 & 755 & 798 & 267 \\
 2023/03/20 04:01:49 & 2.33 & 0.51 - 0.94 & 579 & 573 & 192 \\
 \hline \\
 \label{t:optical obs}
 \end{tabular}
 \end{table*}

\begin{table}
	\caption{Parameter estimation from optical modelling of PSR~J1346$-$2610.}
	\label{T:j1346 modelling}
	\centering
    \renewcommand{\arraystretch}{1.25}
    \begin{tabular}{lc}
        \hline Parameter & Value \\ 
    \hline
    \multicolumn{2}{c}{Fit parameters}\\
    \hline
        Neutron star mass, $M_\text{NS}$ ($M_\odot$) & $1.67^{+0.5}_{-0.4}$ \\
        Reddening, E(B-V) & $0.0507^{+0.003}_{-0.003}$ \\
        Inclination, $i$ ($\degr$) & $49^{+9}_{-3}$ \\
        Filling factor, $f_\text{RL}$ & $0.79^{+0.02}_{-0.03}$ \\
        Base temperature, $T_B$ (K) & $2400^{+800}_{-700}$ \\
        Irradiation temperature, $T_I$ (K) & $6600^{+100}_{-100}$ \\
        Distance, $d$ (kpc) & $5.3^{+0.4}_{-0.4}$ \\
    \hline
    \multicolumn{2}{c}{Derived parameters}\\
    \hline
        Observed temperature: inferior conjunction, $T_\text{Inf}$ (K) & $2700^{+500}_{-300}$ \\
        Observed temperature: superior conjunction, $T_\text{Sup}$ (K) & $5000^{+200}_{-100}$ \\
        Nightside temperature, $T_N$ (K) & $2300^{+800}_{-700}$ \\
        Dayside temperature, $T_D$ (K) & $5300^{+100}_{-100}$ \\
        Volume averaged filling factor, $f_\text{v}$ & $0.93^{+0.02}_{-0.02}$ \\
	Mass ratio, $q$ & $7^{+1}_{-0.9}$ \\
        Companion projected radial velocity, $K_{\rm c}$ (km\,s$^{-1}$) & $330^{+50}_{-40}$ \\
        Companion mass, $M_{\rm c}$ ($M_\odot$) & $0.23^{+0.04}_{-0.04}$ \\ \hline
	\end{tabular}
\end{table}

\subsection{Other MSPs}
\label{S:other_pulsars}

For PSRs~J0657$-$4657 and J1823+1208, the orbital periods are 84.1 and 7.8\,d, with companion masses of 0.28 and 0.23$\,M_{\odot}$, respectively. These masses are consistent with those of low-mass white dwarfs \citep[0.09 -- 0.29$\,M_{\odot}$;][]{Mata2020+J1012,Swihart2022+J1408}. The relatively long orbital periods further suggest a binary system with a He white dwarf \citep[$P_{b} \gtrsim$ 2 d;][]{Tauris2023} rather than a spider binary. In contrast, PSR~J2029$-$4239 appears to be an isolated pulsar, as there is no evidence of orbital acceleration from the radio timing observations or over the $\sim 16.5$ years data span used in gamma-ray timing. This lack of acceleration suggests that it does not have a close companion. Its rapid spin period ($P\sim5.3$\,ms) implies that it was likely recycled in the past through accretion in a binary system, but has since lost or ablated its companion, leaving it isolated.

\subsection{Gamma-ray Pulsations and Timing}
\label{S:gamma-ray pulsations}

All MSPs discovered in this survey were found targeting a \textit{Fermi} unidentified source, thus we performed follow-up searches for gamma-ray pulsations.  Depending on the system, an initial radio timing solution spanning just a few weeks may already be sufficient to guide a successful gamma-ray follow-up search \citep[see, e.g.,][]{Nieder2019+J0952,Ray2022+J1555}.  In more complicated cases, longer radio timing solutions are required to reveal the gamma-ray pulsations \citep[see, e.g.,][]{Thongmeearkom2024+RBs,Burgay2024+timing}.

We analysed these data following LAT standard procedures. We used ``Pass 8'' \texttt{P8R3\_SOURCE\_V3} instrument response functions \citep{Pass8,Bruel2018+P305}. Photons were selected based on their energy as described in the 4FGL-DR4 catalogue \citep{4FGLDR4}. We assigned probability weights to these photons derived from spectral and spatial models of the gamma-ray sky \citep{Bickel2008+weights,Kerr2011,Bruel2019+weights}, following the \texttt{gtsrcprob} routine with source positions and spectra from 4FGL DR4 \citep{4FGLDR4}. The sky model used the \texttt{iso\_P8R3\_SOURCE\_V3\_v1.txt} isotropic diffuse emission model\footnote{\url{https://fermi.gsfc.nasa.gov/ssc/data/access/lat/BackgroundModels.html}} and the rescaled\footnote{\url{https://fermi.gsfc.nasa.gov/ssc/data/analysis/software/aux/4fgl/Galactic_Diffuse_Emission_Model_for_the_4FGL_Catalog_Analysis.pdf}} \texttt{gll\_iem\_v07} interstellar emission model \citep{gll_iem_v07} associated to the 4FGL-DR4 catalogue.

For our analyses, we chose a cut on probability weights to save computing power in searches and timing. The test statistics and especially the weighted \textit{H}-test \citep{deJaeger1989+Htest,Kerr2011} depend on these weights and scale linearly with $W^2 = \sum_j w_j^2$ \citep{Nieder2020+Methods}. The proportionality factors are source-specific, depending on the pulse profile and the background level, but are typically of order unity (0.5 to 2). The cutoff weight is chosen so that most of the low-weight photons are removed while securing $99\%$ of the expected signal power typically corresponding to $w_{\rm min} \approx 0.01 - 0.03$.

We found gamma-ray pulsations from the two pulsars, PSRs~J1823$+$1208 and J2029$-$4239 in a single fold. In these cases, the radio-timing ephemeris revealed gamma-ray pulsations over most of the \textit{Fermi} mission without a search being required. Further tweaking of parameters like the spin-down rate $\dot{\nu}$ showed pulsations over the entire mission starting in $2008$.

We found gamma-ray pulsations from the three pulsars, PSRs~J0657$-$4657, J1259$-$8148, and J1356$+$0230 in follow-up searches. Using the radio ephemerides as starting positions, we searched for gamma-ray pulsations in the \textit{Fermi}-LAT data over a 5-dimensional parameter space ($\alpha, \delta, \nu, \dot{\nu}$ and $P_{\rm b}$).  For computational reasons, we truncated $H$ at the $3$rd out of $20$ harmonics as pulsars have most of their Fourier power in the lowest harmonics \citep{Pletsch2014+searchmethods}.

These five radio and gamma-ray pulsars were timed following the methods described in \citet{Burgay2024+timing}.  The joint radio and gamma-ray timing code\footnote{\url{https://fermi.gsfc.nasa.gov/ssc/data/analysis/user/jrag-timing.py}} uses the pulsar timing software \texttt{PINT}\footnote{\url{https://github.com/nanograv/PINT}} \citep{Luo2021+PINT} to fold the radio and gamma-ray data individually and joins the two standard test statistics ($\chi^2_{\rm radio}$ for radio ToAs, $\log \mathcal{L}_{\gamma}$ for gamma-ray photons) into a new log-likelihood function
\begin{equation}
    \log \mathcal{L}_{\rm joint} = \log \mathcal{L}_{\gamma} - 0.5 \chi^2_{\rm radio} \,.
\end{equation}
The Affine Invariant Markov chain Monte Carlo (MCMC) Ensemble sampler \texttt{emcee} \citep{Foreman-Mackey2013+emcee} is used to explore the parameter space.  The timing solutions are presented in Tables \ref{T:timing} and \ref{T:timing2}.

Gamma-ray pulsations were only detected from the black-widow pulsar J1831$-$6503, after a longer radio timing campaign which revealed orbital period variations. A sliding window method revealed pulsations over the entire mission span and served as starting solution for a joint radio and gamma-ray timing analysis, treating the variations as a stationary Gaussian process \citep{Clark2026+Gibbs}, as has been done for previous TRAPUM spider binaries \citep{Thongmeearkom2024+RBs,Burgay2024+timing,Belmonte2025+J1544}. We use a Mat\'ern covariance function, a power-law function parametrised by three hyperparameters, amplitude ($A_{\rm opv}$), cutoff frequency ($\nu_{\rm c,opv}$) and spectral index ($\gamma_{\rm opv}$).  
%The method and code will be described in more detail (Clark~et~al.,~in preparation). 
The measured pulsar parameters are shown in Table~\ref{T:timing2}, with the phase-folded gamma-ray data and the orbital period variations being depicted in Figure~\ref{F:J1831_opv}. We did not find evidence for orbital period variations in the two other new black widows, PSRs~J1259$-$8148 and J1356$+$0230.

The analysis of the three redback pulsars is hampered by their faintness in gamma rays, and the likelihood that they also exhibit orbital period variations. PSRs~J1346$-$2610 and J1910$-$5320 have $W^2 = 178$ and $W^2 = 181$, respectively, which are low enough to prevent a detection if orbital period variations are present. This is certainly the case for PSR~J1910$-$5320 where two orbital frequency derivatives were required to fit the seven months of radio ToAs. We did not detect significant gamma-ray pulsations from either pulsar when folding only data during the interval spanned by radio ToAs. For these pulsars, significant detections with $H \approx 25$ are only possible over substantially longer segments, of around $1.1 - 4.7$~years (depending on the source-specific proportionality factor), over which time the orbital period could vary substantially. For the redback pulsar PSR~J1712$-$1920, folding the gamma-ray data spanned by the radio ephemeris resulted in faint, but significant pulsations with $H=33$. We have not yet found a phase-connected timing solution spanning much more time than this. This is not unusual as it is difficult to find a timing solution for a faint ($W^2 = 140$) redback with orbital-period variations. The radio-only timing solution is shown in Table~\ref{T:timing2}. All of these will require continued monitoring with radio telescopes to enable gamma-ray pulsation detections and/or a joint timing analysis.

\subsection{Slow Pulsars}
\label{S:slow_pulsars}
Six slow pulsars have been found (PSRs~J1207$-$45, J1303$-$4714, J1816$-$64, J1845$-$25, J1924$-$10, J2355$-$66). Five of these have been found in the incoherent beam, which has a larger field of view (FWHM $\sim$114 arcminutes) compared to the tiling of the small coherent beams ($\sim$7 arcminutes). As a result, it is impossible to localise them like other MSP discoveries from the coherent beams. Hence, we report them with names indicating their coarse positional uncertainties. The information of these pulsars can be found in Table~\ref{T:slow_pulsars}. 

We find no evidence that these slow pulsars emit gamma-ray pulsations. Associations with the gamma-ray sources that were targeted during these observations can be ruled out due to the lack of detection in any of the coherent beams that were tiling the 99\% confidence regions. We also found no other unassociated 4FGL-DR4 sources within the FWHM of the incoherent beams around these sources. Furthermore, PSRs J1303$-$4713 and J1816$-$64 have spin periods ($P=2.56$\,s and $P=1.22$\,s, respectively) longer than that of the slowest known gamma-ray pulsar \citep[PSR~J1731$-$4744 with $P=0.83$\,s,][]{Smith2019+Edot}, while only 3 and 5 known gamma-ray pulsars are slower than PSRs~J1845$-$25 and J1207$-$45, respectively \citep{Smith2023+3PC}. 
PSRs~J1924$-$10 and J2355$-$66 have spin periods of about 114 and 121\,ms, respectively, and could either be energetic young pulsars, or could be mildly recycled pulsars similar to PSR~J1744$-$3922 \citep{Breton2007}. However, because of the lack of precise positions, and the very large number of observations that would be required to tile the incoherent beam FWHM with coherent beams, we deemed a dedicated follow-up investigation of these pulsars to be beyond the scope of this work.

\subsection{Distance to PSR~J0657$-$4657}
\label{S:distances}

The distance to a pulsar may be estimated through the measured DM and an electron density model of our galaxy. For PSR~J0657$-$4657, the measured DM of 126\,pc\,cm$^{-3}$ corresponds to a distance of 0.5\,kpc in the YMW16 model \citep{YMW16}. However, the NE2001 model \citep{ne2001} saturates at a DM of 81.5\,pc\,cm$^{-3}$ at $14$\,kpc along this line of sight and jumps to a distance of 50\,kpc for any larger value of DM. This clearly demonstrates a limitation of the NE2001 model along this particular direction. One possible reason is that the line of sight to PSR~J0657$-$4657 intersects the Gum Nebula, a large H II region with a distance of $\sim$450 pc \citep{Purcell2015}, which might not be correctly accounted for in NE2001. Similar discrepancies have been noted in other MSPs whose sight lines pass near structured ionised regions \citep{Thongmeearkom2024+RBs}, although the mismatch for PSR~J0657$-$4657 appears noticeably larger.

An independent distance constraint on the distance comes from gamma-ray data. The gamma-ray flux of $F_{\gamma} = (2.63 \pm 0.36) \times 10^{-12}$\,erg\,cm$^{-2}$\,s$^{-1}$ is measured in the energy range between 100\,MeV and 100\,GeV from 4FGL DR4 \citep{4FGLDR4}. Assuming a gamma-ray efficiency $\eta_{\gamma} = \mathcal{L}_{\gamma} / \dot{E} = 4 \pi d^2 F_{\gamma} / \dot{E}$ of $100\%$ (i.e., the entire spin-down power is converted into gamma rays) gives a conservative distance upper limit of $3.8$\,kpc. A more realistic value is obtained assuming the canonical heuristic value of $\eta_{\gamma} = \sqrt{10^{33}\,\mathrm{erg\,s^{-1}}/\dot{E}}$ \citep[e.g.,][]{Smith2023+3PC}, resulting in an estimate of $2.6$\,kpc.

Another method exploits the correction factors on the spin-down rate $\dot{\nu}$, the Shklovskii effect \citep{Shklovskii1970} induced by proper motion and acceleration in the Galactic potential. These effects depend on sky position and distance and as under normal circumstances we would assume a negative $\dot{\nu}$, this may be used to set a limit.  Here, this results in a limit of $32$\,kpc, which is less constraining than the estimates using the YMW16 model and the gamma-ray efficiency.

The range of distance estimates for PSR~J0657$-$4657 are wide and without other options available (e.g., parallax) a reliable estimate cannot be reported.  However, it is safe to assume that the NE2001 result can be ruled out and that the distance is likely smaller than $3.8$\,kpc.

\subsection{Continuous Gravitational Wave Searches}
\label{S:cw}
Gravitational wave emission is expected from non-axisymmetric neutron stars with the dominant signal frequency being equal to twice the rotational frequency of the star $(\nu_{\rm GW} = 2\nu)$ \citep{Jones:2015zma}. The intrinsic continuous gravitational wave amplitude, $h_{0}$, crucially depends on the distance to the source and the fraction of the rotational energy being channeled into gravitational wave radiation \citep{Jaranowski:1998qm}. In this regard, of the pulsar discoveries, PSRs~J0657$-$4657 and J2029$-$4239 are the most promising sources to produce detectable continuous gravitational wave signals. We search for such signals using data from the first, second, third and fourth observing runs of the Advanced LIGO detectors \citep{LIGOScientific:2019lzm, KAGRA:2023pio, LIGOScientific:2025snk} from both sources.

Since the timing solutions span all epochs from the first through to the fourth observation run, we perform a single-template targeted-search for continuous gravitational wave emission centered at twice the star's rotational frequency, using a frequentist analysis procedure based on the multi-detector maximum-likelihood $\mathcal{F}$-statistic \citep{Cutler:2005hc}. This method coherently combines data from the Hanford and Livingston Advanced LIGO detectors. The results of this search are consistent with a non-detection and we set 95\% confidence upper limits on the intrinsic gravitational wave amplitude $h_{0}$. We obtain upper limits on $h_{0}$ of $3.0 \times 10^{-27}$ and $6.3 \times 10^{-27}$ for PSRs~J0657$-$4657 and J2029$-$4239, respectively.

Additionally, the gravitational wave emission may be mismatched from exactly twice the rotational frequency, motivating an additional band search around each pulsar's nominal frequency and spin-down. In the present analysis, however, the nominal gravitational wave frequency of PSR~J0657$-$4657 lies within a frequency range for which the sensitivity of the Advanced LIGO detectors is limited due to the mirror's suspension violin modes that render a reliable band search for this pulsar infeasible. We therefore restrict this extended analysis to PSR~J2029$-$4239 only and search in a range of $\Delta \nu_{\rm GW} = 4 \times 10^{-3} \nu_{\rm GW}$ centered on the nominal frequency and correspondingly for the spin-down $\dot{\nu}_{\rm GW}$ consistent with previous searches \citep{Ashok:2021fnj, Clark:2025lai}.

The band search is divided into multiple $10\rm\,mHz$ wide sub-bands and upper limits on $h_{0}$ are set in each sub-band. We obtain a mean upper limit value of $h_{0}$ across all $10\rm\,mHz$ bands of $1.4 \times 10^{-26}$. This is a larger than the targeted search upper limit due to the trials-factor of searching over many waveform templates. We find no evidence of gravitational wave emission in these additional extended parameter space searches as the results are consistent with Gaussian noise.

Our upper limits on $h_0$ can be expressed as upper limits on the ellipticity $\epsilon$ of the pulsars, modelled as a triaxial ellipsoid spinning around a principal moment of inertia axis. This is \citep{Gao:2020zcd},
\begin{align}
\epsilon =\;& 2.36 \times 10^{-6} \times\left( \frac{h_0}{10^{-25}} \right)\left( \frac{10^{45}\mathrm{g\,cm^2}}{I_{\rm zz}} \right) \nonumber \\
&\times\left( \frac{100\,\mathrm{Hz}}{\nu} \right)^2
\left( \frac{d}{1\,\mathrm{kpc}} \right),
\end{align}
where $I_{\rm zz}$ is the moment of inertia about the spin axis, $d$ is the distance to the pulsar and $\nu$ is the rotational frequency. Using the canonical value of $I_{\mathrm{zz}} = 10^{45}{\rm g \, cm^{2}}$ and the estimated distance using the YMW16 electron density model, gives an ellipticity upper limit of $5.5 \times 10^{-9}$ and $2.1 \times 10^{-8}$ at the $95\%$ confidence level for PSRs~J0657$-$4657 and J2029$-$4239, respectively. 

It is unlikely that our searches could have detected gravitational waves from PSRs~J0657$-$4657 and J2029$-$4239, as our upper limits do not beat the spin-down lower limit $h_{0}^{\rm spdwn}$, assuming that all rotational energy loss is converted into gravitational waves.

\section{Discussion}
\label{S:discussion_UHF}
We report results from the first MeerKAT UHF shallow survey targeting the \textit{Fermi}-LAT sources. In this section, we evaluate the number of discoverable pulsars in both \textit{L} band and UHF from the total discoveries and compare them to determine the most effective frequency band to find new pulsars with the MeerKAT setup we used. We also discuss various effects that can cause difficulty finding pulsars.

\subsection{Expected Discoverable Sources}
\label{S:Discoverable_pulsars}
The first full TRAPUM shallow survey was proposed as 10-minute observations of 79 \textit{Fermi}-LAT unassociated sources with four different epochs, starting with two epochs at \textit{L} band \citep{Clark2023+Lband}. We then removed 13 sources for confirmed \textit{L}-band discoveries to save telescope time. We added an additional 13 \textit{Fermi}-LAT targets to match the proposed observing time for two later epochs at UHF. The UHF survey discovered nine new MSPs and six slow pulsars, while the \textit{L}-band survey \citep{Clark2023+Lband} found nine new MSPs and no new slow pulsars. To fairly compare the performance of the two bands, we accounted for sources that would have already been detected in one band before observations were conducted in the other, in order to avoid biasing the comparison.

For the \textit{L}-band discoveries from \citet{Clark2023+Lband}, we have the confirmation campaign where we pointed MeerKAT at these sources using the UHF receiver. All of them have been searched through the pipeline and were blindly detected at UHF. This result is also supported by a number of studies \citep[e.g.,][]{Bates2013}, which shows that pulsars tend to be brighter at lower frequency \citep{Johnston2020+PTA}. Therefore, we conclude that all of the \textit{L}-band pulsars are also discoverable at UHF. For the UHF discoveries, we examined nine MSPs that have also been observed at \textit{L} band as part of the timing campaign. Of these, five sources (PSRs~J1259$-$8148, J1356+0230, J1823+1208, J1910$-$5320, and J2029$-$4239) were not included in previous \textit{L}-band searches, while the remaining four had been part of earlier \textit{L}-band survey pointings but not detected. Although these earlier observations yielded no detections, we cannot rule out the possibility that the pulsars were in eclipse or otherwise obscured during those observations. Therefore, we re-evaluated their detectability using the timing observations. We used the 5-minute \textit{L}-band timing observations of these UHF discoveries for folding comparing with UHF discovery folds and 5-minute UHF timing to understand S/N differences between the two bands. We then multiplied by root two (from the radiometer equation) to predict 5-minute observations if they were observed for 10 minutes. PSRs~J1259$-$8148, J1712$-$1920, J1910$-$5320, and J2029$-$4239 are clearly discoverable at \textit{L} band as their S/Ns are above the threshold of 9 that we use in our \texttt{Peasoup} searching programme. PSR~J0657$-$4658 is also detectable at \textit{L} band (see Section \ref{S:cross_match} for more information). In contrast, PSRs~J1346$-$2610, J1356+0230, J1823+1208, and J1831$-$6503 have low S/Ns of about 5--6 in the 5-minute timing observations, depending on the observation day. Even after scaling to a 10-minute survey observation, their S/Ns remain below our detection threshold. It is important to note that we use the known parameters from the discoveries as a starting point, and these pulsars also show signs of scintillation in frequency bands. 

\begin{figure}
  \centering
  	\includegraphics[width=\columnwidth]{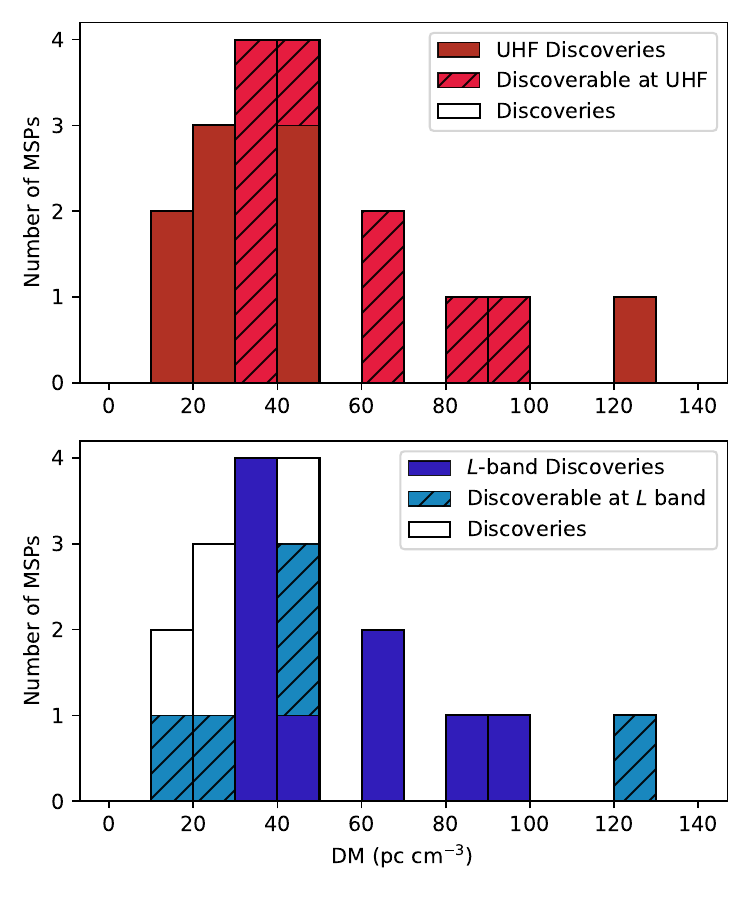}
        \caption{Histograms of discovered MSPs at \textit{L} band \citep{Clark2023+Lband} and UHF. The blue and red colours represent \textit{L} band and UHF, respectively, while the red hatched histogram represents the pulsars that would have been discovered with UHF if the survey had begun at UHF instead of \textit{L} band. This illustrates the advantage of using UHF in our setup for discovering pulsars.} 
    \label{F:DM_L_UHF}
\end{figure}

Therefore, we conclude that MeerKAT \textit{L}-band observations with our search parameters and observation settings can discover five UHF pulsars: PSRs~J0657$-$4658, J1259$-$8148, J1712$-$1920, J1910$-$5320, and J2029$-$4239. The histograms of the shallow survey pulsars are shown with \textit{L}-band pulsars, UHF pulsars, and expected discoverable sources (Figure \ref{F:DM_L_UHF}).

\subsection{Comparing \textit{L} band and UHF}
\label{S:compare}
We then compared the discoveries from both \textit{L} band and UHF based on their DMs (see Figure \ref{F:DM_L_UHF}). The figure shows that many MSPs found in either band have DMs below 50\,pc\,cm$^{-3}$, as our search campaign targets off-plane sources. Since these sources are located outside the Galactic plane, we anticipate lower DMs, which are consistent with expectations from electron density models such as NE2001 \citep{ne2001} and YMW16 \citep{YMW16} along their respective lines of sight.

In terms of pulsar types, the \textit{L}-band discoveries are all MSPs, comprising eight binaries and one isolated pulsar, with DMs ranging between 30 and 100 pc cm$^{-3}$; five of these pulsars fall within a narrower DM range of 30 to 50 pc cm$^{-3}$. In contrast, the UHF survey found nine MSPs and six slow pulsars. Notably, two of the UHF MSPs have DMs below 20 pc cm$^{-3}$, while only one, PSR~J0657$-$4657, has a DM exceeding 100 pc cm$^{-3}$. Unlike the \textit{L}-band survey, which did not detect any new slow pulsars, the UHF survey identified six, primarily via the incoherent beam. This difference is expected since the UHF incoherent beam size is approximately 50\% larger than that of the \textit{L} band, allowing it to cover a larger area and thus detect slow pulsars unassociated with \textit{Fermi} sources. After removing the slow pulsars that have been detected from the incoherent beam, there are three MSPs at low DM (lower than 30 pc cm$^{-3}$) that have been discovered at UHF but are not expected to be discoverable at \textit{L} band. In addition, pulsars generally have steep spectra, making them intrinsically brighter at lower frequencies.

However, UHF has limitations, particularly due to its greater susceptibility to DM smearing. This effect, where the signal broadens and blends with noise, can make it challenging or even impossible to detect very fast-spinning pulsars, since the high time resolution required to resolve their narrow pulses is increasingly difficult to achieve at lower frequencies. The expected pulse broadening due to intra-channel DM smearing is quantified with
\begin{equation}
    \Delta t = 4.15~\mathrm{ms} 
    \left(\frac{\rm DM}{\mathrm{pc~cm^{-3}}}\right)
    \left[
        \left(\frac{\nu}{\mathrm{MHz}}\right)^{-2} - 
        \left(\frac{\nu+\Delta \nu}{\mathrm{MHz}}\right)^{-2}
    \right]\,.
\label{E:DM}
\end{equation}

\begin{figure}
  \centering
  	\includegraphics[width=\columnwidth]{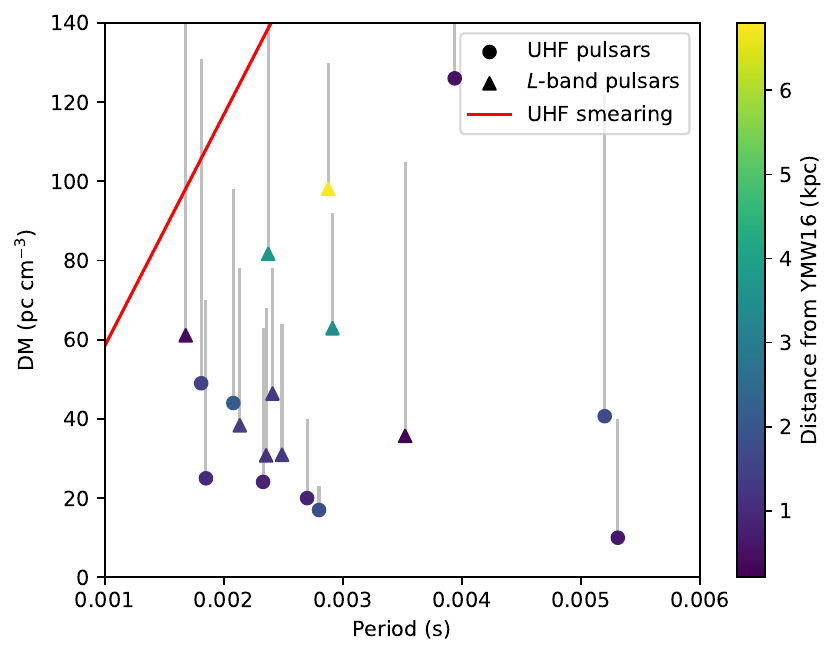}
        \caption{The distribution of discovered MSPs from the \textit{L}-band survey \citep[triangles;][]{Clark2023+Lband} and the UHF survey (circles) over pulse period and dispersion measure (DM). The color map represents the expected distance of each MSP according to the YMW16 electron density model \citep{YMW16}, while the grey lines display the difference between DM of each MSP and the maximum DM along the line of sight. The red line is expected discoverable pulsar periods corresponding to the DM smearing effect at UHF, while the \textit{L}-band smearing line lies beyond the plotted range.}
    \label{F:UHF_P_DM}
\end{figure}

$\Delta t$ is the time smearing in a single frequency channel, $\nu$ is channel centre frequency, and $\Delta \nu$ is the channel width \citep{PSRHandbook}. The pulse broadening due to DM smearing effectively sets the detection limits for pulsars in different frequency bands. Using equation~\eqref{E:DM} and assuming a 15\% duty cycle pulsar, we calculated the theoretical fastest discoverable pulsar, the shortest spin period detectable given the pulse broadening from DM smearing within a channel, as a function of DM for both \textit{L} band and UHF (see Figure \ref{F:UHF_P_DM}). It is important to note that RFI conditions vary between telescope sites, some sites may experience telecommunication signals overlapping with UHF, making observations at this frequency more challenging. At the MeerKAT site, UHF actually exhibits less RFI than \textit{L} band. However, comparing RFI fractions across different observatories is non-trivial because the measured RFI environment depends strongly on the local interference conditions, even though the number of masked channels can provide a site-specific measure.

\subsection{Cross-Matching with the \textit{L}-band Data}
\label{S:cross_match}
\texttt{Peasoup} drastically reduces the number of candidates produced by the search pipeline. Despite this, human interaction is a part of distinguishing the candidates to be either a pulsar or not. In some cases, weak signals can be seen in the resulting diagnostic plots, which is difficult to classify as a pulsar, a tier-two\footnote{We define tier-two candidates as ambiguous detections, such as those with signal-to-noise ratios close to our threshold of 9, or candidates whose pulse profiles are comparable in width to the expected DM smearing, making classification more difficult.} candidate, or noise. Hence, we developed the cross-matching programme\footnote{\url{https://github.com/01tinn/cross_matching}} to assist the candidate classification. For example, if two similar weak signals (i.e., the same candidate) show in two or more observations, there is a higher chance that the candidate is real. The cross-matching script is a Python-based programme developed to check the similarity of three properties for two or more candidates from distinct observations. Those aspects are the position, the period, and the DM. For position matching, if the sky separation between two candidates is smaller than twice the assumed coherent beam FWHM, the two candidates are considered to match. However, the beam size varies with the observing angle. We use 0.007$\degr$ for the beam value ($\sim$25.2\,arcseconds, about the size of one coherent beam, as a rough estimation for simplification in the cross-matching script). The programme then checks the DM difference, which needs to be lower than the smearing of the full channel (\textit{L} band at 856--1712 MHz and UHF at 544--1088 MHz) by using the DM smearing formula \citep{PSRHandbook}. Lastly, for a successful match, the two candidates must have similar spin periods. This condition is influenced by each pulsar's spin-down rate as well as Doppler shifts caused by binary motion. The spin-down rate varies from pulsar to pulsar. Typical MSP spin-down rates are below 10$^{-19}$ \citep[$P-\dot{P}$ diagram;][]{PSRHandbook}. We increase this to 10$^{-16}$ to cover almost every known MSPs including those with unusually high spin-down rate. The programme calculated the change of the period by multiplying the spin-down rate and time between two epochs of an observation. Doppler shifts result from a combination of the Earth's rotation ($\sim$460 m s$^{-1}$), the Earth orbiting around the Sun ($\sim$30 km s$^{-1}$), and a pulsar orbiting within its binary system ($\sim$300 km s$^{-1}$). We selected large values for each parameter to get loosely constrained results by the notion that getting a higher number of matches is better than missing potential matches. With the three conditions are satisfied, the programme flags the candidates as a potential match. We then carry out final confirmation manually. We applied this programme to data from two \textit{L}-band epochs \citep{Clark2023+Lband} and two UHF epochs. However, some targets had only two or three usable epochs due to corrupted files from cluster downtimes. The programme identified 85 pairs of candidates, though we cannot estimate the total number of potentially missed pairs due to incomplete or corrupted input data. Upon reviewing the results, we found some pairs represented genuine discoveries, either matches within \textit{L}-band or UHF data, while others were RFI. The results demonstrated that the programme could reliably identify plausible candidate matches.

We identified a weak pulsar candidate associated with 4FGL~J0657.4$-$4658 from the \textit{L}-band survey \citep{Clark2023+Lband}. This candidate had a S/N of approximately $\sim$11 at \textit{L} band, compared to a higher S/N of $\sim$19 at UHF. The \textit{L}-band detection may be classified as a tier-two candidate, as it is relatively faint and affected by smearing, warranting follow-up observations for confirmation. Using our cross-matching script, we found that this \textit{L}-band candidate corresponded to one of the UHF discoveries, confirming it as a real pulsar and demonstrating the effectiveness of the cross-matching method.

\section{Conclusions and Future work}
\label{S:conclusion_UHF}

The paper described the first UHF shallow (10 minutes) survey of 79 unassociated \textit{Fermi} gamma-ray sources with MeerKAT. Using MeerKAT beam forming, up to 480 coherent beams were formed to cover the \textit{Fermi} localisation region for each target.

We found 15 new pulsars. Seven of them are MSPs associated with \textit{Fermi}-LAT gamma-ray sources, two additional MSPs are candidate associations that require confirmation from gamma-ray pulsations, and six slow pulsars not associated with \textit{Fermi}-LAT sources. Among the nine MSPs, eight are in binary systems and one new MSP (PSR~J2029$-$4239) is an isolated pulsar. Further investigation indicated that PSRs~J1346$-$2610, J1712$-$1920, and J1910$-$5320 are redbacks; PSRs~J1259$-$8148, J1356+0230, and J1831$-$6503 are black widows due to their orbital periods, companion masses, and eclipses. Two other systems (PSRs~J0657$-$4657 and J1823+1208) are He white dwarf binaries.

For future work, we have started the expanded UHF shallow survey targeting 145 \textit{Fermi}-LAT sources that have generally not been previously observed from the latest 4FGL-DR4 catalogue \citep{4FGLDR4}. The upcoming \textit{Fermi}-LAT catalogue will provide improved source localisations for future targeted radio pulsar searches. The UHF survey presented in this paper is part of a larger programme that also includes \textit{L}-band observations \citep{Clark2023+Lband}, this new survey is conducted exclusively at UHF, as our results show that UHF with our observing parameters is more effective than \textit{L} band for finding pulsars. At the time of writing this paper, we found 23 new pulsars from the expanded UHF survey (Thongmeearkom et al., in preparation).

\section*{Acknowledgements}
The MeerKAT telescope is operated by the South African Radio Astronomy Observatory (SARAO), which is a facility of the National Research Foundation, an agency of the Department of Science and Innovation. We thank staff at SARAO for their help with observations and commissioning. TRAPUM observations used the FBFUSE and APSUSE computing clusters for data acquisition, storage and analysis. These clusters were funded and installed by the Max-Planck-Institut f\"{u}r Radioastronomie (MPIfR) and the Max-Planck-Gesellschaft. The National Radio Astronomy Observatory is a facility of the National Science Foundation operated under cooperative agreement by Associated Universities, Inc. The Parkes radio telescope is part of the Australia Telescope National Facility (https://ror.org/05qajvd42) which is funded by the Australian Government for operation as a National Facility managed by CSIRO. We acknowledge the Wiradjuri people as the traditional owners of the Observatory site. The Nan\c cay Radio Observatory is operated by the Paris Observatory, associated with the French Centre National de la Recherche Scientifique (CNRS) and Universit\'{e} d'Orl\'{e}ans. It is partially supported by the Region Centre Val de Loire in France.

The \textit{Fermi} LAT Collaboration acknowledges generous ongoing support from a number of agencies and institutes that have supported both the
development and the operation of the LAT as well as scientific data analysis. These include the National Aeronautics and Space Administration and the Department of Energy in the United States, the Commissariat \`a l'Energie Atomique and the Centre National de la Recherche Scientifique / Institut National de Physique Nucl\'eaire et de Physique des Particules in France, the Agenzia Spaziale Italiana and the Istituto Nazionale di Fisica Nucleare in Italy, the Ministry of Education, Culture, Sports, Science and Technology (MEXT), High Energy Accelerator Research Organization (KEK) and Japan Aerospace Exploration Agency (JAXA) in Japan, and the K.~A.~Wallenberg Foundation, the Swedish Research Council and the Swedish National Space Board in Sweden.

Additional support for science analysis during the operations phase is gratefully acknowledged from the Istituto Nazionale di Astrofisica in Italy and the Centre National d'\'Etudes Spatiales in France. This work performed in part under DOE Contract DE-AC02-76SF00515.

This research has made use of data or software obtained from the Gravitational Wave Open Science Center (gwosc.org), a service of the LIGO Scientific Collaboration, the Virgo Collaboration, and KAGRA. This material is based upon work supported by NSF's LIGO Laboratory which is a major facility fully funded by the National Science Foundation, as well as the Science and Technology Facilities Council (STFC) of the United Kingdom, the Max-Planck-Society (MPS), and the State of Niedersachsen/Germany for support of the construction of Advanced LIGO and construction and operation of the GEO600 detector. Additional support for Advanced LIGO was provided by the Australian Research Council.

This manuscript is based on material presented in the first author’s doctoral thesis \citep{Thongmeearkom2025}, which was written in journal format and contains an earlier draft of this paper. Parts of the text have been revised and expanded for publication.

TT is grateful to the National Astronomical Research Institute of Thailand (NARIT) for awarding a student scholarship. MB acknowledges resources from the research grant “iPeska” (PI: Andrea Possenti), funded under the INAF national call Prin-SKA/CTA approved with the Presidential Decree 70/2016. JB is supported by NASA under award number 80GSFC21M0002. ECF is supported by NASA under award number 80GSFC24M0006. LV acknowledges financial support from the Dean’s Doctoral Scholar Award from the University of Manchester and partial support from NSF grant AST-1816492. SBD acknowledges the support of a Science and Technology Facilities Council (STFC) stipend (grant number: ST/X001229/1) to permit work as a postgraduate researcher. VSD and ULTRACAM are supported by STFC grant ST/Z000033/1.

We thank Matthew Kerr for reviewing this manuscript on behalf of the \textit{Fermi}-LAT collaboration. We are grateful to the scientific editor, Timothy J. Pearson, for helpful suggestions and to the referee, David A. Smith, for valuable comments.

%%%%%%%%%%%%%%%%%%%%%%%%%%%%%%%%%%%%%%%%%%%%%%%%%%
\section*{Data Availability}

TRAPUM data products are available upon reasonable request to the TRAPUM collaboration. The \textit{Fermi}-LAT data are available from the \textit{Fermi} Science Support Center (\url{http://fermi.gsfc.nasa.gov/ssc}).

%%%%%%%%%%%%%%%%%%%% REFERENCES %%%%%%%%%%%%%%%%%%

% The best way to enter references is to use BibTeX:

\bibliographystyle{mnras}
\bibliography{main} % if your bibtex file is called example.bib

@ARTICLE{Atwood2009+LAT,
       author = {{Atwood}, W.~B. and {Abdo}, A.~A. and {Ackermann}, M. and {Althouse}, W. and {Anderson}, B. and {Axelsson}, M. and {Baldini}, L. and {Ballet}, J. and {Band}, D.~L. and {Barbiellini}, G. and {Bartelt}, J. and {Bastieri}, D. and {Baughman}, B.~M. and {Bechtol}, K. and {B{\'e}d{\'e}r{\`e}de}, D. and {Bellardi}, F. and {Bellazzini}, R. and {Berenji}, B. and {Bignami}, G.~F. and {Bisello}, D. and {Bissaldi}, E. and {Blandford}, R.~D. and {Bloom}, E.~D. and {Bogart}, J.~R. and {Bonamente}, E. and {Bonnell}, J. and {Borgland}, A.~W. and {Bouvier}, A. and {Bregeon}, J. and {Brez}, A. and {Brigida}, M. and {Bruel}, P. and {Burnett}, T.~H. and {Busetto}, G. and {Caliandro}, G.~A. and {Cameron}, R.~A. and {Caraveo}, P.~A. and {Carius}, S. and {Carlson}, P. and {Casandjian}, J.~M. and {Cavazzuti}, E. and {Ceccanti}, M. and {Cecchi}, C. and {Charles}, E. and {Chekhtman}, A. and {Cheung}, C.~C. and {Chiang}, J. and {Chipaux}, R. and {Cillis}, A.~N. and {Ciprini}, S. and {Claus}, R. and {Cohen-Tanugi}, J. and {Condamoor}, S. and {Conrad}, J. and {Corbet}, R. and {Corucci}, L. and {Costamante}, L. and {Cutini}, S. and {Davis}, D.~S. and {Decotigny}, D. and {DeKlotz}, M. and {Dermer}, C.~D. and {de Angelis}, A. and {Digel}, S.~W. and {do Couto e Silva}, E. and {Drell}, P.~S. and {Dubois}, R. and {Dumora}, D. and {Edmonds}, Y. and {Fabiani}, D. and {Farnier}, C. and {Favuzzi}, C. and {Flath}, D.~L. and {Fleury}, P. and {Focke}, W.~B. and {Funk}, S. and {Fusco}, P. and {Gargano}, F. and {Gasparrini}, D. and {Gehrels}, N. and {Gentit}, F. -X. and {Germani}, S. and {Giebels}, B. and {Giglietto}, N. and {Giommi}, P. and {Giordano}, F. and {Glanzman}, T. and {Godfrey}, G. and {Grenier}, I.~A. and {Grondin}, M. -H. and {Grove}, J.~E. and {Guillemot}, L. and {Guiriec}, S. and {Haller}, G. and {Harding}, A.~K. and {Hart}, P.~A. and {Hays}, E. and {Healey}, S.~E. and {Hirayama}, M. and {Hjalmarsdotter}, L. and {Horn}, R. and {Hughes}, R.~E. and {J{\'o}hannesson}, G. and {Johansson}, G. and {Johnson}, A.~S. and {Johnson}, R.~P. and {Johnson}, T.~J. and {Johnson}, W.~N. and {Kamae}, T. and {Katagiri}, H. and {Kataoka}, J. and {Kavelaars}, A. and {Kawai}, N. and {Kelly}, H. and {Kerr}, M. and {Klamra}, W. and {Kn{\"o}dlseder}, J. and {Kocian}, M.~L. and {Komin}, N. and {Kuehn}, F. and {Kuss}, M. and {Landriu}, D. and {Latronico}, L. and {Lee}, B. and {Lee}, S. -H. and {Lemoine-Goumard}, M. and {Lionetto}, A.~M. and {Longo}, F. and {Loparco}, F. and {Lott}, B. and {Lovellette}, M.~N. and {Lubrano}, P. and {Madejski}, G.~M. and {Makeev}, A. and {Marangelli}, B. and {Massai}, M.~M. and {Mazziotta}, M.~N. and {McEnery}, J.~E. and {Menon}, N. and {Meurer}, C. and {Michelson}, P.~F. and {Minuti}, M. and {Mirizzi}, N. and {Mitthumsiri}, W. and {Mizuno}, T. and {Moiseev}, A.~A. and {Monte}, C. and {Monzani}, M.~E. and {Moretti}, E. and {Morselli}, A. and {Moskalenko}, I.~V. and {Murgia}, S. and {Nakamori}, T. and {Nishino}, S. and {Nolan}, P.~L. and {Norris}, J.~P. and {Nuss}, E. and {Ohno}, M. and {Ohsugi}, T. and {Omodei}, N. and {Orlando}, E. and {Ormes}, J.~F. and {Paccagnella}, A. and {Paneque}, D. and {Panetta}, J.~H. and {Parent}, D. and {Pearce}, M. and {Pepe}, M. and {Perazzo}, A. and {Pesce-Rollins}, M. and {Picozza}, P. and {Pieri}, L. and {Pinchera}, M. and {Piron}, F. and {Porter}, T.~A. and {Poupard}, L. and {Rain{\`o}}, S. and {Rando}, R. and {Rapposelli}, E. and {Razzano}, M. and {Reimer}, A. and {Reimer}, O. and {Reposeur}, T. and {Reyes}, L.~C. and {Ritz}, S. and {Rochester}, L.~S. and {Rodriguez}, A.~Y. and {Romani}, R.~W. and {Roth}, M. and {Russell}, J.~J. and {Ryde}, F. and {Sabatini}, S. and {Sadrozinski}, H.~F. -W. and {Sanchez}, D. and {Sander}, A. and {Sapozhnikov}, L. and {Parkinson}, P.~M. Saz and {Scargle}, J.~D. and {Schalk}, T.~L. and {Scolieri}, G. and {Sgr{\`o}}, C. and {Share}, G.~H. and {Shaw}, M. and {Shimokawabe}, T. and {Shrader}, C. and {Sierpowska-Bartosik}, A. and {Siskind}, E.~J. and {Smith}, D.~A. and {Smith}, P.~D. and {Spandre}, G. and {Spinelli}, P. and {Starck}, J. -L. and {Stephens}, T.~E. and {Strickman}, M.~S. and {Strong}, A.~W. and {Suson}, D.~J. and {Tajima}, H. and {Takahashi}, H. and {Takahashi}, T. and {Tanaka}, T. and {Tenze}, A. and {Tether}, S. and {Thayer}, J.~B. and {Thayer}, J.~G. and {Thompson}, D.~J. and {Tibaldo}, L. and {Tibolla}, O. and {Torres}, D.~F. and {Tosti}, G. and {Tramacere}, A. and {Turri}, M. and {Usher}, T.~L. and {Vilchez}, N. and {Vitale}, V. and {Wang}, P. and {Watters}, K. and {Winer}, B.~L. and {Wood}, K.~S. and {Ylinen}, T. and {Ziegler}, M.},
        title = "{The Large Area Telescope on the Fermi Gamma-Ray Space Telescope Mission}",
      journal = {\apj},
         year = 2009,
        month = jun,
       volume = {697},
       number = {2},
        pages = {1071-1102},
          doi = {10.1088/0004-637X/697/2/1071},
archivePrefix = {arXiv},
       eprint = {0902.1089},
 primaryClass = {astro-ph.IM},
       adsurl = {https://ui.adsabs.harvard.edu/abs/2009ApJ...697.1071A}
}

@ARTICLE{4FGL,
       author = {{Abdollahi}, S. and {Acero}, F. and {Ackermann}, M. and {Ajello}, M. and {Atwood}, W.~B. and {Axelsson}, M. and {Baldini}, L. and {Ballet}, J. and {Barbiellini}, G. and {Bastieri}, D. and {Becerra Gonzalez}, J. and {Bellazzini}, R. and {Berretta}, A. and {Bissaldi}, E. and {Blandford}, R.~D. and {Bloom}, E.~D. and {Bonino}, R. and {Bottacini}, E. and {Brandt}, T.~J. and {Bregeon}, J. and {Bruel}, P. and {Buehler}, R. and {Burnett}, T.~H. and {Buson}, S. and {Cameron}, R.~A. and {Caputo}, R. and {Caraveo}, P.~A. and {Casandjian}, J.~M. and {Castro}, D. and {Cavazzuti}, E. and {Charles}, E. and {Chaty}, S. and {Chen}, S. and {Cheung}, C.~C. and {Chiaro}, G. and {Ciprini}, S. and {Cohen-Tanugi}, J. and {Cominsky}, L.~R. and {Coronado-Bl{\'a}zquez}, J. and {Costantin}, D. and {Cuoco}, A. and {Cutini}, S. and {D'Ammando}, F. and {DeKlotz}, M. and {de la Torre Luque}, P. and {de Palma}, F. and {Desai}, A. and {Digel}, S.~W. and {Di Lalla}, N. and {Di Mauro}, M. and {Di Venere}, L. and {Dom{\'\i}nguez}, A. and {Dumora}, D. and {Fana Dirirsa}, F. and {Fegan}, S.~J. and {Ferrara}, E.~C. and {Franckowiak}, A. and {Fukazawa}, Y. and {Funk}, S. and {Fusco}, P. and {Gargano}, F. and {Gasparrini}, D. and {Giglietto}, N. and {Giommi}, P. and {Giordano}, F. and {Giroletti}, M. and {Glanzman}, T. and {Green}, D. and {Grenier}, I.~A. and {Griffin}, S. and {Grondin}, M. -H. and {Grove}, J.~E. and {Guiriec}, S. and {Harding}, A.~K. and {Hayashi}, K. and {Hays}, E. and {Hewitt}, J.~W. and {Horan}, D. and {J{\'o}hannesson}, G. and {Johnson}, T.~J. and {Kamae}, T. and {Kerr}, M. and {Kocevski}, D. and {Kovac'evic'}, M. and {Kuss}, M. and {Landriu}, D. and {Larsson}, S. and {Latronico}, L. and {Lemoine-Goumard}, M. and {Li}, J. and {Liodakis}, I. and {Longo}, F. and {Loparco}, F. and {Lott}, B. and {Lovellette}, M.~N. and {Lubrano}, P. and {Madejski}, G.~M. and {Maldera}, S. and {Malyshev}, D. and {Manfreda}, A. and {Marchesini}, E.~J. and {Marcotulli}, L. and {Mart{\'\i}-Devesa}, G. and {Martin}, P. and {Massaro}, F. and {Mazziotta}, M.~N. and {McEnery}, J.~E. and {Mereu}, I. and {Meyer}, M. and {Michelson}, P.~F. and {Mirabal}, N. and {Mizuno}, T. and {Monzani}, M.~E. and {Morselli}, A. and {Moskalenko}, I.~V. and {Negro}, M. and {Nuss}, E. and {Ojha}, R. and {Omodei}, N. and {Orienti}, M. and {Orlando}, E. and {Ormes}, J.~F. and {Palatiello}, M. and {Paliya}, V.~S. and {Paneque}, D. and {Pei}, Z. and {Pe{\~n}a-Herazo}, H. and {Perkins}, J.~S. and {Persic}, M. and {Pesce-Rollins}, M. and {Petrosian}, V. and {Petrov}, L. and {Piron}, F. and {Poon}, H. and {Porter}, T.~A. and {Principe}, G. and {Rain{\`o}}, S. and {Rando}, R. and {Razzano}, M. and {Razzaque}, S. and {Reimer}, A. and {Reimer}, O. and {Remy}, Q. and {Reposeur}, T. and {Romani}, R.~W. and {Saz Parkinson}, P.~M. and {Schinzel}, F.~K. and {Serini}, D. and {Sgr{\`o}}, C. and {Siskind}, E.~J. and {Smith}, D.~A. and {Spandre}, G. and {Spinelli}, P. and {Strong}, A.~W. and {Suson}, D.~J. and {Tajima}, H. and {Takahashi}, M.~N. and {Tak}, D. and {Thayer}, J.~B. and {Thompson}, D.~J. and {Tibaldo}, L. and {Torres}, D.~F. and {Torresi}, E. and {Valverde}, J. and {Van Klaveren}, B. and {van Zyl}, P. and {Wood}, K. and {Yassine}, M. and {Zaharijas}, G.},
        title = "{Fermi Large Area Telescope Fourth Source Catalog}",
      journal = {\apjs},
         year = 2020,
        month = mar,
       volume = {247},
       number = {1},
          eid = {33},
        pages = {33},
          doi = {10.3847/1538-4365/ab6bcb},
archivePrefix = {arXiv},
       eprint = {1902.10045},
 primaryClass = {astro-ph.HE},
       adsurl = {https://ui.adsabs.harvard.edu/abs/2020ApJS..247...33A}
}

@ARTICLE{4FGLDR4,
       author = {{Ballet}, J. and {Bruel}, P. and {Burnett}, T.~H. and {Lott}, B. and {The Fermi-LAT collaboration}},
        title = "{Fermi Large Area Telescope Fourth Source Catalog Data Release 4 (4FGL-DR4)}",
      journal = {arXiv e-prints},
         year = 2023,
        month = jul,
          eid = {arXiv:2307.12546},
        pages = {arXiv:2307.12546},
          doi = {10.48550/arXiv.2307.12546},
archivePrefix = {arXiv},
       eprint = {2307.12546},
 primaryClass = {astro-ph.HE},
       adsurl = {https://ui.adsabs.harvard.edu/abs/2023arXiv230712546B}
}

@ARTICLE{Ackermann2012+classification,
       author = {{Ackermann}, M. and {Ajello}, M. and {Allafort}, A. and {Antolini}, E. and {Baldini}, L. and {Ballet}, J. and {Barbiellini}, G. and {Bastieri}, D. and {Bellazzini}, R. and {Berenji}, B. and {Blandford}, R.~D. and {Bloom}, E.~D. and {Bonamente}, E. and {Borgland}, A.~W. and {Bouvier}, A. and {Brandt}, T.~J. and {Bregeon}, J. and {Brigida}, M. and {Bruel}, P. and {Buehler}, R. and {Burnett}, T.~H. and {Buson}, S. and {Caliandro}, G.~A. and {Cameron}, R.~A. and {Caraveo}, P.~A. and {Casandjian}, J.~M. and {Cavazzuti}, E. and {Cecchi}, C. and {{\c{C}}elik}, {\"O}. and {Charles}, E. and {Chekhtman}, A. and {Chen}, A.~W. and {Cheung}, C.~C. and {Chiang}, J. and {Ciprini}, S. and {Claus}, R. and {Cohen-Tanugi}, J. and {Conrad}, J. and {Cutini}, S. and {de Angelis}, A. and {DeCesar}, M.~E. and {De Luca}, A. and {de Palma}, F. and {Dermer}, C.~D. and {Silva}, E. do Couto e. and {Drell}, P.~S. and {Drlica-Wagner}, A. and {Dubois}, R. and {Enoto}, T. and {Favuzzi}, C. and {Fegan}, S.~J. and {Ferrara}, E.~C. and {Focke}, W.~B. and {Fortin}, P. and {Fukazawa}, Y. and {Funk}, S. and {Fusco}, P. and {Gargano}, F. and {Gasparrini}, D. and {Gehrels}, N. and {Germani}, S. and {Giglietto}, N. and {Giordano}, F. and {Giroletti}, M. and {Glanzman}, T. and {Godfrey}, G. and {Grenier}, I.~A. and {Grondin}, M. -H. and {Grove}, J.~E. and {Guillemot}, L. and {Guiriec}, S. and {Gustafsson}, M. and {Hadasch}, D. and {Hanabata}, Y. and {Harding}, A.~K. and {Hayashida}, M. and {Hays}, E. and {Healey}, S.~E. and {Hill}, A.~B. and {Horan}, D. and {Hou}, X. and {J{\'o}hannesson}, G. and {Johnson}, A.~S. and {Johnson}, T.~J. and {Kamae}, T. and {Katagiri}, H. and {Kataoka}, J. and {Kerr}, M. and {Kn{\"o}dlseder}, J. and {Kuss}, M. and {Lande}, J. and {Latronico}, L. and {Lee}, S. -H. and {Lemoine-Goumard}, M. and {Longo}, F. and {Loparco}, F. and {Lott}, B. and {Lovellette}, M.~N. and {Lubrano}, P. and {Madejski}, G.~M. and {Mazziotta}, M.~N. and {McEnery}, J.~E. and {Mehault}, J. and {Michelson}, P.~F. and {Mignani}, R.~P. and {Mitthumsiri}, W. and {Mizuno}, T. and {Monte}, C. and {Monzani}, M.~E. and {Morselli}, A. and {Moskalenko}, I.~V. and {Murgia}, S. and {Nakamori}, T. and {Naumann-Godo}, M. and {Nolan}, P.~L. and {Norris}, J.~P. and {Nuss}, E. and {Ohsugi}, T. and {Okumura}, A. and {Omodei}, N. and {Orlando}, E. and {Ormes}, J.~F. and {Ozaki}, M. and {Paneque}, D. and {Panetta}, J.~H. and {Parent}, D. and {Pelassa}, V. and {Pesce-Rollins}, M. and {Pierbattista}, M. and {Piron}, F. and {Pivato}, G. and {Porter}, T.~A. and {Rain{\`o}}, S. and {Rando}, R. and {Ray}, P.~S. and {Razzano}, M. and {Reimer}, A. and {Reimer}, O. and {Reposeur}, T. and {Romani}, R.~W. and {Sadrozinski}, H.~F. -W. and {Salvetti}, D. and {Saz Parkinson}, P.~M. and {Schalk}, T.~L. and {Sgr{\`o}}, C. and {Shaw}, M.~S. and {Siskind}, E.~J. and {Smith}, P.~D. and {Spandre}, G. and {Spinelli}, P. and {Suson}, D.~J. and {Takahashi}, H. and {Tanaka}, T. and {Thayer}, J.~G. and {Thayer}, J.~B. and {Thompson}, D.~J. and {Tibaldo}, L. and {Tibolla}, O. and {Torres}, D.~F. and {Tosti}, G. and {Tramacere}, A. and {Troja}, E. and {Usher}, T.~L. and {Vandenbroucke}, J. and {Vasileiou}, V. and {Vianello}, G. and {Vilchez}, N. and {Vitale}, V. and {Waite}, A.~P. and {Wallace}, E. and {Wang}, P. and {Winer}, B.~L. and {Wolff}, M.~T. and {Wood}, D.~L. and {Wood}, K.~S. and {Yang}, Z. and {Zimmer}, S.},
        title = "{A Statistical Approach to Recognizing Source Classes for Unassociated Sources in the First Fermi-LAT Catalog}",
      journal = {\apj},
         year = 2012,
        month = jul,
       volume = {753},
       number = {1},
          eid = {83},
        pages = {83},
          doi = {10.1088/0004-637X/753/1/83},
archivePrefix = {arXiv},
       eprint = {1108.1202},
 primaryClass = {astro-ph.HE},
       adsurl = {https://ui.adsabs.harvard.edu/abs/2012ApJ...753...83A}
}

@ARTICLE{SazParkinson2016+ML,
       author = {{Saz Parkinson}, P.~M. and {Xu}, H. and {Yu}, P.~L.~H. and {Salvetti}, D. and {Marelli}, M. and {Falcone}, A.~D.},
        title = "{Classification and Ranking of Fermi LAT Gamma-ray Sources from the 3FGL Catalog using Machine Learning Techniques}",
      journal = {\apj},
         year = 2016,
        month = mar,
       volume = {820},
       number = {1},
          eid = {8},
        pages = {8},
          doi = {10.3847/0004-637X/820/1/8},
archivePrefix = {arXiv},
       eprint = {1602.00385},
 primaryClass = {astro-ph.HE},
       adsurl = {https://ui.adsabs.harvard.edu/abs/2016ApJ...820....8S}
}

@ARTICLE{Ray2012+PSC,
       author = {{Ray}, P.~S. and {Abdo}, A.~A. and {Parent}, D. and {Bhattacharya}, D. and {Bhattacharyya}, B. and {Camilo}, F. and {Cognard}, I. and {Theureau}, G. and {Ferrara}, E.~C. and {Harding}, A.~K. and {Thompson}, D.~J. and {Freire}, P.~C.~C. and {Guillemot}, L. and {Gupta}, Y. and {Roy}, J. and {Hessels}, J.~W.~T. and {Johnston}, S. and {Keith}, M. and {Shannon}, R. and {Kerr}, M. and {Michelson}, P.~F. and {Romani}, R.~W. and {Kramer}, M. and {McLaughlin}, M.~A. and {Ransom}, S.~M. and {Roberts}, M.~S.~E. and {Saz Parkinson}, P.~M. and {Ziegler}, M. and {Smith}, D.~A. and {Stappers}, B.~W. and {Weltevrede}, P. and {Wood}, K.~S.},
        title = "{Radio Searches of Fermi LAT Sources and Blind Search Pulsars: The Fermi Pulsar Search Consortium}",
      journal = {arXiv e-prints},
         year = 2012,
        month = may,
          eid = {arXiv:1205.3089},
        pages = {arXiv:1205.3089},
          doi = {10.48550/arXiv.1205.3089},
archivePrefix = {arXiv},
       eprint = {1205.3089},
 primaryClass = {astro-ph.HE},
}

@ARTICLE{Kerr2012+Parkes,
       author = {{Kerr}, M. and {Camilo}, F. and {Johnson}, T.~J. and {Ferrara}, E.~C. and {Guillemot}, L. and {Harding}, A.~K. and {Hessels}, J. and {Johnston}, S. and {Keith}, M. and {Kramer}, M. and {Ransom}, S.~M. and {Ray}, P.~S. and {Reynolds}, J.~E. and {Sarkissian}, J. and {Wood}, K.~S.},
        title = "{Five New Millisecond Pulsars from a Radio Survey of 14 Unidentified Fermi-LAT Gamma-Ray Sources}",
      journal = {\apjl},
         year = 2012,
        month = mar,
       volume = {748},
       number = {1},
          eid = {L2},
        pages = {L2},
          doi = {10.1088/2041-8205/748/1/L2},
archivePrefix = {arXiv},
       eprint = {1201.5160},
 primaryClass = {astro-ph.HE},
       adsurl = {https://ui.adsabs.harvard.edu/abs/2012ApJ...748L...2K}
}

@ARTICLE{Cromartie2016+Arecibo,
       author = {{Cromartie}, H.~T. and {Camilo}, F. and {Kerr}, M. and {Deneva}, J.~S. and {Ransom}, S.~M. and {Ray}, P.~S. and {Ferrara}, E.~C. and {Michelson}, P.~F. and {Wood}, K.~S.},
        title = "{Six New Millisecond Pulsars from Arecibo Searches of Fermi Gamma-Ray Sources}",
      journal = {\apj},
         year = 2016,
        month = mar,
       volume = {819},
       number = {1},
          eid = {34},
        pages = {34},
          doi = {10.3847/0004-637X/819/1/34},
archivePrefix = {arXiv},
       eprint = {1601.05343},
 primaryClass = {astro-ph.HE},
       adsurl = {https://ui.adsabs.harvard.edu/abs/2016ApJ...819...34C}
}

@ARTICLE{Wang2021+FASTMSP,
       author = {{Wang}, Pei and {Li}, Di and {Clark}, Colin J. and {Parkinson}, Pablo M. Saz and {Hou}, Xian and {Zhu}, Weiwei and {Qian}, Lei and {Yue}, Youling and {Pan}, Zhichen and {Liu}, Zhijie and {Yu}, Xuhong and {You}, Shanping and {Xie}, Xiaoyao and {Zhi}, Qijun and {Zhang}, Hui and {Yao}, Jumei and {Yan}, Jun and {Zhang}, Chengmin and {Fan}, Kwok Lung and {Ray}, Paul S. and {Kerr}, Matthew and {Smith}, David A. and {Michelson}, Peter F. and {Ferrara}, Elizabeth C. and {Thompson}, David J. and {Shen}, Zhiqiang and {Wang}, Na and {FAST} and {Fermi-LAT Collaboration}},
        title = "{FAST discovery of an extremely radio-faint millisecond pulsar from the Fermi-LAT unassociated source 3FGL J0318.1+0252}",
      journal = {Science China Physics, Mechanics, and Astronomy},
         year = 2021,
        month = dec,
       volume = {64},
       number = {12},
          eid = {129562},
        pages = {129562},
          doi = {10.1007/s11433-021-1757-5},
archivePrefix = {arXiv},
       eprint = {2109.00715},
 primaryClass = {astro-ph.HE},
       adsurl = {https://ui.adsabs.harvard.edu/abs/2021SCPMA..6429562W}
}

@article{Roberts2013,
    title={Surrounded by spiders! New black widows and redbacks in the Galactic field}, 
    volume={8}, 
    DOI={10.1017/S174392131202337X}, 
    number={S291}, journal={Proceedings of the International Astronomical Union}, 
    publisher={Cambridge University Press}, 
    author={Roberts, Mallory S. E.}, 
    year={2012}, pages={127–132}}

@ARTICLE{Chen2013,
       author = {{Chen}, Hai-Liang and {Chen}, Xuefei and {Tauris}, Thomas M. and {Han}, Zhanwen},
        title = "{Formation of Black Widows and Redbacks{\textemdash}Two Distinct Populations of Eclipsing Binary Millisecond Pulsars}",
      journal = {\apj},
         year = 2013,
        month = sep,
       volume = {775},
       number = {1},
          eid = {27},
        pages = {27},
          doi = {10.1088/0004-637X/775/1/27},
archivePrefix = {arXiv},
       eprint = {1308.4107},
 primaryClass = {astro-ph.SR},
       adsurl = {https://ui.adsabs.harvard.edu/abs/2013ApJ...775...27C}
}

@ARTICLE{Ghosh2024+J1242,
       author = {{Ghosh}, Ankita and {Bhattacharyya}, Bhaswati and {Lyne}, Andrew and {Kaplan}, David L. and {Roy}, Jayanta and {Ray}, Paul S. and {Stappers}, Ben and {Kumari}, Sangita and {Singh}, Shubham and {Sharan}, Rahul},
        title = "{The GMRT High-resolution Southern Sky Survey for Pulsars and Transients. VII. Timing of the Spider Millisecond Pulsar PSR J1242{\textendash}4712, a Bridge between Redback and Black Widow Pulsars}",
      journal = {\apj},
         year = 2024,
        month = apr,
       volume = {965},
       number = {1},
          eid = {64},
        pages = {64},
          doi = {10.3847/1538-4357/ad31ab},
archivePrefix = {arXiv},
       eprint = {2403.02646},
 primaryClass = {astro-ph.HE},
       adsurl = {https://ui.adsabs.harvard.edu/abs/2024ApJ...965...64G}
}

@ARTICLE{Benvenuto2014,
       author = {{Benvenuto}, O.~G. and {De Vito}, M.~A. and {Horvath}, J.~E.},
        title = "{Understanding the Evolution of Close Binary Systems with Radio Pulsars}",
      journal = {\apjl},
         year = 2014,
        month = may,
       volume = {786},
       number = {1},
          eid = {L7},
        pages = {L7},
          doi = {10.1088/2041-8205/786/1/L7},
archivePrefix = {arXiv},
       eprint = {1402.7338},
 primaryClass = {astro-ph.SR},
       adsurl = {https://ui.adsabs.harvard.edu/abs/2014ApJ...786L...7B}
}

@ARTICLE{Polzin2018+J1810,
       author = {{Polzin}, E.~J. and {Breton}, R.~P. and {Clarke}, A.~O. and {Kondratiev}, V.~I. and {Stappers}, B.~W. and {Hessels}, J.~W.~T. and {Bassa}, C.~G. and {Broderick}, J.~W. and {Grie{\ss}meier}, J. -M. and {Sobey}, C. and {ter Veen}, S. and {van Leeuwen}, J. and {Weltevrede}, P.},
        title = "{The low-frequency radio eclipses of the black widow pulsar J1810+1744}",
      journal = {\mnras},
         year = 2018,
        month = may,
       volume = {476},
       number = {2},
        pages = {1968-1981},
          doi = {10.1093/mnras/sty349},
archivePrefix = {arXiv},
       eprint = {1802.02594},
 primaryClass = {astro-ph.HE},
       adsurl = {https://ui.adsabs.harvard.edu/abs/2018MNRAS.476.1968P}
}

@ARTICLE{Polzin2019+J2051,
       author = {{Polzin}, E.~J. and {Breton}, R.~P. and {Stappers}, B.~W. and {Bhattacharyya}, B. and {Janssen}, G.~H. and {Os{\l}owski}, S. and {Roberts}, M.~S.~E. and {Sobey}, C.},
        title = "{Long-term variability of a black widow's eclipses - A decade of PSR J2051-0827}",
      journal = {\mnras},
         year = 2019,
        month = nov,
       volume = {490},
       number = {1},
        pages = {889-908},
          doi = {10.1093/mnras/stz2579},
archivePrefix = {arXiv},
       eprint = {1909.06130},
 primaryClass = {astro-ph.HE},
       adsurl = {https://ui.adsabs.harvard.edu/abs/2019MNRAS.490..889P}
}

@ARTICLE{Shang2024+J1816,
       author = {{Shang}, Lunhua and {Yu}, Yan and {Dang}, Shijun and {Bai}, Juntao and {Xu}, Xin and {Wang}, Shuangqiang and {Zhi}, Qijun and {Dong}, Aijun and {Pang}, Lijun and {Li}, Qingying and {Qiao}, Guojun},
        title = "{Studying the Radio Eclipse of Spider Pulsar J1816+4510 with the FAST}",
      journal = {\apj},
         year = 2024,
        month = jul,
       volume = {969},
       number = {1},
          eid = {62},
        pages = {62},
          doi = {10.3847/1538-4357/ad4961},
       adsurl = {https://ui.adsabs.harvard.edu/abs/2024ApJ...969...62S}
}

@ARTICLE{Nieder2020+J1653,
       author = {{Nieder}, L. and {Clark}, C.~J. and {Kandel}, D. and {Romani}, R.~W. and {Bassa}, C.~G. and {Allen}, B. and {Ashok}, A. and {Cognard}, I. and {Fehrmann}, H. and {Freire}, P. and {Karuppusamy}, R. and {Kramer}, M. and {Li}, D. and {Machenschalk}, B. and {Pan}, Z. and {Papa}, M.~A. and {Ransom}, S.~M. and {Ray}, P.~S. and {Roy}, J. and {Wang}, P. and {Wu}, J. and {Aulbert}, C. and {Barr}, E.~D. and {Beheshtipour}, B. and {Behnke}, O. and {Bhattacharyya}, B. and {Breton}, R.~P. and {Camilo}, F. and {Choquet}, C. and {Dhillon}, V.~S. and {Ferrara}, E.~C. and {Guillemot}, L. and {Hessels}, J.~W.~T. and {Kerr}, M. and {Kwang}, S.~A. and {Marsh}, T.~R. and {Mickaliger}, M.~B. and {Pleunis}, Z. and {Pletsch}, H.~J. and {Roberts}, M.~S.~E. and {Sanpa-arsa}, S. and {Steltner}, B.},
        title = "{Discovery of a Gamma-Ray Black Widow Pulsar by GPU-accelerated Einstein@Home}",
      journal = {\apjl},
         year = 2020,
        month = oct,
       volume = {902},
       number = {2},
          eid = {L46},
        pages = {L46},
          doi = {10.3847/2041-8213/abbc02},
archivePrefix = {arXiv},
       eprint = {2009.01513},
 primaryClass = {astro-ph.HE},
       adsurl = {https://ui.adsabs.harvard.edu/abs/2020ApJ...902L..46N}
}

@ARTICLE{Deneva2016+J1048,
       author = {{Deneva}, J.~S. and {Ray}, P.~S. and {Camilo}, F. and {Halpern}, J.~P. and {Wood}, K. and {Cromartie}, H.~T. and {Ferrara}, E. and {Kerr}, M. and {Ransom}, S.~M. and {Wolff}, M.~T. and {Chambers}, K.~C. and {Magnier}, E.~A.},
        title = "{Multiwavelength Observations of the Redback Millisecond Pulsar J1048+2339}",
      journal = {\apj},
         year = 2016,
        month = jun,
       volume = {823},
       number = {2},
          eid = {105},
        pages = {105},
          doi = {10.3847/0004-637X/823/2/105},
archivePrefix = {arXiv},
       eprint = {1601.03681},
 primaryClass = {astro-ph.HE},
       adsurl = {https://ui.adsabs.harvard.edu/abs/2016ApJ...823..105D}
}

@ARTICLE{Perez2023+J0212,
       author = {{Perez}, Karen I. and {Bogdanov}, Slavko and {Halpern}, Jules P. and {Gajjar}, Vishal},
        title = "{Green Bank Telescope Discovery of the Redback Binary Millisecond Pulsar PSR J0212+5321}",
      journal = {\apj},
         year = 2023,
        month = aug,
       volume = {952},
       number = {2},
          eid = {150},
        pages = {150},
          doi = {10.3847/1538-4357/acdc23},
archivePrefix = {arXiv},
       eprint = {2306.04951},
 primaryClass = {astro-ph.HE},
       adsurl = {https://ui.adsabs.harvard.edu/abs/2023ApJ...952..150P}
}

@ARTICLE{Thongmeearkom2024+RBs,
       author = {{Thongmeearkom}, T. and {Clark}, C.~J. and {Breton}, R.~P. and {Burgay}, M. and {Nieder}, L. and {Freire}, P.~C.~C. and {Barr}, E.~D. and {Stappers}, B.~W. and {Ransom}, S.~M. and {Buchner}, S. and {Calore}, F. and {Champion}, D.~J. and {Cognard}, I. and {Grie{\ss}meier}, J. -M. and {Kramer}, M. and {Levin}, L. and {Padmanabh}, P.~V. and {Possenti}, A. and {Ridolfi}, A. and {Krishnan}, V. Venkatraman and {Vleeschower}, L.},
        title = "{A targeted radio pulsar survey of redback candidates with MeerKAT}",
      journal = {\mnras},
         year = 2024,
        month = jun,
       volume = {530},
       number = {4},
        pages = {4676-4694},
          doi = {10.1093/mnras/stae787},
archivePrefix = {arXiv},
       eprint = {2403.09553},
 primaryClass = {astro-ph.HE},
       adsurl = {https://ui.adsabs.harvard.edu/abs/2024MNRAS.530.4676T}
}

@article{Stappers2018+TRAPUM,
  author = "Stappers, Ben  and  Kramer, Michael",
  title = "{An Update on TRAPUM}",
  doi = "10.22323/1.277.0009",
  journal = "PoS",
  year = 2018,
  volume = "MeerKAT2016",
  pages = "009"
}

@ARTICLE{Vleeschower2024+M62,
       author = {{Vleeschower}, L. and {Corongiu}, A. and {Stappers}, B.~W. and {Freire}, P.~C.~C. and {Ridolfi}, A. and {Abbate}, F. and {Ransom}, S.~M. and {Possenti}, A. and {Padmanabh}, P.~V. and {Balakrishnan}, V. and {Kramer}, M. and {Venkatraman Krishnan}, V. and {Zhang}, L. and {Bailes}, M. and {Barr}, E.~D. and {Buchner}, S. and {Chen}, W.},
        title = "{Discoveries and timing of pulsars in M62}",
      journal = {\mnras},
         year = 2024,
        month = may,
       volume = {530},
       number = {2},
        pages = {1436-1456},
          doi = {10.1093/mnras/stae816},
archivePrefix = {arXiv},
       eprint = {2403.12137},
 primaryClass = {astro-ph.HE},
       adsurl = {https://ui.adsabs.harvard.edu/abs/2024MNRAS.530.1436V}
}

@ARTICLE{Carli2024+SMC,
       author = {{Carli}, E. and {Levin}, L. and {Stappers}, B.~W. and {Barr}, E.~D. and {Breton}, R.~P. and {Buchner}, S. and {Burgay}, M. and {Geyer}, M. and {Kramer}, M. and {Padmanabh}, P.~V. and {Possenti}, A. and {Venkatraman Krishnan}, V. and {Becker}, W. and {Filipovi{\'c}}, M.~D. and {Maitra}, C. and {Behrend}, J. and {Champion}, D.~J. and {Chen}, W. and {Men}, Y.~P. and {Ridolfi}, A.},
        title = "{The TRAPUM Small Magellanic Cloud pulsar survey with MeerKAT - I. Discovery of seven new pulsars and two Pulsar Wind Nebula associations}",
      journal = {\mnras},
         year = 2024,
        month = jun,
       volume = {531},
       number = {2},
        pages = {2835-2863},
          doi = {10.1093/mnras/stae1310},
archivePrefix = {arXiv},
       eprint = {2405.12029},
 primaryClass = {astro-ph.HE},
       adsurl = {https://ui.adsabs.harvard.edu/abs/2024MNRAS.531.2835C}
}

@ARTICLE{Prayag2024+LMC,
       author = {{Prayag}, V. and {Levin}, L. and {Geyer}, M. and {Stappers}, B.~W. and {Carli}, E. and {Barr}, E.~D. and {Breton}, R.~P. and {Buchner}, S. and {Burgay}, M. and {Kramer}, M. and {Possenti}, A. and {Krishnan}, V. Venkatraman and {Venter}, C. and {Behrend}, J. and {Chen}, W. and {Horn}, D.~M. and {Padmanabh}, P.~V. and {Ridolfi}, A.},
        title = "{The TRAPUM Large Magellanic Cloud pulsar survey with MeerKAT - I. Survey set-up and first seven pulsar discoveries}",
      journal = {\mnras},
         year = 2024,
        month = sep,
       volume = {533},
       number = {3},
        pages = {2570-2581},
          doi = {10.1093/mnras/stae1917},
archivePrefix = {arXiv},
       eprint = {2408.04899},
 primaryClass = {astro-ph.HE},
       adsurl = {https://ui.adsabs.harvard.edu/abs/2024MNRAS.533.2570P}
}

@ARTICLE{Turner2024+SNR,
       author = {{Turner}, J.~D. and {Stappers}, B.~W. and {Carli}, E. and {Barr}, E.~D. and {Becker}, W. and {Behrend}, J. and {Breton}, R.~P. and {Buchner}, S. and {Burgay}, M. and {Champion}, D.~J. and {Chen}, W. and {Clark}, C.~J. and {Horn}, D.~M. and {Keane}, E.~F. and {Kramer}, M. and {K{\"u}nkel}, L. and {Levin}, L. and {Men}, Y.~P. and {Padmanabh}, P.~V. and {Ridolfi}, A. and {Venkatraman Krishnan}, V.},
        title = "{TRAPUM search for pulsars in supernova remnants and pulsar wind nebulae - I. Survey description and initial discoveries}",
      journal = {\mnras},
         year = 2024,
        month = jul,
       volume = {531},
       number = {3},
        pages = {3579-3594},
          doi = {10.1093/mnras/stae1300},
archivePrefix = {arXiv},
       eprint = {2405.11899},
 primaryClass = {astro-ph.HE},
       adsurl = {https://ui.adsabs.harvard.edu/abs/2024MNRAS.531.3579T}
}

@ARTICLE{Clark2023+Lband,
       author = {{Clark}, C.~J. and {Breton}, R.~P. and {Barr}, E.~D. and {Burgay}, M. and {Thongmeearkom}, T. and {Nieder}, L. and {Buchner}, S. and {Stappers}, B. and {Kramer}, M. and {Becker}, W. and {Mayer}, M. and {Phosrisom}, A. and {Ashok}, A. and {Bezuidenhout}, M.~C. and {Calore}, F. and {Cognard}, I. and {Freire}, P.~C.~C. and {Geyer}, M. and {Grie{\ss}meier}, J. -M. and {Karuppusamy}, R. and {Levin}, L. and {Padmanabh}, P.~V. and {Possenti}, A. and {Ransom}, S. and {Serylak}, M. and {Venkatraman Krishnan}, V. and {Vleeschower}, L. and {Behrend}, J. and {Champion}, D.~J. and {Chen}, W. and {Horn}, D. and {Keane}, E.~F. and {K{\"u}nkel}, L. and {Men}, Y. and {Ridolfi}, A. and {Dhillon}, V.~S. and {Marsh}, T.~R. and {Papa}, M.~A.},
        title = "{The TRAPUM L-band survey for pulsars in Fermi-LAT gamma-ray sources}",
      journal = {\mnras},
         year = 2023,
        month = mar,
       volume = {519},
       number = {4},
        pages = {5590-5606},
          doi = {10.1093/mnras/stac3742},
archivePrefix = {arXiv},
       eprint = {2212.08528},
 primaryClass = {astro-ph.HE},
       adsurl = {https://ui.adsabs.harvard.edu/abs/2023MNRAS.519.5590C}
}

@ARTICLE{Booth2009+MeerKAT,
       author = {{Booth}, R.~S. and {de Blok}, W.~J.~G. and {Jonas}, J.~L. and {Fanaroff}, B.},
        title = "{MeerKAT Key Project Science, Specifications, and Proposals}",
      journal = {arXiv e-prints},
         year = 2009,
        month = oct,
          eid = {arXiv:0910.2935},
        pages = {arXiv:0910.2935},
archivePrefix = {arXiv},
       eprint = {0910.2935},
 primaryClass = {astro-ph.IM},
       adsurl = {https://ui.adsabs.harvard.edu/abs/2009arXiv0910.2935B}
}

@INPROCEEDINGS{Jonas2016+MeerKAT,
       author = {{Jonas}, J. and {MeerKAT Team}},
        title = "{The MeerKAT Radio Telescope}",
    booktitle = {MeerKAT Science: On the Pathway to the SKA},
         year = 2016,
        month = jan,
          eid = {1},
        pages = {1},
       adsurl = {https://ui.adsabs.harvard.edu/abs/2016mks..confE...1J}
}

@INPROCEEDINGS{Barr2018+FBFUSE,
       author = {{Barr}, Ewan D.},
        title = "{An S-band Receiver and Backend System for MeerKAT}",
    booktitle = {Pulsar Astrophysics the Next Fifty Years, Jodrell Bank Observatory, UK, 2017},
         year = 2018,
       editor = {{Weltevrede}, P. and {Perera}, B.~B.~P. and {Preston}, L.~L. and {Sanidas}, S.},
       volume = {337},
        month = aug,
       series = {},
        pages = {175-178},
          doi = {10.1017/S1743921317009036},
       adsurl = {https://ui.adsabs.harvard.edu/abs/2018IAUS..337..175B}
}

@ARTICLE{Chen2021+Beamformer,
       author = {{Chen}, Weiwei and {Barr}, Ewan and {Karuppusamy}, Ramesh and {Kramer}, Michael and {Stappers}, Benjamin},
        title = "{Wide Field Beamformed Observation with MeerKAT}",
      journal = {Journal of Astronomical Instrumentation},
         year = 2021,
        month = jan,
       volume = {10},
       number = {3},
          eid = {2150013-178},
        pages = {2150013-178},
          doi = {10.1142/S2251171721500136},
archivePrefix = {arXiv},
       eprint = {2110.01667},
 primaryClass = {astro-ph.IM},
       adsurl = {https://ui.adsabs.harvard.edu/abs/2021JAI....1050013C}
}

@ARTICLE{Bailes2020+MeerTIME,
       author = {{Bailes}, M. and {Jameson}, A. and {Abbate}, F. and {Barr}, E.~D. and {Bhat}, N.~D.~R. and {Bondonneau}, L. and {Burgay}, M. and {Buchner}, S.~J. and {Camilo}, F. and {Champion}, D.~J. and {Cognard}, I. and {Demorest}, P.~B. and {Freire}, P.~C.~C. and {Gautam}, T. and {Geyer}, M. and {Griessmeier}, J. -M. and {Guillemot}, L. and {Hu}, H. and {Jankowski}, F. and {Johnston}, S. and {Karastergiou}, A. and {Karuppusamy}, R. and {Kaur}, D. and {Keith}, M.~J. and {Kramer}, M. and {van Leeuwen}, J. and {Lower}, M.~E. and {Maan}, Y. and {McLaughlin}, M.~A. and {Meyers}, B.~W. and {Os{\l}owski}, S. and {Oswald}, L.~S. and {Parthasarathy}, A. and {Pennucci}, T. and {Posselt}, B. and {Possenti}, A. and {Ransom}, S.~M. and {Reardon}, D.~J. and {Ridolfi}, A. and {Schollar}, C.~T.~G. and {Serylak}, M. and {Shaifullah}, G. and {Shamohammadi}, M. and {Shannon}, R.~M. and {Sobey}, C. and {Song}, X. and {Spiewak}, R. and {Stairs}, I.~H. and {Stappers}, B.~W. and {van Straten}, W. and {Szary}, A. and {Theureau}, G. and {Venkatraman Krishnan}, V. and {Weltevrede}, P. and {Wex}, N. and {Abbott}, T.~D. and {Adams}, G.~B. and {Burger}, J.~P. and {Gamatham}, R.~R.~G. and {Gouws}, M. and {Horn}, D.~M. and {Hugo}, B. and {Joubert}, A.~F. and {Manley}, J.~R. and {McAlpine}, K. and {Passmoor}, S.~S. and {Peens-Hough}, A. and {Ramudzuli}, Z.~R. and {Rust}, A. and {Salie}, S. and {Schwardt}, L.~C. and {Siebrits}, R. and {Van Tonder}, G. and {Van Tonder}, V. and {Welz}, M.~G.},
        title = "{The MeerKAT telescope as a pulsar facility: System verification and early science results from MeerTime}",
      journal = {\pasa},
         year = 2020,
        month = jul,
       volume = {37},
          eid = {e028},
        pages = {e028},
          doi = {10.1017/pasa.2020.19},
archivePrefix = {arXiv},
       eprint = {2005.14366},
 primaryClass = {astro-ph.IM},
       adsurl = {https://ui.adsabs.harvard.edu/abs/2020PASA...37...28B}
}

@ARTICLE{Haslam1982,
       author = {{Haslam}, C.~G.~T. and {Salter}, C.~J. and {Stoffel}, H. and {Wilson}, W.~E.},
        title = "{A 408-MHZ All-Sky Continuum Survey. II. The Atlas of Contour Maps}",
      journal = {\aaps},
         year = 1982,
        month = jan,
       volume = {47},
        pages = {1},
       adsurl = {https://ui.adsabs.harvard.edu/abs/1982A&AS...47....1H}
}

@ARTICLE{Remazeilles2015,
       author = {{Remazeilles}, M. and {Dickinson}, C. and {Banday}, A.~J. and {Bigot-Sazy}, M. -A. and {Ghosh}, T.},
        title = "{An improved source-subtracted and destriped 408-MHz all-sky map}",
      journal = {\mnras},
         year = 2015,
        month = aug,
       volume = {451},
       number = {4},
        pages = {4311-4327},
          doi = {10.1093/mnras/stv1274},
archivePrefix = {arXiv},
       eprint = {1411.3628},
 primaryClass = {astro-ph.IM},
       adsurl = {https://ui.adsabs.harvard.edu/abs/2015MNRAS.451.4311R}
}

@ARTICLE{Keith2011+Parkes,
       author = {{Keith}, M.~J. and {Johnston}, S. and {Ray}, P.~S. and {Ferrara}, E.~C. and {Saz Parkinson}, P.~M. and {{\c{C}}elik}, {\"O}. and {Belfiore}, A. and {Donato}, D. and {Cheung}, C.~C. and {Abdo}, A.~A. and {Camilo}, F. and {Freire}, P.~C.~C. and {Guillemot}, L. and {Harding}, A.~K. and {Kramer}, M. and {Michelson}, P.~F. and {Ransom}, S.~M. and {Romani}, R.~W. and {Smith}, D.~A. and {Thompson}, D.~J. and {Weltevrede}, P. and {Wood}, K.~S.},
        title = "{Discovery of millisecond pulsars in radio searches of southern Fermi Large Area Telescope sources}",
      journal = {\mnras},
         year = 2011,
        month = jun,
       volume = {414},
       number = {2},
        pages = {1292-1300},
          doi = {10.1111/j.1365-2966.2011.18464.x},
archivePrefix = {arXiv},
       eprint = {1102.0648},
 primaryClass = {astro-ph.HE},
       adsurl = {https://ui.adsabs.harvard.edu/abs/2011MNRAS.414.1292K},
}

@ARTICLE{Ransom2011+GBT,
       author = {{Ransom}, S.~M. and {Ray}, P.~S. and {Camilo}, F. and {Roberts}, M.~S.~E. and {{\c{C}}elik}, {\"O}. and {Wolff}, M.~T. and {Cheung}, C.~C. and {Kerr}, M. and {Pennucci}, T. and {DeCesar}, M.~E. and {Cognard}, I. and {Lyne}, A.~G. and {Stappers}, B.~W. and {Freire}, P.~C.~C. and {Grove}, J.~E. and {Abdo}, A.~A. and {Desvignes}, G. and {Donato}, D. and {Ferrara}, E.~C. and {Gehrels}, N. and {Guillemot}, L. and {Gwon}, C. and {Harding}, A.~K. and {Johnston}, S. and {Keith}, M. and {Kramer}, M. and {Michelson}, P.~F. and {Parent}, D. and {Saz Parkinson}, P.~M. and {Romani}, R.~W. and {Smith}, D.~A. and {Theureau}, G. and {Thompson}, D.~J. and {Weltevrede}, P. and {Wood}, K.~S. and {Ziegler}, M.},
        title = "{Three Millisecond Pulsars in Fermi LAT Unassociated Bright Sources}",
      journal = {\apjl},
         year = 2011,
        month = jan,
       volume = {727},
       number = {1},
          eid = {L16},
        pages = {L16},
          doi = {10.1088/2041-8205/727/1/L16},
archivePrefix = {arXiv},
       eprint = {1012.2862},
 primaryClass = {astro-ph.HE},
       adsurl = {https://ui.adsabs.harvard.edu/abs/2011ApJ...727L..16R}
}

@ARTICLE{Camilo2015+Parkes,
       author = {{Camilo}, F. and {Kerr}, M. and {Ray}, P.~S. and {Ransom}, S.~M. and {Sarkissian}, J. and {Cromartie}, H.~T. and {Johnston}, S. and {Reynolds}, J.~E. and {Wolff}, M.~T. and {Freire}, P.~C.~C. and {Bhattacharyya}, B. and {Ferrara}, E.~C. and {Keith}, M. and {Michelson}, P.~F. and {Saz Parkinson}, P.~M. and {Wood}, K.~S.},
        title = "{Parkes Radio Searches of Fermi Gamma-Ray Sources and Millisecond Pulsar Discoveries}",
      journal = {\apj},
         year = 2015,
        month = sep,
       volume = {810},
       number = {2},
          eid = {85},
        pages = {85},
          doi = {10.1088/0004-637X/810/2/85},
archivePrefix = {arXiv},
       eprint = {1507.04451},
 primaryClass = {astro-ph.HE},
       adsurl = {https://ui.adsabs.harvard.edu/abs/2015ApJ...810...85C}
}

@article{Polzin2020,
    author = {Polzin, E J and Breton, R P and Bhattacharyya, B and Scholte, D and Sobey, C and Stappers, B W},
    title = "{Study of spider pulsar binary eclipses and discovery of an eclipse mechanism transition}",
    journal = {\mnras},
    volume = {494},
    number = {2},
    pages = {2948-2968},
    year = {2020},
    month = {03},
    issn = {0035-8711},
    doi = {10.1093/mnras/staa596},
    url = {https://doi.org/10.1093/mnras/staa596},
    eprint = {https://academic.oup.com/mnras/article-pdf/494/2/2948/33129109/staa596.pdf},
}

@ARTICLE{Jankowski2018,
       author = {{Jankowski}, F. and {van Straten}, W. and {Keane}, E.~F. and {Bailes}, M. and {Barr}, E.~D. and {Johnston}, S. and {Kerr}, M.},
        title = "{Spectral properties of 441 radio pulsars}",
      journal = {\mnras},
         year = 2018,
        month = feb,
       volume = {473},
       number = {4},
        pages = {4436-4458},
          doi = {10.1093/mnras/stx2476},
archivePrefix = {arXiv},
       eprint = {1709.08864},
 primaryClass = {astro-ph.HE},
       adsurl = {https://ui.adsabs.harvard.edu/abs/2018MNRAS.473.4436J}
}

@ARTICLE{Morello2019+HTRUrepro,
       author = {{Morello}, V. and {Barr}, E.~D. and {Cooper}, S. and {Bailes}, M. and {Bates}, S. and {Bhat}, N.~D.~R. and {Burgay}, M. and {Burke-Spolaor}, S. and {Cameron}, A.~D. and {Champion}, D.~J. and {Eatough}, R.~P. and {Flynn}, C.~M.~L. and {Jameson}, A. and {Johnston}, S. and {Keith}, M.~J. and {Keane}, E.~F. and {Kramer}, M. and {Levin}, L. and {Ng}, C. and {Petroff}, E. and {Possenti}, A. and {Stappers}, B.~W. and {van Straten}, W. and {Tiburzi}, C.},
        title = "{The High Time Resolution Universe survey - XIV. Discovery of 23 pulsars through GPU-accelerated reprocessing}",
      journal = {\mnras},
         year = 2019,
        month = mar,
       volume = {483},
       number = {3},
        pages = {3673-3685},
          doi = {10.1093/mnras/sty3328},
archivePrefix = {arXiv},
       eprint = {1811.04929},
 primaryClass = {astro-ph.IM},
       adsurl = {https://ui.adsabs.harvard.edu/abs/2019MNRAS.483.3673M}
}

@MISC{Barr2020+peasoup,
       author = {{Barr}, Ewen},
        title = "{Peasoup: C++/CUDA GPU pulsar searching library}",
         year = 2020,
        month = jan,
          eid = {ascl:2001.014},
        pages = {ascl:2001.014},
archivePrefix = {ascl},
       eprint = {2001.014},
       adsurl = {https://ui.adsabs.harvard.edu/abs/2020ascl.soft01014B}
}

@ARTICLE{YMW16,
       author = {{Yao}, J.~M. and {Manchester}, R.~N. and {Wang}, N.},
        title = "{A New Electron-density Model for Estimation of Pulsar and FRB Distances}",
      journal = {\apj},
         year = 2017,
        month = jan,
       volume = {835},
       number = {1},
          eid = {29},
        pages = {29},
          doi = {10.3847/1538-4357/835/1/29},
archivePrefix = {arXiv},
       eprint = {1610.09448},
 primaryClass = {astro-ph.GA},
       adsurl = {https://ui.adsabs.harvard.edu/abs/2017ApJ...835...29Y}
}

@ARTICLE{Strader2019+RBs,
       author = {{Strader}, Jay and {Swihart}, Samuel and {Chomiuk}, Laura and {Bahramian}, Arash and {Britt}, Chris and {Cheung}, C.~C. and {Dage}, Kristen and {Halpern}, Jules and {Li}, Kwan-Lok and {Mignani}, Roberto P. and {Orosz}, Jerome A. and {Peacock}, Mark and {Salinas}, Ricardo and {Shishkovsky}, Laura and {Tremou}, Evangelia},
        title = "{Optical Spectroscopy and Demographics of Redback Millisecond Pulsar Binaries}",
      journal = {\apj},
         year = 2019,
        month = feb,
       volume = {872},
       number = {1},
          eid = {42},
        pages = {42},
          doi = {10.3847/1538-4357/aafbaa},
archivePrefix = {arXiv},
       eprint = {1812.04626},
 primaryClass = {astro-ph.HE},
       adsurl = {https://ui.adsabs.harvard.edu/abs/2019ApJ...872...42S}
}

@ARTICLE{Zhu2014+PICS,
       author = {{Zhu}, W.~W. and {Berndsen}, A. and {Madsen}, E.~C. and {Tan}, M. and {Stairs}, I.~H. and {Brazier}, A. and {Lazarus}, P. and {Lynch}, R. and {Scholz}, P. and {Stovall}, K. and {Ransom}, S.~M. and {Banaszak}, S. and {Biwer}, C.~M. and {Cohen}, S. and {Dartez}, L.~P. and {Flanigan}, J. and {Lunsford}, G. and {Martinez}, J.~G. and {Mata}, A. and {Rohr}, M. and {Walker}, A. and {Allen}, B. and {Bhat}, N.~D.~R. and {Bogdanov}, S. and {Camilo}, F. and {Chatterjee}, S. and {Cordes}, J.~M. and {Crawford}, F. and {Deneva}, J.~S. and {Desvignes}, G. and {Ferdman}, R.~D. and {Freire}, P.~C.~C. and {Hessels}, J.~W.~T. and {Jenet}, F.~A. and {Kaplan}, D.~L. and {Kaspi}, V.~M. and {Knispel}, B. and {Lee}, K.~J. and {van Leeuwen}, J. and {Lyne}, A.~G. and {McLaughlin}, M.~A. and {Siemens}, X. and {Spitler}, L.~G. and {Venkataraman}, A.},
        title = "{Searching for Pulsars Using Image Pattern Recognition}",
      journal = {\apj},
         year = 2014,
        month = feb,
       volume = {781},
       number = {2},
          eid = {117},
        pages = {117},
          doi = {10.1088/0004-637X/781/2/117},
archivePrefix = {arXiv},
       eprint = {1309.0776},
 primaryClass = {astro-ph.IM},
       adsurl = {https://ui.adsabs.harvard.edu/abs/2014ApJ...781..117Z}
}

@ARTICLE{Men2023+PulsarX,
       author = {{Men}, Yunpeng and {Barr}, Ewan and {Clark}, Colin J. and {Carli}, Emma and {Desvignes}, Gregory},
        title = "{PulsarX: A new pulsar searching package. I. A high performance folding program for pulsar surveys}",
      journal = {\aap},
         year = 2023,
        month = nov,
       volume = {679},
          eid = {A20},
        pages = {A20},
          doi = {10.1051/0004-6361/202347356},
archivePrefix = {arXiv},
       eprint = {2309.02544},
 primaryClass = {astro-ph.IM},
       adsurl = {https://ui.adsabs.harvard.edu/abs/2023A&A...679A..20M}
}

@ARTICLE{Lee2012+GMM,
       author = {{Lee}, K.~J. and {Guillemot}, L. and {Yue}, Y.~L. and {Kramer}, M. and {Champion}, D.~J.},
        title = "{Application of the Gaussian mixture model in pulsar astronomy - pulsar classification and candidates ranking for the Fermi 2FGL catalogue}",
      journal = {\mnras},
         year = 2012,
        month = aug,
       volume = {424},
       number = {4},
        pages = {2832-2840},
          doi = {10.1111/j.1365-2966.2012.21413.x},
archivePrefix = {arXiv},
       eprint = {1205.6221},
 primaryClass = {astro-ph.IM},
       adsurl = {https://ui.adsabs.harvard.edu/abs/2012MNRAS.424.2832L}
}

@ARTICLE{Luo2020+ML,
       author = {{Luo}, Shengda and {Leung}, Alex P. and {Hui}, C.~Y. and {Li}, K.~L.},
        title = "{An investigation on the factors affecting machine learning classifications in gamma-ray astronomy}",
      journal = {\mnras},
         year = 2020,
        month = mar,
       volume = {492},
       number = {4},
        pages = {5377-5390},
          doi = {10.1093/mnras/staa166},
archivePrefix = {arXiv},
       eprint = {2001.04081},
 primaryClass = {astro-ph.IM},
       adsurl = {https://ui.adsabs.harvard.edu/abs/2020MNRAS.492.5377L}
}

@ARTICLE{Finke2021+DNN,
       author = {{Finke}, Thorben and {Kr{\"a}mer}, Michael and {Manconi}, Silvia},
        title = "{Classification of Fermi-LAT sources with deep learning using energy and time spectra}",
      journal = {\mnras},
         year = 2021,
        month = nov,
       volume = {507},
       number = {3},
        pages = {4061-4073},
          doi = {10.1093/mnras/stab2389},
archivePrefix = {arXiv},
       eprint = {2012.05251},
 primaryClass = {astro-ph.HE},
       adsurl = {https://ui.adsabs.harvard.edu/abs/2021MNRAS.507.4061F}
}

@ARTICLE{Padmanabh2023+MGPS,
       author = {{Padmanabh}, P.~V. and {Barr}, E.~D. and {Sridhar}, S.~S. and {Rugel}, M.~R. and {Damas-Segovia}, A. and {Jacob}, A.~M. and {Balakrishnan}, V. and {Berezina}, M. and {Bernadich}, M.~C. and {Brunthaler}, A. and {Champion}, D.~J. and {Freire}, P.~C.~C. and {Khan}, S. and {Kl{\"o}ckner}, H. -R. and {Kramer}, M. and {Ma}, Y.~K. and {Mao}, S.~A. and {Men}, Y.~P. and {Menten}, K.~M. and {Sengupta}, S. and {Venkatraman Krishnan}, V. and {Wucknitz}, O. and {Wyrowski}, F. and {Bezuidenhout}, M.~C. and {Buchner}, S. and {Burgay}, M. and {Chen}, W. and {Clark}, C.~J. and {K{\"u}nkel}, L. and {Nieder}, L. and {Stappers}, B. and {Legodi}, L.~S. and {Nyamai}, M.~M.},
        title = "{The MPIfR-MeerKAT Galactic Plane Survey - I. System set-up and early results}",
      journal = {\mnras},
         year = 2023,
        month = sep,
       volume = {524},
       number = {1},
        pages = {1291-1315},
          doi = {10.1093/mnras/stad1900},
archivePrefix = {arXiv},
       eprint = {2303.09231},
 primaryClass = {astro-ph.HE},
       adsurl = {https://ui.adsabs.harvard.edu/abs/2023MNRAS.524.1291P}
}

@ARTICLE{Au2023+J1910,
       author = {{Au}, Ka-Yui and {Strader}, Jay and {Swihart}, Samuel J. and {Lin}, Lupin C.~C. and {Kong}, Albert K.~H. and {Takata}, Jumpei and {Hui}, Chung-Yue and {Panurach}, Teresa and {Molina}, Isabella and {Aydi}, Elias and {Sokolovsky}, Kirill and {Li}, Kwan-Lok},
        title = "{Multiwavelength Observations of a New Redback Millisecond Pulsar 4FGL J1910.7-5320}",
      journal = {\apj},
         year = 2023,
        month = feb,
       volume = {943},
       number = {2},
          eid = {103},
        pages = {103},
          doi = {10.3847/1538-4357/acae8a},
archivePrefix = {arXiv},
       eprint = {2212.11618},
 primaryClass = {astro-ph.HE},
       adsurl = {https://ui.adsabs.harvard.edu/abs/2023ApJ...943..103A}
}

@ARTICLE{Djorgovski2011+CRTS,
       author = {{Djorgovski}, S.~G. and {Drake}, A.~J. and {Mahabal}, A.~A. and {Graham}, M.~J. and {Donalek}, C. and {Williams}, R. and {Beshore}, E.~C. and {Larson}, S.~M. and {Prieto}, J. and {Catelan}, M. and {Christensen}, E. and {McNaught}, R.~H.},
        title = "{The Catalina Real-Time Transient Survey (CRTS)}",
      journal = {arXiv e-prints},
         year = 2011,
        month = feb,
          eid = {arXiv:1102.5004},
        pages = {arXiv:1102.5004},
          doi = {10.48550/arXiv.1102.5004},
archivePrefix = {arXiv},
       eprint = {1102.5004},
 primaryClass = {astro-ph.IM},
       adsurl = {https://ui.adsabs.harvard.edu/abs/2011arXiv1102.5004D}
}

@INPROCEEDINGS{Clemens2004+SOAR,
       author = {{Clemens}, J. Christopher and {Crain}, J. Adam and {Anderson}, Robert},
        title = "{The Goodman spectrograph}",
    booktitle = {Ground-based Instrumentation for Astronomy},
         year = 2004,
       editor = {{Moorwood}, Alan F.~M. and {Iye}, Masanori},
       series = {Society of Photo-Optical Instrumentation Engineers (SPIE) Conference Series},
       volume = {5492},
        month = sep,
        pages = {331-340},
          doi = {10.1117/12.550069},
       adsurl = {https://ui.adsabs.harvard.edu/abs/2004SPIE.5492..331C}
}

@ARTICLE{Hobbs2020+UWL,
       author = {{Hobbs}, George and {Manchester}, Richard N. and {Dunning}, Alex and {Jameson}, Andrew and {Roberts}, Paul and {George}, Daniel and {Green}, J.~A. and {Tuthill}, John and {Toomey}, Lawrence and {Kaczmarek}, Jane F. and {Mader}, Stacy and {Marquarding}, Malte and {Ahmed}, Azeem and {Amy}, Shaun W. and {Bailes}, Matthew and {Beresford}, Ron and {Bhat}, N.~D.~R. and {Bock}, Douglas C. -J. and {Bourne}, Michael and {Bowen}, Mark and {Brothers}, Michael and {Cameron}, Andrew D. and {Carretti}, Ettore and {Carter}, Nick and {Castillo}, Santy and {Chekkala}, Raji and {Cheng}, Wan and {Chung}, Yoon and {Craig}, Daniel A. and {Dai}, Shi and {Dawson}, Joanne and {Dempsey}, James and {Doherty}, Paul and {Dong}, Bin and {Edwards}, Philip and {Ergesh}, Tuohutinuer and {Gao}, Xuyang and {Han}, JinLin and {Hayman}, Douglas and {Indermuehle}, Balthasar and {Jeganathan}, Kanapathippillai and {Johnston}, Simon and {Kanoniuk}, Henry and {Kesteven}, Michael and {Kramer}, Michael and {Leach}, Mark and {Mcintyre}, Vince and {Moss}, Vanessa and {Os{\l}owski}, Stefan and {Phillips}, Chris and {Pope}, Nathan and {Preisig}, Brett and {Price}, Daniel and {Reeves}, Ken and {Reilly}, Les and {Reynolds}, John and {Robishaw}, Tim and {Roush}, Peter and {Ruckley}, Tim and {Sadler}, Elaine and {Sarkissian}, John and {Severs}, Sean and {Shannon}, Ryan and {Smart}, Ken and {Smith}, Malcolm and {Smith}, Stephanie and {Sobey}, Charlotte and {Staveley-Smith}, Lister and {Tzioumis}, Anastasios and {van Straten}, Willem and {Wang}, Nina and {Wen}, Linqing and {Whiting}, Matthew},
        title = "{An ultra-wide bandwidth (704 to 4 032 MHz) receiver for the Parkes radio telescope}",
      journal = {\pasa},
         year = 2020,
        month = apr,
       volume = {37},
          eid = {e012},
        pages = {e012},
          doi = {10.1017/pasa.2020.2},
archivePrefix = {arXiv},
       eprint = {1911.00656},
 primaryClass = {astro-ph.IM},
       adsurl = {https://ui.adsabs.harvard.edu/abs/2020PASA...37...12H}
}

@ARTICLE{Obrocka2015+reloc,
       author = {{Obrocka}, M. and {Stappers}, B. and {Wilkinson}, P.},
        title = "{Localising fast radio bursts and other transients using interferometric arrays}",
      journal = {\aap},
         year = 2015,
        month = jul,
       volume = {579},
          eid = {A69},
        pages = {A69},
          doi = {10.1051/0004-6361/201425538},
archivePrefix = {arXiv},
       eprint = {1502.06825},
 primaryClass = {astro-ph.IM},
       adsurl = {https://ui.adsabs.harvard.edu/abs/2015A&A...579A..69O}
}

@ARTICLE{SeeKAT,
       author = {{Bezuidenhout}, M.~C. and {Clark}, C.~J. and {Breton}, R.~P. and {Stappers}, B.~W. and {Barr}, E.~D. and {Caleb}, M. and {Chen}, W. and {Jankowski}, F. and {Kramer}, M. and {Rajwade}, K. and {Surnis}, M.},
        title = "{Tied-array beam localization of radio transients and pulsars}",
      journal = {RAS Techniques and Instruments},
         year = 2023,
        month = jan,
       volume = {2},
       number = {1},
        pages = {114-128},
          doi = {10.1093/rasti/rzad007},
archivePrefix = {arXiv},
       eprint = {2302.09812},
 primaryClass = {astro-ph.HE},
       adsurl = {https://ui.adsabs.harvard.edu/abs/2023RASTI...2..114B}
}

@ARTICLE{Burgay2024+timing,
       author = {{Burgay}, M. and {Nieder}, L. and {Clark}, C.~J. and {Freire}, P.~C.~C. and {Buchner}, S. and {Thongmeearkom}, T. and {Turner}, J.~D. and {Carli}, E. and {Cognard}, I. and {Grie{\ss}meier}, J. -M. and {Karuppusamy}, R. and {i Bernadich}, M.~C. and {Possenti}, A. and {Venkatraman Krishnan}, V. and {Breton}, R.~P. and {Barr}, E.~D. and {Stappers}, B.~W. and {Kramer}, M. and {Levin}, L. and {Ransom}, S.~M. and {Padmanabh}, P.~V.},
        title = "{Radio and gamma-ray timing of TRAPUM L-band Fermi pulsar survey discoveries}",
      journal = {\aap},
         year = 2024,
        month = nov,
       volume = {691},
          eid = {A315},
        pages = {A315},
          doi = {10.1051/0004-6361/202451530},
archivePrefix = {arXiv},
       eprint = {2411.14895},
 primaryClass = {astro-ph.HE},
       adsurl = {https://ui.adsabs.harvard.edu/abs/2024A&A...691A.315B}
}

@misc{Ransom2011+PRESTO,
       author = {{Ransom}, Scott},
        title = "{PRESTO: PulsaR Exploration and Search TOolkit}",
 howpublished = {Astrophysics Source Code Library, record ascl:1107.017},
         year = 2011,
        month = jul,
          eid = {ascl:1107.017},
       adsurl = {https://ui.adsabs.harvard.edu/abs/2011ascl.soft07017R}
}

@ARTICLE{tempo2,
       author = {{Hobbs}, G.~B. and {Edwards}, R.~T. and {Manchester}, R.~N.},
        title = "{TEMPO2, a new pulsar-timing package - I. An overview}",
      journal = {\mnras},
         year = 2006,
        month = jun,
       volume = {369},
       number = {2},
        pages = {655-672},
          doi = {10.1111/j.1365-2966.2006.10302.x},
archivePrefix = {arXiv},
       eprint = {astro-ph/0603381},
 primaryClass = {astro-ph},
       adsurl = {https://ui.adsabs.harvard.edu/abs/2006MNRAS.369..655H}
}

@ARTICLE{Freire2018+DRACULA,
       author = {{Freire}, Paulo C.~C. and {Ridolfi}, Alessandro},
        title = "{An algorithm for determining the rotation count of pulsars}",
      journal = {\mnras},
         year = 2018,
        month = jun,
       volume = {476},
       number = {4},
        pages = {4794-4805},
          doi = {10.1093/mnras/sty524},
archivePrefix = {arXiv},
       eprint = {1802.07211},
 primaryClass = {astro-ph.IM},
       adsurl = {https://ui.adsabs.harvard.edu/abs/2018MNRAS.476.4794F}
}

@ARTICLE{Swihart2022+J1408,
       author = {{Swihart}, Samuel J. and {Strader}, Jay and {Chomiuk}, Laura and {Aydi}, Elias and {Sokolovsky}, Kirill V. and {Ray}, Paul S. and {Kerr}, Matthew},
        title = "{A New Flaring Black Widow Candidate and Demographics of Black Widow Millisecond Pulsars in the Galactic Field}",
      journal = {\apj},
         year = 2022,
        month = dec,
       volume = {941},
       number = {2},
          eid = {199},
        pages = {199},
          doi = {10.3847/1538-4357/aca2ac},
archivePrefix = {arXiv},
       eprint = {2210.16295},
 primaryClass = {astro-ph.HE},
       adsurl = {https://ui.adsabs.harvard.edu/abs/2022ApJ...941..199S}
}

@ARTICLE{Callanan1995+B1957,
       author = {{Callanan}, Paul J. and {van Paradijs}, Jan and {Rengelink}, Roeland},
        title = "{The Orbital Light Curve of PSR 1957+20}",
      journal = {\apj},
         year = 1995,
        month = feb,
       volume = {439},
        pages = {928},
          doi = {10.1086/175229},
       adsurl = {https://ui.adsabs.harvard.edu/abs/1995ApJ...439..928C}
}

@ARTICLE{Breton2013,
       author = {{Breton}, R.~P. and {van Kerkwijk}, M.~H. and {Roberts}, M.~S.~E. and {Hessels}, J.~W.~T. and {Camilo}, F. and {McLaughlin}, M.~A. and {Ransom}, S.~M. and {Ray}, P.~S. and {Stairs}, I.~H.},
        title = "{Discovery of the Optical Counterparts to Four Energetic Fermi Millisecond Pulsars}",
      journal = {\apj},
         year = 2013,
        month = jun,
       volume = {769},
       number = {2},
          eid = {108},
        pages = {108},
          doi = {10.1088/0004-637X/769/2/108},
archivePrefix = {arXiv},
       eprint = {1302.1790},
 primaryClass = {astro-ph.HE},
       adsurl = {https://ui.adsabs.harvard.edu/abs/2013ApJ...769..108B}
}

@ARTICLE{Dodge2024+J1910,
       author = {{Dodge}, O.~G. and {Breton}, R.~P. and {Clark}, C.~J. and {Burgay}, M. and {Strader}, J. and {Au}, K. -Y. and {Barr}, E.~D. and {Buchner}, S. and {Dhillon}, V.~S. and {Ferrara}, E.~C. and {Freire}, P.~C.~C. and {Griessmeier}, J. -M. and {Kennedy}, M.~R. and {Kramer}, M. and {Li}, K. -L. and {Padmanabh}, P.~V. and {Phosrisom}, A. and {Stappers}, B.~W. and {Swihart}, S.~J. and {Thongmeearkom}, T.},
        title = "{Mass estimates from optical modelling of the new TRAPUM redback PSR J1910-5320}",
      journal = {\mnras},
         year = 2024,
        month = mar,
       volume = {528},
       number = {3},
        pages = {4337-4353},
          doi = {10.1093/mnras/stae211},
archivePrefix = {arXiv},
       eprint = {2401.09928},
 primaryClass = {astro-ph.HE},
       adsurl = {https://ui.adsabs.harvard.edu/abs/2024MNRAS.528.4337D}
}

@ARTICLE{Mata2020+J1012,
       author = {{Mata S{\'a}nchez}, D. and {Istrate}, A.~G. and {van Kerkwijk}, M.~H. and {Breton}, R.~P. and {Kaplan}, D.~L.},
        title = "{PSR J1012+5307: a millisecond pulsar with an extremely low-mass white dwarf companion}",
      journal = {\mnras},
         year = 2020,
        month = may,
       volume = {494},
       number = {3},
        pages = {4031-4042},
          doi = {10.1093/mnras/staa983},
archivePrefix = {arXiv},
       eprint = {2004.02901},
 primaryClass = {astro-ph.HE},
       adsurl = {https://ui.adsabs.harvard.edu/abs/2020MNRAS.494.4031M}
}

@ARTICLE{Nieder2020+Methods,
       author = {{Nieder}, L. and {Allen}, B. and {Clark}, C.~J. and {Pletsch}, H.~J.},
        title = "{Exploiting Orbital Constraints from Optical Data to Detect Binary Gamma-Ray Pulsars}",
      journal = {\apj},
         year = 2020,
        month = oct,
       volume = {901},
       number = {2},
          eid = {156},
        pages = {156},
          doi = {10.3847/1538-4357/abaf53},
archivePrefix = {arXiv},
       eprint = {2004.11740},
 primaryClass = {astro-ph.HE},
       adsurl = {https://ui.adsabs.harvard.edu/abs/2020ApJ...901..156N}
}

@ARTICLE{deJaeger1989+Htest,
       author = {{de Jager}, O.~C. and {Raubenheimer}, B.~C. and {Swanepoel}, J.~W.~H.},
        title = "{A powerful test for weak periodic signals with unknown light curve shape in sparse data.}",
      journal = {\aap},
         year = 1989,
        month = aug,
       volume = {221},
        pages = {180-190},
       adsurl = {https://ui.adsabs.harvard.edu/abs/1989A&A...221..180D}
}

@article{Kerr2011,
   author = {{Kerr}, M.},
    title = "{Improving Sensitivity to Weak Pulsations with Photon Probability Weighting}",
  journal = {\apj},
archivePrefix = "arXiv",
   eprint = {1103.2128},
 primaryClass = "astro-ph.IM",
     year = 2011,
    month = may,
   volume = 732,
      eid = {38},
    pages = {38},
      doi = {10.1088/0004-637X/732/1/38},
   adsurl = {http://adsabs.harvard.edu/abs/2011ApJ...732...38K}
}

@inproceedings{Pass8,
  author = {{Atwood}, W. and {Albert}, A. and {Baldini}, L. and {Tinivella}, M. and 
	{Bregeon}, J. and {Pesce-Rollins}, M. and {Sgr{\`o}}, C. and 
	{Bruel}, P. and {Charles}, E. and {Drlica-Wagner}, A. and {Franckowiak}, A. and 
	{Jogler}, T. and {Rochester}, L. and {Usher}, T. and {Wood}, M. and 
	{Cohen-Tanugi}, J. and {S.~Zimmer for the Fermi-LAT Collaboration}
	},
  title = "{Pass 8: Toward the Full Realization of the Fermi-LAT Scientific Potential}",
  booktitle = {Proceedings of the 4th Fermi Symposium, Monterey, California, 2012},
  editor = {{Brandt}, T.~J. and {Omodei}, N. and {Wilson-Hodge}, C.},
  year = {2013},
  series = {eConf C121028},
  pages = {8},
  journal = {ArXiv e-prints},
  archivePrefix = "arXiv",
  note = {arXiv:1303.3514},
  primaryClass = "astro-ph.IM",
}

@ARTICLE{Bruel2018+P305,
       author = {{Bruel}, P. and {Burnett}, T.~H. and {Digel}, S.~W. and {Johannesson},
        G. and {Omodei}, N. and {Wood}, M.},
        title = "{Fermi-LAT improved Pass\textasciitilde8 event selection}",
      journal = {arXiv e-prints},
         year = 2018,
        month = Oct,
          eid = {arXiv:1810.11394},
        pages = {arXiv:1810.11394},
archivePrefix = {arXiv},
       eprint = {1810.11394},
 primaryClass = {astro-ph.IM},
       adsurl = {https://ui.adsabs.harvard.edu/\#abs/2018arXiv181011394B}
}

@ARTICLE{Smith2019+edot,
       author = {{Smith}, D.~A. and {Bruel}, P. and {Cognard}, I. and {Cameron}, A.~D. and {Camilo}, F. and {Dai}, S. and {Guillemot}, L. and {Johnson}, T.~J. and {Johnston}, S. and {Keith}, M.~J. and {Kerr}, M. and {Kramer}, M. and {Lyne}, A.~G. and {Manchester}, R.~N. and {Shannon}, R. and {Sobey}, C. and {Stappers}, B.~W. and {Weltevrede}, P.},
        title = "{Searching a Thousand Radio Pulsars for Gamma-Ray Emission}",
      journal = {\apj},
         year = 2019,
        month = jan,
       volume = {871},
       number = {1},
          eid = {78},
        pages = {78},
          doi = {10.3847/1538-4357/aaf57d},
archivePrefix = {arXiv},
       eprint = {1812.00719},
 primaryClass = {astro-ph.HE},
       adsurl = {https://ui.adsabs.harvard.edu/abs/2019ApJ...871...78S}
}

@ARTICLE{Bates2013,
       author = {{Bates}, S.~D. and {Lorimer}, D.~R. and {Verbiest}, J.~P.~W.},
        title = "{The pulsar spectral index distribution}",
      journal = {\mnras},
         year = 2013,
        month = may,
       volume = {431},
       number = {2},
        pages = {1352-1358},
          doi = {10.1093/mnras/stt257},
archivePrefix = {arXiv},
       eprint = {1302.2053},
 primaryClass = {astro-ph.SR},
       adsurl = {https://ui.adsabs.harvard.edu/abs/2013MNRAS.431.1352B}
}

@ARTICLE{ne2001,
       author = {{Cordes}, J.~M. and {Lazio}, T.~J.~W.},
        title = "{NE2001.I. A New Model for the Galactic Distribution of Free Electrons and its Fluctuations}",
      journal = {arXiv e-prints},
         year = 2002,
        month = jul,
          eid = {astro-ph/0207156},
        pages = {astro-ph/0207156},
archivePrefix = {arXiv},
       eprint = {astro-ph/0207156},
 primaryClass = {astro-ph},
       adsurl = {https://ui.adsabs.harvard.edu/abs/2002astro.ph..7156C}
}

@BOOK{PSRHandbook,
       author = {{Lorimer}, D.~R. and {Kramer}, M.},
        title = "{Handbook of Pulsar Astronomy}",
        publisher = {Cambridge University Press},
        series = {},
         year = 2004,
       volume = {4},
       adsurl = {https://ui.adsabs.harvard.edu/abs/2004hpa..book.....L}
}

@ARTICLE{Smith2023+3PC,
       author = {{Smith}, D.~A. and {Abdollahi}, S. and {Ajello}, M. and {Bailes}, M. and {Baldini}, L. and {Ballet}, J. and {Baring}, M.~G. and {Bassa}, C. and {Gonzalez}, J. Becerra and {Bellazzini}, R. and {Berretta}, A. and {Bhattacharyya}, B. and {Bissaldi}, E. and {Bonino}, R. and {Bottacini}, E. and {Bregeon}, J. and {Bruel}, P. and {Burgay}, M. and {Burnett}, T.~H. and {Cameron}, R.~A. and {Camilo}, F. and {Caputo}, R. and {Caraveo}, P.~A. and {Cavazzuti}, E. and {Chiaro}, G. and {Ciprini}, S. and {Clark}, C.~J. and {Cognard}, I. and {Corongiu}, A. and {Orestano}, P. Cristarella and {Crnogorcevic}, M. and {Cuoco}, A. and {Cutini}, S. and {D'Ammando}, F. and {de Angelis}, A. and {DeCesar}, M.~E. and {De Gaetano}, S. and {de Menezes}, R. and {Deneva}, J. and {de Palma}, F. and {Di Lalla}, N. and {Dirirsa}, F. and {Di Venere}, L. and {Dom{\'\i}nguez}, A. and {Dumora}, D. and {Fegan}, S.~J. and {Ferrara}, E.~C. and {Fiori}, A. and {Fleischhack}, H. and {Flynn}, C. and {Franckowiak}, A. and {Freire}, P.~C.~C. and {Fukazawa}, Y. and {Fusco}, P. and {Galanti}, G. and {Gammaldi}, V. and {Gargano}, F. and {Gasparrini}, D. and {Giacchino}, F. and {Giglietto}, N. and {Giordano}, F. and {Giroletti}, M. and {Green}, D. and {Grenier}, I.~A. and {Guillemot}, L. and {Guiriec}, S. and {Gustafsson}, M. and {Harding}, A.~K. and {Hays}, E. and {Hewitt}, J.~W. and {Horan}, D. and {Hou}, X. and {Jankowski}, F. and {Johnson}, R.~P. and {Johnson}, T.~J. and {Johnston}, S. and {Kataoka}, J. and {Keith}, M.~J. and {Kerr}, M. and {Kramer}, M. and {Kuss}, M. and {Latronico}, L. and {Lee}, S. -H. and {Li}, D. and {Li}, J. and {Limyansky}, B. and {Longo}, F. and {Loparco}, F. and {Lorusso}, L. and {Lovellette}, M.~N. and {Lower}, M. and {Lubrano}, P. and {Lyne}, A.~G. and {Maan}, Y. and {Maldera}, S. and {Manchester}, R.~N. and {Manfreda}, A. and {Marelli}, M. and {Mart{\'\i}-Devesa}, G. and {Mazziotta}, M.~N. and {McEnery}, J.~E. and {Mereu}, I. and {Michelson}, P.~F. and {Mickaliger}, M. and {Mitthumsiri}, W. and {Mizuno}, T. and {Moiseev}, A.~A. and {Monzani}, M.~E. and {Morselli}, A. and {Negro}, M. and {Nemmen}, R. and {Nieder}, L. and {Nuss}, E. and {Omodei}, N. and {Orienti}, M. and {Orlando}, E. and {Ormes}, J.~F. and {Palatiello}, M. and {Paneque}, D. and {Panzarini}, G. and {Parthasarathy}, A. and {Persic}, M. and {Pesce-Rollins}, M. and {Pillera}, R. and {Poon}, H. and {Porter}, T.~A. and {Possenti}, A. and {Principe}, G. and {Rain{\`o}}, S. and {Rando}, R. and {Ransom}, S.~M. and {Ray}, P.~S. and {Razzano}, M. and {Razzaque}, S. and {Reimer}, A. and {Reimer}, O. and {Renault-Tinacci}, N. and {Romani}, R.~W. and {S{\'a}nchez-Conde}, M. and {Parkinson}, P.~M. Saz and {Scotton}, L. and {Serini}, D. and {Sgr{\`o}}, C. and {Shannon}, R. and {Sharma}, V. and {Shen}, Z. and {Siskind}, E.~J. and {Spandre}, G. and {Spinelli}, P. and {Stappers}, B.~W. and {Stephens}, T.~E. and {Suson}, D.~J. and {Tabassum}, S. and {Tajima}, H. and {Tak}, D. and {Theureau}, G. and {Thompson}, D.~J. and {Tibolla}, O. and {Torres}, D.~F. and {Valverde}, J. and {Venter}, C. and {Wadiasingh}, Z. and {Wang}, N. and {Wang}, N. and {Wang}, P. and {Weltevrede}, P. and {Wood}, K. and {Yan}, J. and {Zaharijas}, G. and {Zhang}, C. and {Zhu}, W.},
        title = "{The Third Fermi Large Area Telescope Catalog of Gamma-Ray Pulsars}",
      journal = {\apj},
         year = 2023,
        month = dec,
       volume = {958},
       number = {2},
          eid = {191},
        pages = {191},
          doi = {10.3847/1538-4357/acee67},
archivePrefix = {arXiv},
       eprint = {2307.11132},
 primaryClass = {astro-ph.HE},
       adsurl = {https://ui.adsabs.harvard.edu/abs/2023ApJ...958..191S}
}

@ARTICLE{Ridolfi2021+GC,
       author = {{Ridolfi}, A. and {Gautam}, T. and {Freire}, P.~C.~C. and {Ransom}, S.~M. and {Buchner}, S.~J. and {Possenti}, A. and {Venkatraman Krishnan}, V. and {Bailes}, M. and {Kramer}, M. and {Stappers}, B.~W. and {Abbate}, F. and {Barr}, E.~D. and {Burgay}, M. and {Camilo}, F. and {Corongiu}, A. and {Jameson}, A. and {Padmanabh}, P.~V. and {Vleeschower}, L. and {Champion}, D.~J. and {Chen}, W. and {Geyer}, M. and {Karastergiou}, A. and {Karuppusamy}, R. and {Parthasarathy}, A. and {Reardon}, D.~J. and {Serylak}, M. and {Shannon}, R.~M. and {Spiewak}, R.},
        title = "{Eight new millisecond pulsars from the first MeerKAT globular cluster census}",
      journal = {\mnras},
         year = 2021,
        month = jun,
       volume = {504},
       number = {1},
        pages = {1407-1426},
          doi = {10.1093/mnras/stab790},
archivePrefix = {arXiv},
       eprint = {2103.04800},
 primaryClass = {astro-ph.HE},
       adsurl = {https://ui.adsabs.harvard.edu/abs/2021MNRAS.504.1407R}
}

@ARTICLE{Kerr2025,
       author = {{Kerr}, M. and {Johnston}, S. and {Clark}, C.~J. and {Camilo}, F. and {Ferrara}, E.~C. and {Wolff}, M.~T. and {Ransom}, S.~M. and {Dai}, S. and {Ray}, P.~S. and {Reynolds}, J.~E. and {Sarkissian}, J.~M. and {Barr}, E.~D. and {Kramer}, M.~K. and {Stappers}, B.~W.},
        title = "{Discovery and Timing of Four {\ensuremath{\gamma}}-Ray Millisecond Pulsars}",
      journal = {\apj},
         year = 2025,
        month = may,
       volume = {984},
       number = {2},
          eid = {180},
        pages = {180},
          doi = {10.3847/1538-4357/adc7a6},
archivePrefix = {arXiv},
       eprint = {2503.12636},
 primaryClass = {astro-ph.HE},
       adsurl = {https://ui.adsabs.harvard.edu/abs/2025ApJ...984..180K}
}

@ARTICLE{Breton2012,
       author = {{Breton}, R.~P. and {Rappaport}, S.~A. and {van Kerkwijk}, M.~H. and {Carter}, J.~A.},
        title = "{KOI 1224: A Fourth Bloated Hot White Dwarf Companion Found with Kepler}",
      journal = {\apj},
         year = 2012,
        month = apr,
       volume = {748},
       number = {2},
          eid = {115},
        pages = {115},
          doi = {10.1088/0004-637X/748/2/115},
archivePrefix = {arXiv},
       eprint = {1109.6847},
 primaryClass = {astro-ph.SR},
       adsurl = {https://ui.adsabs.harvard.edu/abs/2012ApJ...748..115B}
}

@article{Brown2022,
  title = {Characterizing Eclipsing White Dwarf {{M}} Dwarf Binaries from Multiband Eclipse Photometry},
  author = {Brown, Alex J. and Parsons, Steven G. and Littlefair, Stuart P. and Wild, James F. and Ashley, R. P. and Breedt, E. and Dhillon, V. S. and Dyer, M. J. and Green, M. J. and Kerry, P. and Marsh, T. R. and Pelisoli, I. and Sahman, D. I.},
  year = {2022},
  month = jun,
  journal = {\mnras},
  volume = {513},
  number = {2},
  pages = {3050--3064},
  doi = {10.1093/mnras/stac1047}
}

@ARTICLE{Bangale2024,
       author = {{Bangale}, P. and {Bhattacharyya}, B. and {Camilo}, F. and {Clark}, C.~J. and {Cognard}, I. and {DeCesar}, M.~E. and {Ferrara}, E.~C. and {Gentile}, P. and {Guillemot}, L. and {Hessels}, J.~W.~T. and {Johnson}, T.~J. and {Kerr}, M. and {McLaughlin}, M.~A. and {Nieder}, L. and {Ransom}, S.~M. and {Ray}, P.~S. and {Roberts}, M.~S.~E. and {Roy}, J. and {Sanpa-arsa}, S. and {Theureau}, G. and {Wolff}, M.~T.},
        title = "{A 350 MHz Green Bank Telescope Survey of Unassociated Fermi LAT Sources: Discovery and Timing of 10 Millisecond Pulsars}",
      journal = {\apj},
         year = 2024,
        month = may,
       volume = {966},
       number = {2},
          eid = {161},
        pages = {161},
          doi = {10.3847/1538-4357/ad2994},
archivePrefix = {arXiv},
       eprint = {2402.09366},
 primaryClass = {astro-ph.HE},
       adsurl = {https://ui.adsabs.harvard.edu/abs/2024ApJ...966..161B}
}

@ARTICLE{Dhillon2007,
       author = {{Dhillon}, V.~S. and {Marsh}, T.~R. and {Stevenson}, M.~J. and {Atkinson}, D.~C. and {Kerry}, P. and {Peacocke}, P.~T. and {Vick}, A.~J.~A. and {Beard}, S.~M. and {Ives}, D.~J. and {Lunney}, D.~W. and {McLay}, S.~A. and {Tierney}, C.~J. and {Kelly}, J. and {Littlefair}, S.~P. and {Nicholson}, R. and {Pashley}, R. and {Harlaftis}, E.~T. and {O'Brien}, K.},
        title = "{ULTRACAM: an ultrafast, triple-beam CCD camera for high-speed astrophysics}",
      journal = {\mnras},
         year = 2007,
        month = jul,
       volume = {378},
       number = {3},
        pages = {825-840},
          doi = {10.1111/j.1365-2966.2007.11881.x},
archivePrefix = {arXiv},
       eprint = {0704.2557},
 primaryClass = {astro-ph},
       adsurl = {https://ui.adsabs.harvard.edu/abs/2007MNRAS.378..825D}
}

@ARTICLE{Dhillon2021,
       author = {{Dhillon}, V.~S. and {Bezawada}, N. and {Black}, M. and {Dixon}, S.~D. and {Gamble}, T. and {Gao}, X. and {Henry}, D.~M. and {Kerry}, P. and {Littlefair}, S.~P. and {Lunney}, D.~W. and {Marsh}, T.~R. and {Miller}, C. and {Parsons}, S.~G. and {Ashley}, R.~P. and {Breedt}, E. and {Brown}, A. and {Dyer}, M.~J. and {Green}, M.~J. and {Pelisoli}, I. and {Sahman}, D.~I. and {Wild}, J. and {Ives}, D.~J. and {Mehrgan}, L. and {Stegmeier}, J. and {Dubbeldam}, C.~M. and {Morris}, T.~J. and {Osborn}, J. and {Wilson}, R.~W. and {Casares}, J. and {Mu{\~n}oz-Darias}, T. and {Pall{\'e}}, E. and {Rodr{\'\i}guez-Gil}, P. and {Shahbaz}, T. and {Torres}, M.~A.~P. and {de Ugarte Postigo}, A. and {Cabrera-Lavers}, A. and {Corradi}, R.~L.~M. and {Dom{\'\i}nguez}, R.~D. and {Garc{\'\i}a-Alvarez}, D.},
        title = "{HiPERCAM: a quintuple-beam, high-speed optical imager on the 10.4-m Gran Telescopio Canarias}",
      journal = {\mnras},
         year = 2021,
        month = oct,
       volume = {507},
       number = {1},
        pages = {350-366},
          doi = {10.1093/mnras/stab2130},
archivePrefix = {arXiv},
       eprint = {2107.10124},
 primaryClass = {astro-ph.IM},
       adsurl = {https://ui.adsabs.harvard.edu/abs/2021MNRAS.507..350D}
}

@ARTICLE{PS1,
       author = {{Chambers}, K.~C. and {Magnier}, E.~A. and {Metcalfe}, N. and {Flewelling}, H.~A. and {Huber}, M.~E. and {Waters}, C.~Z. and {Denneau}, L. and {Draper}, P.~W. and {Farrow}, D. and {Finkbeiner}, D.~P. and {Holmberg}, C. and {Koppenhoefer}, J. and {Price}, P.~A. and {Rest}, A. and {Saglia}, R.~P. and {Schlafly}, E.~F. and {Smartt}, S.~J. and {Sweeney}, W. and {Wainscoat}, R.~J. and {Burgett}, W.~S. and {Chastel}, S. and {Grav}, T. and {Heasley}, J.~N. and {Hodapp}, K.~W. and {Jedicke}, R. and {Kaiser}, N. and {Kudritzki}, R. -P. and {Luppino}, G.~A. and {Lupton}, R.~H. and {Monet}, D.~G. and {Morgan}, J.~S. and {Onaka}, P.~M. and {Shiao}, B. and {Stubbs}, C.~W. and {Tonry}, J.~L. and {White}, R. and {Ba{\~n}ados}, E. and {Bell}, E.~F. and {Bender}, R. and {Bernard}, E.~J. and {Boegner}, M. and {Boffi}, F. and {Botticella}, M.~T. and {Calamida}, A. and {Casertano}, S. and {Chen}, W. -P. and {Chen}, X. and {Cole}, S. and {Deacon}, N. and {Frenk}, C. and {Fitzsimmons}, A. and {Gezari}, S. and {Gibbs}, V. and {Goessl}, C. and {Goggia}, T. and {Gourgue}, R. and {Goldman}, B. and {Grant}, P. and {Grebel}, E.~K. and {Hambly}, N.~C. and {Hasinger}, G. and {Heavens}, A.~F. and {Heckman}, T.~M. and {Henderson}, R. and {Henning}, T. and {Holman}, M. and {Hopp}, U. and {Ip}, W. -H. and {Isani}, S. and {Jackson}, M. and {Keyes}, C.~D. and {Koekemoer}, A.~M. and {Kotak}, R. and {Le}, D. and {Liska}, D. and {Long}, K.~S. and {Lucey}, J.~R. and {Liu}, M. and {Martin}, N.~F. and {Masci}, G. and {McLean}, B. and {Mindel}, E. and {Misra}, P. and {Morganson}, E. and {Murphy}, D.~N.~A. and {Obaika}, A. and {Narayan}, G. and {Nieto-Santisteban}, M.~A. and {Norberg}, P. and {Peacock}, J.~A. and {Pier}, E.~A. and {Postman}, M. and {Primak}, N. and {Rae}, C. and {Rai}, A. and {Riess}, A. and {Riffeser}, A. and {Rix}, H.~W. and {R{\"o}ser}, S. and {Russel}, R. and {Rutz}, L. and {Schilbach}, E. and {Schultz}, A.~S.~B. and {Scolnic}, D. and {Strolger}, L. and {Szalay}, A. and {Seitz}, S. and {Small}, E. and {Smith}, K.~W. and {Soderblom}, D.~R. and {Taylor}, P. and {Thomson}, R. and {Taylor}, A.~N. and {Thakar}, A.~R. and {Thiel}, J. and {Thilker}, D. and {Unger}, D. and {Urata}, Y. and {Valenti}, J. and {Wagner}, J. and {Walder}, T. and {Walter}, F. and {Watters}, S.~P. and {Werner}, S. and {Wood-Vasey}, W.~M. and {Wyse}, R.},
        title = "{The Pan-STARRS1 Surveys}",
      journal = {arXiv e-prints},
         year = 2016,
        month = dec,
          eid = {arXiv:1612.05560},
        pages = {arXiv:1612.05560},
          doi = {10.48550/arXiv.1612.05560},
archivePrefix = {arXiv},
       eprint = {1612.05560},
 primaryClass = {astro-ph.IM},
       adsurl = {https://ui.adsabs.harvard.edu/abs/2016arXiv161205560C}
}

@ARTICLE{PS1DB,
       author = {{Flewelling}, H.~A. and {Magnier}, E.~A. and {Chambers}, K.~C. and {Heasley}, J.~N. and {Holmberg}, C. and {Huber}, M.~E. and {Sweeney}, W. and {Waters}, C.~Z. and {Calamida}, A. and {Casertano}, S. and {Chen}, X. and {Farrow}, D. and {Hasinger}, G. and {Henderson}, R. and {Long}, K.~S. and {Metcalfe}, N. and {Narayan}, G. and {Nieto-Santisteban}, M.~A. and {Norberg}, P. and {Rest}, A. and {Saglia}, R.~P. and {Szalay}, A. and {Thakar}, A.~R. and {Tonry}, J.~L. and {Valenti}, J. and {Werner}, S. and {White}, R. and {Denneau}, L. and {Draper}, P.~W. and {Hodapp}, K.~W. and {Jedicke}, R. and {Kaiser}, N. and {Kudritzki}, R.~P. and {Price}, P.~A. and {Wainscoat}, R.~J. and {Chastel}, S. and {McLean}, B. and {Postman}, M. and {Shiao}, B.},
        title = "{The Pan-STARRS1 Database and Data Products}",
      journal = {\apjs},
         year = 2020,
        month = nov,
       volume = {251},
       number = {1},
          eid = {7},
        pages = {7},
          doi = {10.3847/1538-4365/abb82d},
archivePrefix = {arXiv},
       eprint = {1612.05243},
 primaryClass = {astro-ph.IM},
       adsurl = {https://ui.adsabs.harvard.edu/abs/2020ApJS..251....7F}
}

@ARTICLE{SpiderCAT,
       author = {{Koljonen}, Karri I.~I. and {Linares}, Manuel},
        title = "{SpiderCat: A Catalog of Compact Binary Millisecond Pulsars}",
      journal = {\apj},
         year = 2025,
        month = nov,
       volume = {994},
       number = {1},
          eid = {8},
        pages = {8},
          doi = {10.3847/1538-4357/ae08a5},
archivePrefix = {arXiv},
       eprint = {2505.11691},
 primaryClass = {astro-ph.HE},
       adsurl = {https://ui.adsabs.harvard.edu/abs/2025ApJ...994....8K}
}

@ARTICLE{Belmonte2025+J1544,
       author = {{Belmonte D{\'\i}az}, S. and {Thongmeearkom}, T. and {Phosrisom}, A. and {Breton}, R.~P. and {Burgay}, M. and {Clark}, C.~J. and {Nieder}, L. and {Mayer}, M.~G.~F. and {Becker}, W. and {Barr}, E.~D. and {Buchner}, S. and {Das}, K.~K. and {Dhillon}, V.~S. and {Dodge}, O.~G. and {Ferrara}, E.~C. and {Griessmeier}, J.-M. and {Karuppusamy}, R. and {Kennedy}, M.~R. and {Kramer}, M. and {Padmanabh}, P.~V. and {Paice}, J.~A. and {Rodr{\'\i}guez}, A.~C. and {Stappers}, B.~W.},
        title = "{Multiwavelength observations of a new black-widow millisecond pulsar PSR J1544‑2555}",
      journal = {\mnras},
         year = 2025,
        month = nov,
       volume = {543},
       number = {3},
        pages = {3019-3034},
          doi = {10.1093/mnras/staf1544},
archivePrefix = {arXiv},
       eprint = {2509.09605},
 primaryClass = {astro-ph.HE},
       adsurl = {https://ui.adsabs.harvard.edu/abs/2025MNRAS.543.3019B}
}

@ARTICLE{Breton2007,
       author = {{Breton}, R.~P. and {Roberts}, M.~S.~E. and {Ransom}, S.~M. and {Kaspi}, V.~M. and {Durant}, M. and {Bergeron}, P. and {Faulkner}, A.~J.},
        title = "{The Unusual Binary Pulsar PSR J1744-3922: Radio Flux Variability, Near-Infrared Observation, and Evolution}",
      journal = {\apj},
         year = 2007,
        month = jun,
       volume = {661},
       number = {2},
        pages = {1073-1083},
          doi = {10.1086/515392},
archivePrefix = {arXiv},
       eprint = {astro-ph/0702347},
 primaryClass = {astro-ph},
       adsurl = {https://ui.adsabs.harvard.edu/abs/2007ApJ...661.1073B}
}

@phdthesis{Thongmeearkom2021,
  author       = {Thongmeearkom, Tinn},
  year         = {2021},
  title        = {Searching for pulsars with MeerKAT},
  school       = {The University of Manchester},
  type   = {\href{https://research.manchester.ac.uk/en/studentTheses/searching-for-pulsars-with-meerkat/}{{MPhil} thesis}}
}

@ARTICLE{Johnston2020+PTA,
       author = {{Johnston}, Simon and {Karastergiou}, A. and {Keith}, M.~J. and {Song}, X. and {Weltevrede}, P. and {Abbate}, F. and {Bailes}, M. and {Buchner}, S. and {Camilo}, F. and {Geyer}, M. and {Hugo}, B. and {Jameson}, A. and {Kramer}, M. and {Parthasarathy}, A. and {Reardon}, D.~J. and {Ridolfi}, A. and {Serylak}, M. and {Shannon}, R.~M. and {Spiewak}, R. and {van Straten}, W. and {Venkatraman Krishnan}, V. and {Jankowski}, F. and {Meyers}, B.~W. and {Oswald}, L. and {Posselt}, B. and {Sobey}, C. and {Szary}, A. and {van Leeuwen}, J.},
        title = "{The Thousand-Pulsar-Array programme on MeerKAT - I. Science objectives and first results}",
      journal = {\mnras},
         year = 2020,
        month = apr,
       volume = {493},
       number = {3},
        pages = {3608-3615},
          doi = {10.1093/mnras/staa516},
archivePrefix = {arXiv},
       eprint = {2002.10250},
 primaryClass = {astro-ph.HE},
       adsurl = {https://ui.adsabs.harvard.edu/abs/2020MNRAS.493.3608J}
}

@ARTICLE{Tian2025+MeerTRAP,
       author = {{Tian}, J. and {Singh}, S. and {Stappers}, B.~W. and {Turner}, J.~D. and {Rajwade}, K.~M. and {Bezuidenhout}, M.~C. and {Caleb}, M. and {Pastor-Marazuela}, I. and {Jankowski}, F. and {Gupta}, V. and {Flynn}, C. and {Karuppusamy}, R. and {Barr}, E.~D. and {Kramer}, M. and {Breton}, R. and {Clark}, C.~J. and {Champion}, D.~J. and {Thongmeearkom}, T.},
        title = "{Discovery of 30 Galactic radio transient pulsars with MeerTRAP}",
      journal = {\mnras},
         year = 2025,
        month = dec,
       volume = {544},
       number = {2},
        pages = {1843-1860},
          doi = {10.1093/mnras/staf1827},
archivePrefix = {arXiv},
       eprint = {2510.17723},
 primaryClass = {astro-ph.HE},
       adsurl = {https://ui.adsabs.harvard.edu/abs/2025MNRAS.544.1843T}
}

@ARTICLE{Simpson2025,
       author = {{Simpson}, Jordan A. and {Linares}, Manuel and {Casares}, Jorge and {Shahbaz}, Tariq and {Sen}, Bidisha and {Camilo}, Fernando},
        title = "{A GTC spectroscopic study of three spider pulsar companions: line-based temperatures, a new face-on redback, and improved mass constraints}",
      journal = {\mnras},
         year = 2025,
        month = jan,
       volume = {536},
       number = {3},
        pages = {2169-2186},
          doi = {10.1093/mnras/stae2728},
archivePrefix = {arXiv},
       eprint = {2408.11099},
 primaryClass = {astro-ph.HE},
       adsurl = {https://ui.adsabs.harvard.edu/abs/2025MNRAS.536.2169S}
}

@ARTICLE{Purcell2015,
       author = {{Purcell}, C.~R. and {Gaensler}, B.~M. and {Sun}, X.~H. and {Carretti}, E. and {Bernardi}, G. and {Haverkorn}, M. and {Kesteven}, M.~J. and {Poppi}, S. and {Schnitzeler}, D.~H.~F.~M. and {Staveley-Smith}, L.},
        title = "{A Radio-Polarisation and Rotation Measure Study of the Gum Nebula and Its Environment}",
      journal = {\apj},
         year = 2015,
        month = may,
       volume = {804},
       number = {1},
          eid = {22},
        pages = {22},
          doi = {10.1088/0004-637X/804/1/22},
archivePrefix = {arXiv},
       eprint = {1502.06296},
 primaryClass = {astro-ph.GA},
       adsurl = {https://ui.adsabs.harvard.edu/abs/2015ApJ...804...22P}
}

@ARTICLE{Shklovskii1970,
       author = {{Shklovskii}, I.~S.},
        title = "{Possible Causes of the Secular Increase in Pulsar Periods.}",
      journal = {\sovast},
         year = 1970,
        month = feb,
       volume = {13},
        pages = {562},
       adsurl = {https://ui.adsabs.harvard.edu/abs/1970SvA....13..562S}
}

@ARTICLE{Strader2025+J1947,
       author = {{Strader}, Jay and {Ray}, Paul S. and {Urquhart}, Ryan and {Swihart}, Samuel J. and {Chomiuk}, Laura and {Aydi}, Elias and {Bellm}, Eric C. and {Dage}, Kristen C. and {DeCesar}, Megan E. and {Deneva}, Julia S. and {McLaughlin}, Maura A. and {Molina}, Isabella and {Panurach}, Teresa and {Sokolovsky}, Kirill V.},
        title = "{PSR J1947‑1120: A New Huntsman Millisecond Pulsar Binary}",
      journal = {\apj},
         year = 2025,
        month = feb,
       volume = {980},
       number = {1},
          eid = {124},
        pages = {124},
          doi = {10.3847/1538-4357/ada897},
archivePrefix = {arXiv},
       eprint = {2501.05509},
 primaryClass = {astro-ph.HE},
       adsurl = {https://ui.adsabs.harvard.edu/abs/2025ApJ...980..124S}
}

@ARTICLE{Nieder2019+J0952,
       author = {{Nieder}, L. and {Clark}, C.~J. and {Bassa}, C.~G. and {Wu}, J. and {Singh}, A. and {Donner}, J.~Y. and {Allen}, B. and {Breton}, R.~P. and {Dhillon}, V.~S. and {Eggenstein}, H.-B. and {Hessels}, J.~W.~T. and {Kennedy}, M.~R. and {Kerr}, M. and {Littlefair}, S. and {Marsh}, T.~R. and {Mata S{\'a}nchez}, D. and {Papa}, M.~A. and {Ray}, P.~S. and {Steltner}, B. and {Verbiest}, J.~P.~W.},
        title = "{Detection and Timing of Gamma-Ray Pulsations from the 707 Hz Pulsar J0952-0607}",
      journal = {\apj},
         year = 2019,
        month = sep,
       volume = {883},
       number = {1},
          eid = {42},
        pages = {42},
          doi = {10.3847/1538-4357/ab357e},
archivePrefix = {arXiv},
       eprint = {1905.11352},
 primaryClass = {astro-ph.HE},
       adsurl = {https://ui.adsabs.harvard.edu/abs/2019ApJ...883...42N}
}

@ARTICLE{Ray2022+J1555,
       author = {{Ray}, Paul S. and {Nieder}, Lars and {Clark}, Colin J. and {Ransom}, Scott M. and {Cromartie}, H. Thankful and {Frail}, Dale A. and {Mooley}, Kunal P. and {Intema}, Huib and {Jagannathan}, Preshanth and {Demorest}, Paul and {Stovall}, Kevin and {Halpern}, Jules P. and {Deneva}, Julia and {Guillot}, Sebastien and {Kerr}, Matthew and {Swihart}, Samuel J. and {Bruel}, Philippe and {Stappers}, Ben W. and {Lyne}, Andrew and {Mickaliger}, Mitch and {Camilo}, Fernando and {Ferrara}, Elizabeth C. and {Wolff}, Michael T. and {Michelson}, P.~F.},
        title = "{Discovery, Timing, and Multiwavelength Observations of the Black Widow Millisecond Pulsar PSR J1555-2908}",
      journal = {\apj},
         year = 2022,
        month = mar,
       volume = {927},
       number = {2},
          eid = {216},
        pages = {216},
          doi = {10.3847/1538-4357/ac49ef},
archivePrefix = {arXiv},
       eprint = {2202.04783},
 primaryClass = {astro-ph.HE},
       adsurl = {https://ui.adsabs.harvard.edu/abs/2022ApJ...927..216R}
}

@ARTICLE{Bickel2008+weights,
       author = {{Bickel}, Peter and {Kleijn}, Bas and {Rice}, John},
        title = "{Event-Weighted Tests for Detecting Periodicity in Photon Arrival Times}",
      journal = {\apj},
         year = 2008,
        month = sep,
       volume = {685},
       number = {1},
        pages = {384-389},
          doi = {10.1086/590399},
archivePrefix = {arXiv},
       eprint = {0706.4108},
 primaryClass = {stat.ME},
       adsurl = {https://ui.adsabs.harvard.edu/abs/2008ApJ...685..384B}
}

@ARTICLE{Bruel2019+weights,
       author = {{Bruel}, P.},
        title = "{Extending the event-weighted pulsation search to very faint gamma-ray sources}",
      journal = {\aap},
         year = 2019,
        month = feb,
       volume = {622},
          eid = {A108},
        pages = {A108},
          doi = {10.1051/0004-6361/201834555},
archivePrefix = {arXiv},
       eprint = {1812.06681},
 primaryClass = {astro-ph.IM},
       adsurl = {https://ui.adsabs.harvard.edu/abs/2019A&A...622A.108B}
}

@ARTICLE{Pletsch2014+searchmethods,
       author = {{Pletsch}, Holger J. and {Clark}, Colin J.},
        title = "{Optimized Blind Gamma-Ray Pulsar Searches at Fixed Computing Budget}",
      journal = {\apj},
         year = 2014,
        month = nov,
       volume = {795},
       number = {1},
          eid = {75},
        pages = {75},
          doi = {10.1088/0004-637X/795/1/75},
archivePrefix = {arXiv},
       eprint = {1408.6962},
 primaryClass = {astro-ph.HE},
       adsurl = {https://ui.adsabs.harvard.edu/abs/2014ApJ...795...75P}
}

@ARTICLE{Luo2021+PINT,
       author = {{Luo}, Jing and {Ransom}, Scott and {Demorest}, Paul and {Ray}, Paul S. and {Archibald}, Anne and {Kerr}, Matthew and {Jennings}, Ross J. and {Bachetti}, Matteo and {van Haasteren}, Rutger and {Champagne}, Chloe A. and {Colen}, Jonathan and {Phillips}, Camryn and {Zimmerman}, Josef and {Stovall}, Kevin and {Lam}, Michael T. and {Jenet}, Fredrick A.},
        title = "{PINT: A Modern Software Package for Pulsar Timing}",
      journal = {\apj},
         year = 2021,
        month = apr,
       volume = {911},
       number = {1},
          eid = {45},
        pages = {45},
          doi = {10.3847/1538-4357/abe62f},
archivePrefix = {arXiv},
       eprint = {2012.00074},
 primaryClass = {astro-ph.IM},
       adsurl = {https://ui.adsabs.harvard.edu/abs/2021ApJ...911...45L}
}

@ARTICLE{Foreman-Mackey2013+emcee,
       author = {{Foreman-Mackey}, Daniel and {Hogg}, David W. and {Lang}, Dustin and {Goodman}, Jonathan},
        title = "{emcee: The MCMC Hammer}",
      journal = {\pasp},
         year = 2013,
        month = mar,
       volume = {125},
       number = {925},
        pages = {306},
          doi = {10.1086/670067},
archivePrefix = {arXiv},
       eprint = {1202.3665},
 primaryClass = {astro-ph.IM},
       adsurl = {https://ui.adsabs.harvard.edu/abs/2013PASP..125..306F}
}

@ARTICLE{gll_iem_v07,
       author = {{Acero}, F. and {Ackermann}, M. and {Ajello}, M. and {Albert}, A. and {Baldini}, L. and {Ballet}, J. and {Barbiellini}, G. and {Bastieri}, D. and {Bellazzini}, R. and {Bissaldi}, E. and {Bloom}, E.~D. and {Bonino}, R. and {Bottacini}, E. and {Brandt}, T.~J. and {Bregeon}, J. and {Bruel}, P. and {Buehler}, R. and {Buson}, S. and {Caliandro}, G.~A. and {Cameron}, R.~A. and {Caragiulo}, M. and {Caraveo}, P.~A. and {Casandjian}, J.~M. and {Cavazzuti}, E. and {Cecchi}, C. and {Charles}, E. and {Chekhtman}, A. and {Chiang}, J. and {Chiaro}, G. and {Ciprini}, S. and {Claus}, R. and {Cohen-Tanugi}, J. and {Conrad}, J. and {Cuoco}, A. and {Cutini}, S. and {D'Ammando}, F. and {de Angelis}, A. and {de Palma}, F. and {Desiante}, R. and {Digel}, S.~W. and {Di Venere}, L. and {Drell}, P.~S. and {Favuzzi}, C. and {Fegan}, S.~J. and {Ferrara}, E.~C. and {Focke}, W.~B. and {Franckowiak}, A. and {Funk}, S. and {Fusco}, P. and {Gargano}, F. and {Gasparrini}, D. and {Giglietto}, N. and {Giordano}, F. and {Giroletti}, M. and {Glanzman}, T. and {Godfrey}, G. and {Grenier}, I.~A. and {Guiriec}, S. and {Hadasch}, D. and {Harding}, A.~K. and {Hayashi}, K. and {Hays}, E. and {Hewitt}, J.~W. and {Hill}, A.~B. and {Horan}, D. and {Hou}, X. and {Jogler}, T. and {J{\'o}hannesson}, G. and {Kamae}, T. and {Kuss}, M. and {Landriu}, D. and {Larsson}, S. and {Latronico}, L. and {Li}, J. and {Li}, L. and {Longo}, F. and {Loparco}, F. and {Lovellette}, M.~N. and {Lubrano}, P. and {Maldera}, S. and {Malyshev}, D. and {Manfreda}, A. and {Martin}, P. and {Mayer}, M. and {Mazziotta}, M.~N. and {McEnery}, J.~E. and {Michelson}, P.~F. and {Mirabal}, N. and {Mizuno}, T. and {Monzani}, M.~E. and {Morselli}, A. and {Nuss}, E. and {Ohsugi}, T. and {Omodei}, N. and {Orienti}, M. and {Orlando}, E. and {Ormes}, J.~F. and {Paneque}, D. and {Pesce-Rollins}, M. and {Piron}, F. and {Pivato}, G. and {Rain{\`o}}, S. and {Rando}, R. and {Razzano}, M. and {Razzaque}, S. and {Reimer}, A. and {Reimer}, O. and {Remy}, Q. and {Renault}, N. and {S{\'a}nchez-Conde}, M. and {Schaal}, M. and {Schulz}, A. and {Sgr{\`o}}, C. and {Siskind}, E.~J. and {Spada}, F. and {Spandre}, G. and {Spinelli}, P. and {Strong}, A.~W. and {Suson}, D.~J. and {Tajima}, H. and {Takahashi}, H. and {Thayer}, J.~B. and {Thompson}, D.~J. and {Tibaldo}, L. and {Tinivella}, M. and {Torres}, D.~F. and {Tosti}, G. and {Troja}, E. and {Vianello}, G. and {Werner}, M. and {Wood}, K.~S. and {Wood}, M. and {Zaharijas}, G. and {Zimmer}, S.},
        title = "{Development of the Model of Galactic Interstellar Emission for Standard Point-source Analysis of Fermi Large Area Telescope Data}",
      journal = {\apjs},
         year = 2016,
        month = apr,
       volume = {223},
       number = {2},
          eid = {26},
        pages = {26},
          doi = {10.3847/0067-0049/223/2/26},
archivePrefix = {arXiv},
       eprint = {1602.07246},
 primaryClass = {astro-ph.HE},
       adsurl = {https://ui.adsabs.harvard.edu/abs/2016ApJS..223...26A}
}

@ARTICLE{Abdollahi2022+4FGLDR3,
       author = {{Abdollahi}, S. and {Acero}, F. and {Baldini}, L. and {Ballet}, J. and {Bastieri}, D. and {Bellazzini}, R. and {Berenji}, B. and {Berretta}, A. and {Bissaldi}, E. and {Blandford}, R.~D. and {Bloom}, E. and {Bonino}, R. and {Brill}, A. and {Britto}, R.~J. and {Bruel}, P. and {Burnett}, T.~H. and {Buson}, S. and {Cameron}, R.~A. and {Caputo}, R. and {Caraveo}, P.~A. and {Castro}, D. and {Chaty}, S. and {Cheung}, C.~C. and {Chiaro}, G. and {Cibrario}, N. and {Ciprini}, S. and {Coronado-Bl{\'a}zquez}, J. and {Crnogorcevic}, M. and {Cutini}, S. and {D'Ammando}, F. and {De Gaetano}, S. and {Digel}, S.~W. and {Di Lalla}, N. and {Dirirsa}, F. and {Di Venere}, L. and {Dom{\'\i}nguez}, A. and {Fallah Ramazani}, V. and {Fegan}, S.~J. and {Ferrara}, E.~C. and {Fiori}, A. and {Fleischhack}, H. and {Franckowiak}, A. and {Fukazawa}, Y. and {Funk}, S. and {Fusco}, P. and {Galanti}, G. and {Gammaldi}, V. and {Gargano}, F. and {Garrappa}, S. and {Gasparrini}, D. and {Giacchino}, F. and {Giglietto}, N. and {Giordano}, F. and {Giroletti}, M. and {Glanzman}, T. and {Green}, D. and {Grenier}, I.~A. and {Grondin}, M.-H. and {Guillemot}, L. and {Guiriec}, S. and {Gustafsson}, M. and {Harding}, A.~K. and {Hays}, E. and {Hewitt}, J.~W. and {Horan}, D. and {Hou}, X. and {J{\'o}hannesson}, G. and {Karwin}, C. and {Kayanoki}, T. and {Kerr}, M. and {Kuss}, M. and {Landriu}, D. and {Larsson}, S. and {Latronico}, L. and {Lemoine-Goumard}, M. and {Li}, J. and {Liodakis}, I. and {Longo}, F. and {Loparco}, F. and {Lott}, B. and {Lubrano}, P. and {Maldera}, S. and {Malyshev}, D. and {Manfreda}, A. and {Mart{\'\i}-Devesa}, G. and {Mazziotta}, M.~N. and {Mereu}, I. and {Meyer}, M. and {Michelson}, P.~F. and {Mirabal}, N. and {Mitthumsiri}, W. and {Mizuno}, T. and {Moiseev}, A.~A. and {Monzani}, M.~E. and {Morselli}, A. and {Moskalenko}, I.~V. and {Negro}, M. and {Nuss}, E. and {Omodei}, N. and {Orienti}, M. and {Orlando}, E. and {Paneque}, D. and {Pei}, Z. and {Perkins}, J.~S. and {Persic}, M. and {Pesce-Rollins}, M. and {Petrosian}, V. and {Pillera}, R. and {Poon}, H. and {Porter}, T.~A. and {Principe}, G. and {Rain{\`o}}, S. and {Rando}, R. and {Rani}, B. and {Razzano}, M. and {Razzaque}, S. and {Reimer}, A. and {Reimer}, O. and {Reposeur}, T. and {S{\'a}nchez-Conde}, M. and {Saz Parkinson}, P.~M. and {Scotton}, L. and {Serini}, D. and {Sgr{\`o}}, C. and {Siskind}, E.~J. and {Smith}, D.~A. and {Spandre}, G. and {Spinelli}, P. and {Sueoka}, K. and {Suson}, D.~J. and {Tajima}, H. and {Tak}, D. and {Thayer}, J.~B. and {Thompson}, D.~J. and {Torres}, D.~F. and {Troja}, E. and {Valverde}, J. and {Wood}, K. and {Zaharijas}, G.},
        title = "{Incremental Fermi Large Area Telescope Fourth Source Catalog}",
      journal = {\apjs},
         year = 2022,
        month = jun,
       volume = {260},
       number = {2},
          eid = {53},
        pages = {53},
          doi = {10.3847/1538-4365/ac6751},
archivePrefix = {arXiv},
       eprint = {2201.11184},
 primaryClass = {astro-ph.HE},
       adsurl = {https://ui.adsabs.harvard.edu/abs/2022ApJS..260...53A}
}

@ARTICLE{Jaranowski:1998qm,
       author = {{Jaranowski}, Piotr and {Kr{\'o}lak}, Andrzej and {Schutz}, Bernard F.},
        title = "{Data analysis of gravitational-wave signals from spinning neutron stars: The signal and its detection}",
      journal = {\prd},
         year = 1998,
        month = sep,
       volume = {58},
       number = {6},
          eid = {063001},
        pages = {063001},
          doi = {10.1103/PhysRevD.58.063001},
archivePrefix = {arXiv},
       eprint = {gr-qc/9804014},
 primaryClass = {gr-qc},
       adsurl = {https://ui.adsabs.harvard.edu/abs/1998PhRvD..58f3001J}
}

@ARTICLE{Jones:2015zma,
       author = {{Jones}, D.~I.},
        title = "{Parameter choices and ranges for continuous gravitational wave searches for steadily spinning neutron stars}",
      journal = {\mnras},
         year = 2015,
        month = oct,
       volume = {453},
       number = {1},
        pages = {53-66},
          doi = {10.1093/mnras/stv1584},
archivePrefix = {arXiv},
       eprint = {1501.05832},
 primaryClass = {gr-qc},
       adsurl = {https://ui.adsabs.harvard.edu/abs/2015MNRAS.453...53J}
}

@ARTICLE{Cutler:2005hc,
       author = {{Cutler}, Curt and {Schutz}, Bernard F.},
        title = "{Generalized F-statistic: Multiple detectors and multiple gravitational wave pulsars}",
      journal = {\prd},
         year = 2005,
        month = sep,
       volume = {72},
       number = {6},
          eid = {063006},
        pages = {063006},
          doi = {10.1103/PhysRevD.72.063006},
archivePrefix = {arXiv},
       eprint = {gr-qc/0504011},
 primaryClass = {gr-qc},
       adsurl = {https://ui.adsabs.harvard.edu/abs/2005PhRvD..72f3006C}
}

@ARTICLE{Clark:2025lai,
       author = {{Clark}, C.~J. and {Di Mauro}, M. and {Wu}, J. and {Allen}, B. and {Behnke}, O. and {Eggenstein}, H.~B. and {Machenschalk}, B. and {Nieder}, L. and {Saz Parkinson}, P.~M. and {Ashok}, A. and {Bruel}, P. and {McGloughlin}, B. and {Papa}, M.~A. and {Camilo}, F. and {Kerr}, M. and {Padmanabh}, P. Voraganti and {Ransom}, S.~M.},
        title = "{Einstein@Home Searches for Gamma-Ray Pulsars in the Inner Galaxy}",
      journal = {\apj},
         year = 2025,
        month = dec,
       volume = {994},
       number = {2},
          eid = {149},
        pages = {149},
          doi = {10.3847/1538-4357/ae0b5c},
archivePrefix = {arXiv},
       eprint = {2509.21307},
 primaryClass = {astro-ph.HE},
       adsurl = {https://ui.adsabs.harvard.edu/abs/2025ApJ...994..149C}
}

@ARTICLE{Ashok:2021fnj,
       author = {{Ashok}, A. and {Beheshtipour}, B. and {Papa}, M.~A. and {Freire}, P.~C.~C. and {Steltner}, B. and {Machenschalk}, B. and {Behnke}, O. and {Allen}, B. and {Prix}, R.},
        title = "{New Searches for Continuous Gravitational Waves from Seven Fast Pulsars}",
      journal = {\apj},
         year = 2021,
        month = dec,
       volume = {923},
       number = {1},
          eid = {85},
        pages = {85},
          doi = {10.3847/1538-4357/ac2582},
archivePrefix = {arXiv},
       eprint = {2107.09727},
 primaryClass = {astro-ph.HE},
       adsurl = {https://ui.adsabs.harvard.edu/abs/2021ApJ...923...85A}
}

@ARTICLE{Gao:2020zcd,
       author = {{Gao}, Yong and {Shao}, Lijing and {Xu}, Rui and {Sun}, Ling and {Liu}, Chang and {Xu}, Ren-Xin},
        title = "{Triaxially deformed freely precessing neutron stars: continuous electromagnetic and gravitational radiation}",
      journal = {\mnras},
         year = 2020,
        month = oct,
       volume = {498},
       number = {2},
        pages = {1826-1838},
          doi = {10.1093/mnras/staa2476},
archivePrefix = {arXiv},
       eprint = {2007.02528},
 primaryClass = {astro-ph.HE},
       adsurl = {https://ui.adsabs.harvard.edu/abs/2020MNRAS.498.1826G}
}

@ARTICLE{LIGOScientific:2019lzm,
       author = {{Abbott}, R. and {Abbott}, T.~D. and {Abraham}, S. and {Acernese}, F. and {Ackley}, K. and {Adams}, C. and {Adhikari}, R.~X. and {Adya}, V.~B. and {Affeldt}, C. and {Agathos}, M. and {Agatsuma}, K. and {Aggarwal}, N. and {Aguiar}, O.~D. and {Aich}, A. and {Aiello}, L. and {Ain}, A. and {Ajith}, P. and {Allen}, G. and {Allocca}, A. and {Altin}, P.~A. and {Amato}, A. and {Anand}, S. and {Ananyeva}, A. and {Anderson}, S.~B. and {Anderson}, W.~G. and {Angelova}, S.~V. and {Ansoldi}, S. and {Antier}, S. and {Appert}, S. and {Arai}, K. and {Araya}, M.~C. and {Areeda}, J.~S. and {Ar{\`e}ne}, M. and {Arnaud}, N. and {Aronson}, S.~M. and {Arun}, K.~G. and {Ascenzi}, S. and {Ashton}, G. and {Aston}, S.~M. and {Astone}, P. and {Aubin}, F. and {Aufmuth}, P. and {AultONeal}, K. and {Austin}, C. and {Avendano}, V. and {Babak}, S. and {Bacon}, P. and {Badaracco}, F. and {Bader}, M.~K.~M. and {Bae}, S. and {Baer}, A.~M. and {Baird}, J. and {Baldaccini}, F. and {Ballardin}, G. and {Ballmer}, S.~W. and {Bals}, A. and {Balsamo}, A. and {Baltus}, G. and {Banagiri}, S. and {Bankar}, D. and {Bankar}, R.~S. and {Barayoga}, J.~C. and {Barbieri}, C. and {Barish}, B.~C. and {Barker}, D. and {Barkett}, K. and {Barneo}, P. and {Barone}, F. and {Barr}, B. and {Barsotti}, L. and {Barsuglia}, M. and {Barta}, D. and {Bartlett}, J. and {Bartos}, I. and {Bassiri}, R. and {Basti}, A. and {Bawaj}, M. and {Bayley}, J.~C. and {Bazzan}, M. and {B{\'e}csy}, B. and {Bejger}, M. and {Belahcene}, I. and {Bell}, A.~S. and {Beniwal}, D. and {Benjamin}, M.~G. and {Bentley}, J.~D. and {Bergamin}, F. and {Berger}, B.~K. and {Bergmann}, G. and {Bernuzzi}, S. and {Berry}, C.~P.~L. and {Bersanetti}, D. and {Bertolini}, A. and {Betzwieser}, J. and {Bhandare}, R. and {Bhandari}, A.~V. and {Bidler}, J. and {Biggs}, E. and {Bilenko}, I.~A. and {Billingsley}, G. and {Birney}, R. and {Birnholtz}, O. and {Biscans}, S. and {Bischi}, M. and {Biscoveanu}, S. and {Bisht}, A. and {Bissenbayeva}, G. and {Bitossi}, M. and {Bizouard}, M.~A. and {Blackburn}, J.~K. and {Blackman}, J. and {Blair}, C.~D. and {Blair}, D.~G. and {Blair}, R.~M. and {Bobba}, F. and {Bode}, N. and {Boer}, M. and {Boetzel}, Y. and {Bogaert}, G. and {Bondu}, F. and {Bonilla}, E. and {Bonnand}, R. and {Booker}, P. and {Boom}, B.~A. and {Bork}, R. and {Boschi}, V. and {Bose}, S. and {Bossilkov}, V. and {Bosveld}, J. and {Bouffanais}, Y. and {Bozzi}, A. and {Bradaschia}, C. and {Brady}, P.~R. and {Bramley}, A. and {Branchesi}, M. and {Brau}, J.~E. and {Breschi}, M. and {Briant}, T. and {Briggs}, J.~H. and {Brighenti}, F. and {Brillet}, A. and {Brinkmann}, M. and {Brockill}, P. and {Brooks}, A.~F. and {Brooks}, J. and {Brown}, D.~D. and {Brunett}, S. and {Bruno}, G. and {Bruntz}, R. and {Buikema}, A. and {Bulik}, T. and {Bulten}, H.~J. and {Buonanno}, A. and {Buskulic}, D. and {Byer}, R.~L. and {Cabero}, M. and {Cadonati}, L. and {Cagnoli}, G. and {Cahillane}, C. and {Bustillo}, J. Calder{\'o}n and {Callaghan}, J.~D. and {Callister}, T.~A. and {Calloni}, E. and {Camp}, J.~B. and {Canepa}, M. and {Cannon}, K.~C. and {Cao}, H. and {Cao}, J. and {Carapella}, G. and {Carbognani}, F. and {Caride}, S. and {Carney}, M.~F. and {Carullo}, G. and {Diaz}, J. Casanueva and {Casentini}, C. and {Casta{\~n}eda}, J. and {Caudill}, S. and {Cavagli{\`a}}, M. and {Cavalier}, F. and {Cavalieri}, R. and {Cella}, G. and {Cerd{\'a}-Dur{\'a}n}, P. and {Cesarini}, E. and {Chaibi}, O. and {Chakravarti}, K. and {Chan}, C. and {Chan}, M. and {Chao}, S. and {Charlton}, P. and {Chase}, E.~A. and {Chassande-Mottin}, E. and {Chatterjee}, D. and {Chaturvedi}, M. and {Chen}, H.~Y. and {Chen}, X. and {Chen}, Y. and {Cheng}, H.-P. and {Cheong}, C.~K. and {Chia}, H.~Y. and {Chiadini}, F.},
        title = "{Open data from the first and second observing runs of Advanced LIGO and Advanced Virgo}",
      journal = {SoftwareX},
         year = 2021,
        month = jan,
       volume = {13},
          eid = {100658},
        pages = {100658},
          doi = {10.1016/j.softx.2021.100658},
archivePrefix = {arXiv},
       eprint = {1912.11716},
 primaryClass = {gr-qc},
       adsurl = {https://ui.adsabs.harvard.edu/abs/2021SoftX..1300658A}
}

@ARTICLE{KAGRA:2023pio,
       author = {{Abbott}, R. and {Abe}, H. and {Acernese}, F. and {Ackley}, K. and {Adhicary}, S. and {Adhikari}, N. and {Adhikari}, R.~X. and {Adkins}, V.~K. and {Adya}, V.~B. and {Affeldt}, C. and {Agarwal}, D. and {Agathos}, M. and {Aguiar}, O.~D. and {Aiello}, L. and {Ain}, A. and {Ajith}, P. and {Akutsu}, T. and {Albanesi}, S. and {Alfaidi}, R.~A. and {Al-Jodah}, A. and {All{\'e}n{\'e}}, C. and {Allocca}, A. and {Almualla}, M. and {Altin}, P.~A. and {Amato}, A. and {Amez-Droz}, L. and {Amorosi}, A. and {Anand}, S. and {Ananyeva}, A. and {Andersen}, R. and {Anderson}, S.~B. and {Anderson}, W.~G. and {Andia}, M. and {Ando}, M. and {Andrade}, T. and {Andres}, N. and {Andr{\'e}s-Carcasona}, M. and {Andri{\'c}}, T. and {Ansoldi}, S. and {Antelis}, J.~M. and {Antier}, S. and {Aoumi}, M. and {Apostolatos}, T. and {Appavuravther}, E.~Z. and {Appert}, S. and {Apple}, S.~K. and {Arai}, K. and {Araya}, A. and {Araya}, M.~C. and {Areeda}, J.~S. and {Ar{\`e}ne}, M. and {Aritomi}, N. and {Arnaud}, N. and {Arogeti}, M. and {Aronson}, S.~M. and {Arun}, K.~G. and {Asada}, H. and {Ashton}, G. and {Aso}, Y. and {Assiduo}, M. and {Assis de Souza Melo}, S. and {Aston}, S.~M. and {Astone}, P. and {Aubin}, F. and {Aultoneal}, K. and {Babak}, S. and {Badalyan}, A. and {Badaracco}, F. and {Badger}, C. and {Bae}, S. and {Bagnasco}, S. and {Bai}, Y. and {Baier}, J.~G. and {Baiotti}, L. and {Baird}, J. and {Bajpai}, R. and {Baka}, T. and {Ball}, M. and {Ballardin}, G. and {Ballmer}, S.~W. and {Baltus}, G. and {Banagiri}, S. and {Banerjee}, B. and {Bankar}, D. and {Baral}, P. and {Barayoga}, J.~C. and {Barber}, J. and {Barish}, B.~C. and {Barker}, D. and {Barneo}, P. and {Barone}, F. and {Barr}, B. and {Barsotti}, L. and {Barsuglia}, M. and {Barta}, D. and {Barthelmy}, S.~D. and {Barton}, M.~A. and {Bartos}, I. and {Basak}, S. and {Basalaev}, A. and {Bassiri}, R. and {Basti}, A. and {Bawaj}, M. and {Bayley}, J.~C. and {Baylor}, A.~C. and {Bazzan}, M. and {B{\'e}csy}, B. and {Bedakihale}, V.~M. and {Beirnaert}, F. and {Bejger}, M. and {Bell}, A.~S. and {Benedetto}, V. and {Beniwal}, D. and {Benoit}, W. and {Bentley}, J.~D. and {Yaala}, M. Ben and {Bera}, S. and {Berbel}, M. and {Bergamin}, F. and {Berger}, B.~K. and {Bernuzzi}, S. and {Beroiz}, M. and {Berry}, C.~P.~L. and {Bersanetti}, D. and {Bertolini}, A. and {Betzwieser}, J. and {Beveridge}, D. and {Bevins}, N. and {Bhandare}, R. and {Bhandari}, A.~V. and {Bhardwaj}, U. and {Bhatt}, R. and {Bhattacharjee}, D. and {Bhaumik}, S. and {Bianchi}, A. and {Bilenko}, I.~A. and {Bilicki}, M. and {Billingsley}, G. and {Bini}, S. and {Birnholtz}, O. and {Biscans}, S. and {Bischi}, M. and {Biscoveanu}, S. and {Bisht}, A. and {Biswas}, B. and {Bitossi}, M. and {Bizouard}, M.-A. and {Blackburn}, J.~K. and {Blair}, C.~D. and {Blair}, D.~G. and {Blair}, R.~M. and {Bobba}, F. and {Bode}, N. and {Bo{\"e}r}, M. and {Bogaert}, G. and {Boileau}, G. and {Boldrini}, M. and {Bolingbroke}, G.~N. and {Bonavena}, L.~D. and {Bondarescu}, R. and {Bondu}, F. and {Bonilla}, E. and {Bonilla}, G.~S. and {Bonnand}, R. and {Booker}, P. and {Bork}, R. and {Boschi}, V. and {Bose}, N. and {Bose}, S. and {Bossilkov}, V. and {Boudart}, V. and {Bouffanais}, Y. and {Bozzi}, A. and {Bradaschia}, C. and {Brady}, P.~R. and {Braglia}, M. and {Branch}, A. and {Branchesi}, M. and {Brau}, J.~E. and {Breschi}, M. and {Briant}, T. and {Brillet}, A. and {Brinkmann}, M. and {Brockill}, P. and {Brooks}, A.~F. and {Brooks}, J. and {Brown}, D.~D. and {Brunett}, S. and {Bruno}, G. and {Bruntz}, R. and {Bryant}, J. and {Bucci}, F. and {Buchanan}, J. and {Bulashenko}, O. and {Bulik}, T. and {Bulten}, H.~J. and {Buonanno}, A. and {Burtnyk}, K. and {Buscicchio}, R. and {Buskulic}, D.},
        title = "{Open Data from the Third Observing Run of LIGO, Virgo, KAGRA, and GEO}",
      journal = {\apjs},
         year = 2023,
        month = aug,
       volume = {267},
       number = {2},
          eid = {29},
        pages = {29},
          doi = {10.3847/1538-4365/acdc9f},
archivePrefix = {arXiv},
       eprint = {2302.03676},
 primaryClass = {gr-qc},
       adsurl = {https://ui.adsabs.harvard.edu/abs/2023ApJS..267...29A}
}

@ARTICLE{Damour1991+Pdot,
       author = {{Damour}, Thibault and {Taylor}, J.~H.},
        title = "{On the Orbital Period Change of the Binary Pulsar PSR 1913+16}",
      journal = {\apj},
         year = 1991,
        month = jan,
       volume = {366},
        pages = {501},
          doi = {10.1086/169585},
       adsurl = {https://ui.adsabs.harvard.edu/abs/1991ApJ...366..501D}
}

@BOOK{Tauris2023,
  author    = {Tauris, Thomas M. and van den Heuvel, Edward P.~J.},
  title     = {Physics of Binary Star Evolution. From Stars to X-ray Binaries and Gravitational Wave Sources},
  year      = {2023},
  publisher = {Princeton University Press},
  doi       = {10.48550/arXiv.2305.09388},
  adsurl    = {https://ui.adsabs.harvard.edu/abs/2023pbse.book.....T}
}

@ARTICLE{Clark2026+Gibbs,
       author = {{Clark}, Colin J. and {Valtolina}, Serena and {Nieder}, Lars and {van Haasteren}, Rutger},
        title = "{Timing Gamma-ray Pulsars using Gibbs Sampling}",
      journal = {arXiv e-prints},
         year = 2026,
        month = jan,
          eid = {arXiv:2601.07592},
        pages = {arXiv:2601.07592},
archivePrefix = {arXiv},
       eprint = {2601.07592},
 primaryClass = {astro-ph.HE},
       adsurl = {https://ui.adsabs.harvard.edu/abs/2026arXiv260107592C}
}

@phdthesis{Thongmeearkom2025,
  author  = {Thongmeearkom, Tinn},
  title   = {The Investigation of Pulsars in Binary Systems},
  school  = {The University of Manchester},
  year    = {2025},
  type     = {\href{https://research.manchester.ac.uk/en/studentTheses/the-investigation-of-pulsars-in-binary-systems/}{{PhD} thesis}}
}

% Alternatively you could enter them by hand, like this:
% This method is tedious and prone to error if you have lots of references
%\begin{thebibliography}{99}
%\bibitem[\protect\citeauthoryear{Author}{2012}]{Author2012}
%Author A.~N., 2013, Journal of Improbable Astronomy, 1, 1
%\bibitem[\protect\citeauthoryear{Others}{2013}]{Others2013}
%Others S., 2012, Journal of Interesting Stuff, 17, 198
%\end{thebibliography}

%%%%%%%%%%%%%%%%%%%%%%%%%%%%%%%%%%%%%%%%%%%%%%%%%%

%%%%%%%%%%%%%%%%% APPENDICES %%%%%%%%%%%%%%%%%%%%%

\appendix

\section{Observations and Slow Pulsars}
Tables \ref{T:source_list_1} and \ref{T:source_list_2} summarise the UHF observations of \textit{Fermi}-LAT sources, while Table \ref{T:slow_pulsars} presents initial information on the slow pulsars discussed in this work. The observational tables and the timing parameter (.par) files corresponding to the slow pulsars are available as supplementary material accompanying this paper.

\begin{table*}
	\caption{A list of UHF survey observations of \textit{Fermi}-LAT sources for the first epoch. $S_{\rm min}$, $S_{\rm med}$, and $S_{95}$ are sensitivity estimates from the radiometer equation. $S_{\rm min}$ represents a pulsar located at the centre of a coherent beam under ideal assumptions, while $S_{\rm med}$ (median) and $S_{95}$ (95th percentile) account for the tiling pattern of the beams. nAnt is the number of antennas.$T_{\rm sky}$ and $T_{\rm sys}$ are the sky and system temperatures, respectively. Pulsars discovered in this survey (including both confirmed and unconfirmed \textit{Fermi}-LAT associations) are indicated with an asterisk (*), while those from other \textit{Fermi}-TRAPUM surveys are marked with a dagger ($\dagger$).}
	\label{T:source_list_1}
	\centering
    \renewcommand{\arraystretch}{1.25}
    \begin{tabular}{lccccccccc}
    \hline
    4FGL~NAME & R.A. (J2000) & Decl. (J2000) & MJD & $S_{\rm min}$ ($\mu$Jy) & $S_{\rm med}$ ($\mu$Jy) & $S_{95}$ ($\mu$Jy) & nAnt & $T_{\rm sky}$ (K) & $T_{\rm sys}$ (K)\\
    \hline
J0048.6$-$6347 & $00^{\rm h}48^{\rm m}40\fs44$ & $-63\degr47\arcmin29\farcs0433$ & 59471.7393 & 69 & 148 & 167 & 56 & 3.9 & 22.4 \\
J0139.5$-$2228 & $01^{\rm h}39^{\rm m}35\fs3041$ & $-22\degr28\arcmin39\farcs7174$ & 59471.8307 & 68 & 153 & 183 & 56 & 3.4 & 21.9 \\
J0414.7$-$4300 & $04^{\rm h}14^{\rm m}47\fs4481$ & $-43\degr00\arcmin43\farcs2037$ & 59563.0916 & 62 & 106 & 122 & 60 & 2.6 & 21.1 \\
J0529.9$-$0224 & $05^{\rm h}29^{\rm m}54\fs4794$ & $-02\degr24\arcmin17\farcs2798$ & 59563.0843 & 69 & 121 & 137 & 60 & 4.8 & 23.3 \\
J0540.0$-$7552 & $05^{\rm h}40^{\rm m}01\fs5363$ & $-75\degr52\arcmin41\farcs5082$ & 59563.1874 & 64 & 96 & 115 & 60 & 3.1 & 21.6 \\
J0657.4$-$4658* & $06^{\rm h}57^{\rm m}26\fs3525$ & $-46\degr58\arcmin48\farcs7262$ & 59716.3578 & 76 & 111 & 130 & 56 & 3.4 & 22.0 \\
J0712.0$-$6431 & $07^{\rm h}12^{\rm m}01\fs9208$ & $-64\degr31\arcmin32\farcs1478$ & 59730.3127 & 64 & 115 & 133 & 60 & 3.0 & 21.5 \\
J0906.8$-$2122 & $09^{\rm h}06^{\rm m}49\fs1052$ & $-21\degr22\arcmin20\farcs641$ & 59496.4563 & 63 & 118 & 133 & 60 & 2.6 & 21.2 \\
J0953.6$-$1509 & $09^{\rm h}53^{\rm m}37\fs3206$ & $-15\degr09\arcmin17\farcs6385$ & 59496.4402 & 64 & 81 & 93 & 60 & 3.1 & 21.7 \\
J1036.6$-$4349 & $10^{\rm h}36^{\rm m}36\fs167$ & $-43\degr49\arcmin26\farcs0339$ & 59496.5378 & 68 & 98 & 116 & 60 & 4.2 & 22.7 \\
J1106.7$-$1742 & $11^{\rm h}06^{\rm m}47\fs7612$ & $-17\degr42\arcmin53\farcs2796$ & 59496.5302 & 64 & 91 & 106 & 60 & 3.2 & 21.7 \\
J1120.0$-$2204 & $11^{\rm h}20^{\rm m}00\fs3845$ & $-22\degr04\arcmin40\farcs4398$ & 59563.1075 & 62 & 74 & 82 & 60 & 2.5 & 21.1 \\
J1126.0$-$5007 & $11^{\rm h}26^{\rm m}03\fs479$ & $-50\degr07\arcmin09\farcs8401$ & 59496.5449 & 69 & 116 & 134 & 60 & 4.3 & 22.8 \\
J1204.5$-$5032 & $12^{\rm h}04^{\rm m}35\fs592$ & $-50\degr32\arcmin44\farcs1632$ & 59887.1282 & 67 & 126 & 144 & 60 & 4.8 & 23.3 \\
J1207.4$-$4536* & $12^{\rm h}07^{\rm m}29\fs6155$ & $-45\degr36\arcmin44\farcs9973$ & 59563.0618 & 66 & 115 & 131 & 60 & 3.6 & 22.2 \\
J1213.9$-$4416$^\dagger$ & $12^{\rm h}13^{\rm m}58\fs656$ & $-44\degr16\arcmin46\farcs1966$ & 59563.0688 & 66 & 98 & 115 & 60 & 3.7 & 22.2 \\
J1259.0$-$8148* & $12^{\rm h}59^{\rm m}02\fs9993$ & $-81\degr48\arcmin58\farcs678$ & 59471.7322 & 70 & 156 & 182 & 56 & 3.9 & 22.5 \\
J1303.1$-$4714* & $13^{\rm h}03^{\rm m}09\fs071$ & $-47\degr14\arcmin17\farcs1533$ & 59471.718 & 77 & 172 & 201 & 56 & 5.3 & 23.9 \\
J1345.9$-$2612* & $13^{\rm h}45^{\rm m}55\fs5615$ & $-26\degr12\arcmin41\farcs7577$ & 59563.1001 & 69 & 112 & 130 & 60 & 4.7 & 23.2 \\
J1356.6+0234* & $13^{\rm h}56^{\rm m}39\fs3347$ & $+02\degr34\arcmin38\farcs2803$ & 59563.1296 & 65 & 119 & 132 & 60 & 3.4 & 21.9 \\
J1400.0$-$2415 & $14^{\rm h}00^{\rm m}04\fs9438$ & $-24\degr15\arcmin57\farcs6027$ & 59563.1223 & 69 & 101 & 119 & 60 & 4.6 & 23.2 \\
J1416.7$-$5023 & $14^{\rm h}16^{\rm m}42\fs7441$ & $-50\degr23\arcmin22\farcs1988$ & 59563.076 & 83 & 121 & 141 & 60 & 9.8 & 28.4 \\
J1450.8$-$1424 & $14^{\rm h}50^{\rm m}50\fs592$ & $-14\degr24\arcmin21\farcs24$ & 59563.1368 & 69 & 117 & 135 & 60 & 4.9 & 23.4 \\
J1458.8$-$2120 & $14^{\rm h}58^{\rm m}48\fs7903$ & $-21\degr20\arcmin19\farcs6815$ & 59563.1726 & 71 & 100 & 117 & 60 & 5.6 & 24.1 \\
J1513.7$-$1519 & $15^{\rm h}13^{\rm m}46\fs6809$ & $-15\degr19\arcmin59\farcs1611$ & 59563.1512 & 73 & 105 & 123 & 60 & 6.3 & 24.8 \\
J1517.7$-$4446 & $15^{\rm h}17^{\rm m}42\fs7185$ & $-44\degr46\arcmin36\farcs1166$ & 59563.115 & 90 & 156 & 177 & 60 & 12.2 & 30.8 \\
J1526.6$-$3810 & $15^{\rm h}26^{\rm m}38\fs2544$ & $-38\degr10\arcmin08\farcs3954$ & 59563.1439 & 87 & 154 & 172 & 60 & 11.0 & 29.5 \\
J1527.8+1013 & $15^{\rm h}27^{\rm m}53\fs0164$ & $+10\degr13\arcmin42\farcs9584$ & 59496.4178 & 72 & 138 & 157 & 60 & 6.0 & 24.5 \\
J1530.0$-$1522 & $15^{\rm h}30^{\rm m}00\fs2637$ & $-15\degr22\arcmin33\farcs241$ & 59563.18 & 74 & 137 & 153 & 60 & 6.4 & 24.9 \\
J1539.4$-$3323 & $15^{\rm h}39^{\rm m}24\fs2651$ & $-33\degr23\arcmin55\farcs3226$ & 59563.1654 & 79 & 89 & 96 & 60 & 8.4 & 27.0 \\
J1543.6$-$0244 & $15^{\rm h}43^{\rm m}37\fs8479$ & $-02\degr44\arcmin49\farcs5604$ & 59471.7711 & 72 & 157 & 180 & 56 & 4.5 & 23.0 \\
J1544.2$-$2554$^\dagger$ & $15^{\rm h}44^{\rm m}12\fs5537$ & $-25\degr54\arcmin45\farcs0014$ & 59563.1583 & 75 & 95 & 109 & 60 & 6.9 & 25.5 \\
J1612.1+1407 & $16^{\rm h}12^{\rm m}07\fs511$ & $+14\degr07\arcmin00\farcs4811$ & 59496.4787 & 81 & 121 & 142 & 60 & 8.5 & 27.0 \\
J1622.2$-$7202 & $16^{\rm h}22^{\rm m}12\fs6013$ & $-72\degr02\arcmin23\farcs6462$ & 59471.7859 & 75 & 162 & 185 & 56 & 5.5 & 24.0 \\
J1630.1$-$1049 & $16^{\rm h}30^{\rm m}06\fs936$ & $-10\degr49\arcmin05\farcs8809$ & 59496.4043 & 83 & 122 & 143 & 60 & 7.0 & 25.5 \\
J1641.3$-$2908 & $16^{\rm h}41^{\rm m}21\fs3611$ & $-29\degr08\arcmin21\farcs4778$ & 59471.8387 & 91 & 206 & 245 & 56 & 10.6 & 29.1 \\
J1646.7$-$2154 & $16^{\rm h}46^{\rm m}44\fs425$ & $-21\degr54\arcmin26\farcs9975$ & 59471.8225 & 89 & 202 & 243 & 56 & 10.1 & 28.6 \\
J1656.4$-$0410 & $16^{\rm h}56^{\rm m}28\fs656$ & $-04\degr10\arcmin12\farcs7195$ & 59496.4484 & 76 & 111 & 131 & 60 & 6.8 & 25.4 \\
J1659.0$-$0140 & $16^{\rm h}59^{\rm m}03\fs7207$ & $-01\degr40\arcmin39\arcsec$ & 59496.5004 & 79 & 115 & 134 & 60 & 8.2 & 26.8 \\
J1708.8$-$0923 & $17^{\rm h}08^{\rm m}48\fs0981$ & $-09\degr23\arcmin39\farcs4817$ & 59496.4253 & 81 & 160 & 185 & 60 & 8.4 & 27.0 \\
J1709.4$-$2127 & $17^{\rm h}09^{\rm m}25\fs3711$ & $-21\degr27\arcmin23\farcs0397$ & 59496.4108 & 95 & 168 & 191 & 60 & 12.6 & 31.1 \\
J1709.9$-$0900 & $17^{\rm h}09^{\rm m}56\fs0156$ & $-09\degr00\arcmin51\farcs8383$ & 59496.4325 & 82 & 161 & 185 & 60 & 8.4 & 26.9 \\
    \hline
  \end{tabular}
\end{table*}

\begin{table*}
	\contcaption{A list of UHF survey observations of \textit{Fermi}-LAT sources for the first epoch.}
	\centering
    \renewcommand{\arraystretch}{1.25}
    \begin{tabular}{lccccccccc}
    \hline
    4FGL~NAME & R.A. (J2000) & Decl. (J2000) & MJD & $S_{\rm min}$ ($\mu$Jy) & $S_{\rm med}$ ($\mu$Jy) & $S_{95}$ ($\mu$Jy) & nAnt & $T_{\rm sky}$ (K) & $T_{\rm sys}$ (K)\\
    \hline
%J1709.4$-$2127 & $17^{\rm h}09^{\rm m}25\fs3711$ & $-21\degr27\arcmin23\farcs0397$ & 59496.4108 & 95 & 168 & 191 & 60 & 12.6 & 31.1 \\
%J1709.9$-$0900 & $17^{\rm h}09^{\rm m}56\fs0156$ & $-09\degr00\arcmin51\farcs8383$ & 59496.4325 & 82 & 161 & 185 & 60 & 8.4 & 26.9 \\
J1711.9$-$1922* & $17^{\rm h}11^{\rm m}54\fs624$ & $-19\degr22\arcmin03\farcs3582$ & 59496.4934 & 88 & 124 & 146 & 60 & 10.7 & 29.3 \\
J1717.5$-$5804 & $17^{\rm h}17^{\rm m}30\fs813$ & $-58\degr04\arcmin14\farcs155$ & 59471.7932 & 92 & 200 & 228 & 56 & 10.8 & 29.4 \\
J1720.6+0708 & $17^{\rm h}20^{\rm m}39\fs3091$ & $+07\degr08\arcmin48\farcs8406$ & 59471.8153 & 90 & 196 & 227 & 56 & 10.2 & 28.7 \\
J1722.8$-$0418 & $17^{\rm h}22^{\rm m}50\fs7568$ & $-04\degr18\arcmin11\farcs8797$ & 59496.4718 & 84 & 110 & 128 & 60 & 8.8 & 27.3 \\
J1727.4+0326 & $17^{\rm h}27^{\rm m}27\fs1436$ & $+03\degr26\arcmin59\farcs2801$ & 59496.4644 & 82 & 115 & 135 & 60 & 8.7 & 27.2 \\
J1730.4$-$0359 & $17^{\rm h}30^{\rm m}26\fs0669$ & $-03\degr59\arcmin32\farcs2801$ & 59471.7782 & 88 & 182 & 207 & 56 & 9.8 & 28.3 \\
J1735.3$-$0717 & $17^{\rm h}35^{\rm m}21\fs6211$ & $-07\degr17\arcmin07\farcs0792$ & 59471.8549 & 95 & 211 & 248 & 56 & 11.5 & 30.1 \\
J1739.1$-$1059 & $17^{\rm h}39^{\rm m}08\fs064$ & $-10\degr59\arcmin41\farcs2786$ & 59471.862 & 99 & 230 & 286 & 56 & 12.9 & 31.5 \\
J1747.6+0324 & $17^{\rm h}47^{\rm m}37\fs2217$ & $+03\degr24\arcmin26\farcs28$ & 59496.5152 & 90 & 125 & 147 & 60 & 11.3 & 29.9 \\
J1749.8$-$0303 & $17^{\rm h}49^{\rm m}50\fs9253$ & $-03\degr03\arcmin06\farcs48$ & 59496.4863 & 96 & 137 & 162 & 60 & 12.9 & 31.4 \\
J1813.7$-$6846 & $18^{\rm h}13^{\rm m}47\fs5415$ & $-68\degr46\arcmin42\farcs6123$ & 59786.0961 & 70 & 98 & 117 & 60 & 5.8 & 24.3 \\
J1816.4$-$6405* & $18^{\rm h}16^{\rm m}27\fs5977$ & $-64\degr05\arcmin05\farcs9967$ & 59786.089 & 73 & 95 & 111 & 60 & 7.0 & 25.5 \\
J1816.7+1749 & $18^{\rm h}16^{\rm m}45\fs0513$ & $+17\degr49\arcmin40\farcs4398$ & 59471.8081 & 86 & 187 & 213 & 56 & 9.0 & 27.6 \\
J1818.6+1316 & $18^{\rm h}18^{\rm m}36\fs6504$ & $+13\degr16\arcmin23\farcs1596$ & 59471.7475 & 86 & 188 & 214 & 56 & 9.2 & 27.8 \\
J1822.9$-$4718 & $18^{\rm h}22^{\rm m}56\fs6895$ & $-47\degr18\arcmin16\farcs9162$ & 59786.0742 & 76 & 96 & 111 & 60 & 8.0 & 26.6 \\
J1823.2+1209* & $18^{\rm h}23^{\rm m}14\fs8828$ & $+12\degr09\arcmin26\farcs28$ & 59471.8475 & 90 & 205 & 246 & 56 & 10.5 & 29.0 \\
J1824.2$-$5427 & $18^{\rm h}24^{\rm m}16\fs897$ & $-54\degr27\arcmin05\farcs0427$ & 59471.8003 & 77 & 166 & 188 & 56 & 6.2 & 24.7 \\
J1827.5+1141 & $18^{\rm h}27^{\rm m}30\fs8643$ & $+11\degr41\arcmin10\farcs681$ & 59471.7546 & 94 & 201 & 228 & 56 & 11.3 & 29.9 \\
J1831.1$-$6503* & $18^{\rm h}31^{\rm m}06\fs5552$ & $-65\degr03\arcmin57\farcs2498$ & 59786.1032 & 71 & 81 & 90 & 60 & 6.2 & 24.8 \\
J1845.8$-$2521* & $18^{\rm h}45^{\rm m}51\fs5552$ & $-25\degr21\arcmin30\farcs5983$ & 59786.067 & 90 & 116 & 134 & 60 & 13.0 & 31.5 \\
J1906.0$-$1718 & $19^{\rm h}06^{\rm m}04\fs4604$ & $-17\degr18\arcmin55\farcs08$ & 59786.0523 & 81 & 127 & 147 & 60 & 9.7 & 28.3 \\
J1910.7$-$5320* & $19^{\rm h}10^{\rm m}47\fs7393$ & $-53\degr20\arcmin18\farcs6035$ & 59716.312 & 85 & 137 & 159 & 56 & 5.8 & 24.3 \\
J1913.4$-$1526 & $19^{\rm h}13^{\rm m}24\fs3604$ & $-15\degr26\arcmin58\farcs5608$ & 59496.508 & 85 & 123 & 145 & 60 & 9.5 & 28.1 \\
J1916.8$-$3025 & $19^{\rm h}16^{\rm m}53\fs6865$ & $-30\degr25\arcmin28\farcs5622$ & 59835.7782 & 83 & 98 & 109 & 56 & 8.2 & 26.8 \\
J1924.8$-$1035* & $19^{\rm h}24^{\rm m}49\fs27$ & $-10\degr35\arcmin27\farcs2415$ & 59786.0301 & 83 & 100 & 112 & 60 & 10.2 & 28.8 \\
J1947.6$-$1121 & $19^{\rm h}47^{\rm m}39\fs8657$ & $-11\degr21\arcmin54\farcs7202$ & 59786.0816 & 73 & 96 & 110 & 60 & 6.8 & 25.4 \\
J2026.3+1431 & $20^{\rm h}26^{\rm m}23\fs7085$ & $+14\degr31\arcmin21\farcs0001$ & 59786.0596 & 68 & 88 & 101 & 60 & 5.2 & 23.7 \\
J2029.5$-$4237* & $20^{\rm h}29^{\rm m}30\fs3589$ & $-42\degr37\arcmin42\farcs6068$ & 59496.5226 & 73 & 140 & 162 & 60 & 5.7 & 24.2 \\
J2043.9$-$4802 & $20^{\rm h}43^{\rm m}55\fs5615$ & $-48\degr02\arcmin23\farcs2755$ & 59786.0375 & 67 & 86 & 98 & 60 & 5.0 & 23.5 \\
J2112.5$-$3043 & $21^{\rm h}12^{\rm m}33\fs6035$ & $-30\degr43\arcmin45\farcs4784$ & 59786.0448 & 66 & 71 & 75 & 60 & 4.4 & 23.0 \\
J2121.8$-$3412 & $21^{\rm h}21^{\rm m}53\fs811$ & $-34\degr12\arcmin11\farcs8817$ & 59730.297 & 70 & 140 & 163 & 60 & 4.1 & 22.6 \\
J2133.1$-$6432 & $21^{\rm h}33^{\rm m}10\fs8252$ & $-64\degr32\arcmin17\farcs8784$ & 59786.1105 & 65 & 83 & 95 & 60 & 4.2 & 22.8 \\
J2201.0$-$6928 & $22^{\rm h}01^{\rm m}01\fs6772$ & $-69\degr28\arcmin26\farcs0394$ & 59730.3039 & 70 & 109 & 130 & 60 & 4.2 & 22.7 \\
J2212.4+0708 & $22^{\rm h}12^{\rm m}25\fs9937$ & $+07\degr08\arcmin34\farcs0795$ & 59730.2817 & 70 & 106 & 129 & 60 & 3.8 & 22.3 \\
J2219.7$-$6837 & $22^{\rm h}19^{\rm m}47\fs2559$ & $-68\degr37\arcmin02\farcs287$ & 59730.2893 & 71 & 101 & 120 & 60 & 3.9 & 22.4 \\
J2241.4$-$8327 & $22^{\rm h}41^{\rm m}25\fs6348$ & $-83\degr27\arcmin40\farcs3088$ & 59471.7251 & 72 & 158 & 183 & 56 & 4.2 & 22.7 \\
J2355.5$-$6614* & $23^{\rm h}55^{\rm m}33\fs7939$ & $-66\degr14\arcmin04\farcs1895$ & 59471.7627 & 71 & 159 & 187 & 56 & 3.9 & 22.4 \\
    \hline
  \end{tabular}
\end{table*}

\begin{table*}
	\caption{A list of UHF survey observations of \textit{Fermi}-LAT sources for the second epoch. $S_{\rm min}$, $S_{\rm med}$, and $S_{95}$ are sensitivity estimates from the radiometer equation. $S_{\rm min}$ represents a pulsar located at the centre of a coherent beam under ideal assumptions, while $S_{\rm med}$ (median) and $S_{95}$ (95th percentile) account for the tiling pattern of the beams. nAnt is the number of antennas.$T_{\rm sky}$ and $T_{\rm sys}$ are the sky and system temperatures, respectively. Pulsars discovered in this survey (including both confirmed and unconfirmed \textit{Fermi}-LAT associations) are indicated with an asterisk (*), while those from other \textit{Fermi}-TRAPUM surveys are marked with a dagger ($\dagger$). Dashes (--) indicate missing values corresponding to sources removed from the second epoch after being independently discovered by other surveys.}
    \label{T:source_list_2}
	\centering
    \renewcommand{\arraystretch}{1.25}
    \begin{tabular}{lccccccccc}
    \hline
    4FGL~NAME & R.A. (J2000) & Decl. (J2000) & MJD & $S_{\rm min}$ ($\mu$Jy) & $S_{\rm med}$ ($\mu$Jy) & $S_{95}$ ($\mu$Jy) & nAnt & $T_{\rm sky}$ (K) & $T_{\rm sys}$ (K)\\
    \hline
J0048.6$-$6347 & $00^{\rm h}48^{\rm m}40\fs44$ & $-63\degr47\arcmin29\farcs0433$ & 59882.0926 & 64 & 91 & 108 & 60 & 3.9 & 22.4 \\
J0139.5$-$2228 & $01^{\rm h}39^{\rm m}35\fs3041$ & $-22\degr28\arcmin39\farcs7174$ & 59882.0854 & 64 & 109 & 124 & 60 & 3.4 & 21.9 \\
J0414.7$-$4300 & $04^{\rm h}14^{\rm m}47\fs4481$ & $-43\degr00\arcmin43\farcs2037$ & 59887.1587 & 61 & 120 & 135 & 60 & 2.6 & 21.1 \\
J0529.9$-$0224 & $05^{\rm h}29^{\rm m}54\fs4794$ & $-02\degr24\arcmin17\farcs2798$ & 59887.1435 & 67 & 136 & 151 & 60 & 4.8 & 23.3 \\
J0540.0$-$7552 & $05^{\rm h}40^{\rm m}01\fs5363$ & $-75\degr52\arcmin41\farcs5082$ & 59887.1512 & 62 & 105 & 125 & 60 & 3.1 & 21.6 \\
J0657.4$-$4658* & $06^{\rm h}57^{\rm m}26\fs3525$ & $-46\degr58\arcmin48\farcs7262$ & 59730.3197 & 68 & 111 & 132 & 60 & 3.4 & 22.0 \\
J0712.0$-$6431 & $07^{\rm h}12^{\rm m}01\fs9208$ & $-64\degr31\arcmin32\farcs1478$ & 59786.1495 & 62 & 91 & 107 & 60 & 3.0 & 21.5 \\
J0906.8$-$2122 & $09^{\rm h}06^{\rm m}49\fs1052$ & $-21\degr22\arcmin20\farcs641$ & 59978.1197 & 61 & 112 & 128 & 60 & 2.6 & 21.2 \\
J0953.6$-$1509 & $09^{\rm h}53^{\rm m}37\fs3206$ & $-15\degr09\arcmin17\farcs6385$ & 59887.1066 & 62 & 81 & 94 & 60 & 3.1 & 21.7 \\
J1036.6$-$4349 & $10^{\rm h}36^{\rm m}36\fs167$ & $-43\degr49\arcmin26\farcs0339$ & -- & -- & -- & -- & -- & -- & -- \\
J1106.7$-$1742 & $11^{\rm h}06^{\rm m}47\fs7612$ & $-17\degr42\arcmin53\farcs2796$ & 59887.0924 & 63 & 93 & 110 & 60 & 3.2 & 21.7 \\
J1120.0$-$2204 & $11^{\rm h}20^{\rm m}00\fs3845$ & $-22\degr04\arcmin40\farcs4398$ & 59887.0994 & 60 & 67 & 72 & 60 & 2.5 & 21.1 \\
J1126.0$-$5007 & $11^{\rm h}26^{\rm m}03\fs479$ & $-50\degr07\arcmin09\farcs8401$ & 59886.0715 & 71 & 108 & 124 & 56 & 4.3 & 22.8 \\
J1204.5$-$5032 & $12^{\rm h}04^{\rm m}35\fs592$ & $-50\degr32\arcmin44\farcs1632$ & -- & -- & -- & -- & -- & -- & -- \\
J1207.4$-$4536* & $12^{\rm h}07^{\rm m}29\fs6155$ & $-45\degr36\arcmin44\farcs9973$ & 59887.121 & 63 & 107 & 127 & 60 & 3.6 & 22.2 \\
J1213.9$-$4416$^\dagger$ & $12^{\rm h}13^{\rm m}58\fs656$ & $-44\degr16\arcmin46\farcs1966$ & 59887.114 & 64 & 87 & 103 & 60 & 3.7 & 22.2 \\
J1259.0$-$8148* & $12^{\rm h}59^{\rm m}02\fs9993$ & $-81\degr48\arcmin58\farcs678$ & -- & -- & -- & -- & -- & -- & -- \\
J1303.1$-$4714* & $13^{\rm h}03^{\rm m}09\fs071$ & $-47\degr14\arcmin17\farcs1533$ & 59887.1354 & 68 & 126 & 147 & 60 & 5.3 & 23.9 \\
J1345.9$-$2612* & $13^{\rm h}45^{\rm m}55\fs5615$ & $-26\degr12\arcmin41\farcs7577$ & 59835.6952 & 71 & 102 & 119 & 56 & 4.7 & 23.2 \\
J1356.6+0234* & $13^{\rm h}56^{\rm m}39\fs3347$ & $+02\degr34\arcmin38\farcs2803$ & -- & -- & -- & -- & -- & -- & -- \\
J1400.0$-$2415 & $14^{\rm h}00^{\rm m}04\fs9438$ & $-24\degr15\arcmin57\farcs6027$ & 59835.7255 & 72 & 90 & 102 & 56 & 4.6 & 23.2 \\
J1416.7$-$5023 & $14^{\rm h}16^{\rm m}42\fs7441$ & $-50\degr23\arcmin22\farcs1988$ & 59835.7475 & 88 & 115 & 133 & 56 & 9.8 & 28.4 \\
J1450.8$-$1424 & $14^{\rm h}50^{\rm m}50\fs592$ & $-14\degr24\arcmin21\farcs24$ & 59835.7401 & 72 & 109 & 129 & 56 & 4.9 & 23.4 \\
J1458.8$-$2120 & $14^{\rm h}58^{\rm m}48\fs7903$ & $-21\degr20\arcmin19\farcs6815$ & 59835.7619 & 74 & 90 & 100 & 56 & 5.6 & 24.1 \\
J1513.7$-$1519 & $15^{\rm h}13^{\rm m}46\fs6809$ & $-15\degr19\arcmin59\farcs1611$ & 59835.7548 & 76 & 100 & 115 & 56 & 6.3 & 24.8 \\
J1517.7$-$4446 & $15^{\rm h}17^{\rm m}42\fs7185$ & $-44\degr46\arcmin36\farcs1166$ & 59933.5247 & 88 & 151 & 181 & 60 & 12.2 & 30.8 \\
J1526.6$-$3810 & $15^{\rm h}26^{\rm m}38\fs2544$ & $-38\degr10\arcmin08\farcs3954$ & 59835.809 & 91 & 137 & 161 & 56 & 11.0 & 29.5 \\
J1527.8+1013 & $15^{\rm h}27^{\rm m}53\fs0164$ & $+10\degr13\arcmin42\farcs9584$ & 59835.7329 & 76 & 129 & 151 & 56 & 6.0 & 24.5 \\
J1530.0$-$1522 & $15^{\rm h}30^{\rm m}00\fs2637$ & $-15\degr22\arcmin33\farcs241$ & 59933.5174 & 71 & 119 & 143 & 60 & 6.4 & 24.9 \\
J1539.4$-$3323 & $15^{\rm h}39^{\rm m}24\fs2651$ & $-33\degr23\arcmin55\farcs3226$ & 59933.5031 & 77 & 85 & 91 & 60 & 8.4 & 27.0 \\
J1543.6$-$0244 & $15^{\rm h}43^{\rm m}37\fs8479$ & $-02\degr44\arcmin49\farcs5604$ & 59978.0833 & 67 & 92 & 108 & 60 & 4.5 & 23.0 \\
J1544.2$-$2554$^\dagger$ & $15^{\rm h}44^{\rm m}12\fs5537$ & $-25\degr54\arcmin45\farcs0014$ & 60102.9951 & 30 & 49 & 59 & 60 & 7.0 & 25.5 \\
J1612.1+1407 & $16^{\rm h}12^{\rm m}07\fs511$ & $+14\degr07\arcmin00\farcs4811$ & 59835.7181 & 83 & 110 & 127 & 56 & 8.5 & 27.0 \\
J1622.2$-$7202 & $16^{\rm h}22^{\rm m}12\fs6013$ & $-72\degr02\arcmin23\farcs6462$ & 59933.5967 & 69 & 87 & 100 & 60 & 5.5 & 24.0 \\
J1630.1$-$1049 & $16^{\rm h}30^{\rm m}06\fs936$ & $-10\degr49\arcmin05\farcs8809$ & 59933.532 & 73 & 102 & 121 & 60 & 7.0 & 25.5 \\
J1641.3$-$2908 & $16^{\rm h}41^{\rm m}21\fs3611$ & $-29\degr08\arcmin21\farcs4778$ & 59978.0904 & 84 & 152 & 174 & 60 & 10.6 & 29.1 \\
J1646.7$-$2154 & $16^{\rm h}46^{\rm m}44\fs425$ & $-21\degr54\arcmin26\farcs9975$ & 59978.0975 & 82 & 135 & 160 & 60 & 10.1 & 28.6 \\
J1656.4$-$0410 & $16^{\rm h}56^{\rm m}28\fs656$ & $-04\degr10\arcmin12\farcs7195$ & 59933.5536 & 73 & 95 & 110 & 60 & 6.8 & 25.4 \\
J1659.0$-$0140 & $16^{\rm h}59^{\rm m}03\fs7207$ & $-01\degr40\arcmin39\arcsec$ & 59933.5466 & 77 & 97 & 110 & 60 & 8.2 & 26.8 \\
J1708.8$-$0923 & $17^{\rm h}08^{\rm m}48\fs0981$ & $-09\degr23\arcmin39\farcs4817$ & 59978.1349 & 78 & 157 & 174 & 60 & 8.4 & 27.0 \\
J1709.4$-$2127 & $17^{\rm h}09^{\rm m}25\fs3711$ & $-21\degr27\arcmin23\farcs0397$ & 59933.5893 & 89 & 151 & 182 & 60 & 12.6 & 31.1 \\
J1709.9$-$0900 & $17^{\rm h}09^{\rm m}56\fs0156$ & $-09\degr00\arcmin51\farcs8383$ & 59933.5679 & 77 & 157 & 176 & 60 & 8.4 & 26.9 \\
    \hline
  \end{tabular}
\end{table*}

\begin{table*}
	\contcaption{A list of UHF survey observations of \textit{Fermi}-LAT sources for the second epoch.}
	\centering
    \renewcommand{\arraystretch}{1.25}
    \begin{tabular}{lccccccccc}
    \hline
    4FGL~NAME & R.A. (J2000) & Decl. (J2000) & MJD & $S_{\rm min}$ ($\mu$Jy) & $S_{\rm med}$ ($\mu$Jy) & $S_{95}$ ($\mu$Jy) & nAnt & $T_{\rm sky}$ (K) & $T_{\rm sys}$ (K)\\
    \hline
%J1709.4$-$2127 & $17^{\rm h}09^{\rm m}25\fs3711$ & $-21\degr27\arcmin23\farcs0397$ & 59933.5893 & 89 & 151 & 182 & 60 & 12.6 & 31.1 \\
%J1709.9$-$0900 & $17^{\rm h}09^{\rm m}56\fs0156$ & $-09\degr00\arcmin51\farcs8383$ & 59933.5679 & 77 & 157 & 176 & 60 & 8.4 & 26.9 \\
J1711.9$-$1922* & $17^{\rm h}11^{\rm m}54\fs624$ & $-19\degr22\arcmin03\farcs3582$ & 59978.1277 & 84 & 103 & 116 & 60 & 10.7 & 29.3 \\
J1717.5$-$5804 & $17^{\rm h}17^{\rm m}30\fs813$ & $-58\degr04\arcmin14\farcs155$ & 59933.6039 & 84 & 107 & 123 & 60 & 10.8 & 29.4 \\
J1720.6+0708 & $17^{\rm h}20^{\rm m}39\fs3091$ & $+07\degr08\arcmin48\farcs8406$ & 59835.7943 & 89 & 118 & 137 & 56 & 10.2 & 28.7 \\
J1722.8$-$0418 & $17^{\rm h}22^{\rm m}50\fs7568$ & $-04\degr18\arcmin11\farcs8797$ & 59933.575 & 78 & 94 & 105 & 60 & 8.8 & 27.3 \\
J1727.4+0326 & $17^{\rm h}27^{\rm m}27\fs1436$ & $+03\degr26\arcmin59\farcs2801$ & 59933.5608 & 78 & 103 & 120 & 60 & 8.7 & 27.2 \\
J1730.4$-$0359 & $17^{\rm h}30^{\rm m}26\fs0669$ & $-03\degr59\arcmin32\farcs2801$ & 59933.4959 & 82 & 103 & 117 & 60 & 9.8 & 28.3 \\
J1735.3$-$0717 & $17^{\rm h}35^{\rm m}21\fs6211$ & $-07\degr17\arcmin07\farcs0792$ & 59933.5822 & 86 & 127 & 152 & 60 & 11.5 & 30.1 \\
J1739.1$-$1059 & $17^{\rm h}39^{\rm m}08\fs064$ & $-10\degr59\arcmin41\farcs2786$ & 59978.1496 & 90 & 173 & 195 & 60 & 12.9 & 31.5 \\
J1747.6+0324 & $17^{\rm h}47^{\rm m}37\fs2217$ & $+03\degr24\arcmin26\farcs28$ & 59978.1645 & 86 & 105 & 121 & 60 & 11.3 & 29.9 \\
J1749.8$-$0303 & $17^{\rm h}49^{\rm m}50\fs9253$ & $-03\degr03\arcmin06\farcs48$ & 59978.1716 & 90 & 118 & 137 & 60 & 12.9 & 31.4 \\
J1813.7$-$6846 & $18^{\rm h}13^{\rm m}47\fs5415$ & $-68\degr46\arcmin42\farcs6123$ & 59978.112 & 70 & 103 & 121 & 60 & 5.8 & 24.3 \\
J1816.4$-$6405* & $18^{\rm h}16^{\rm m}27\fs5977$ & $-64\degr05\arcmin05\farcs9967$ & 59978.1049 & 73 & 97 & 113 & 60 & 7.0 & 25.5 \\
J1816.7+1749 & $18^{\rm h}16^{\rm m}45\fs0513$ & $+17\degr49\arcmin40\farcs4398$ & 59835.8015 & 85 & 109 & 126 & 56 & 9.0 & 27.6 \\
J1818.6+1316 & $18^{\rm h}18^{\rm m}36\fs6504$ & $+13\degr16\arcmin23\farcs1596$ & 59933.5393 & 80 & 104 & 121 & 60 & 9.2 & 27.8 \\
J1822.9$-$4718 & $18^{\rm h}22^{\rm m}56\fs6895$ & $-47\degr18\arcmin16\farcs9162$ & 59978.1423 & 76 & 101 & 117 & 60 & 8.0 & 26.6 \\
J1823.2+1209* & $18^{\rm h}23^{\rm m}14\fs8828$ & $+12\degr09\arcmin26\farcs28$ & -- & -- & -- & -- & -- & -- & -- \\
J1824.2$-$5427 & $18^{\rm h}24^{\rm m}16\fs897$ & $-54\degr27\arcmin05\farcs0427$ & 59978.157 & 71 & 88 & 100 & 60 & 6.2 & 24.7 \\
J1827.5+1141 & $18^{\rm h}27^{\rm m}30\fs8643$ & $+11\degr41\arcmin10\farcs681$ & 59978.2075 & 86 & 103 & 121 & 60 & 11.3 & 29.9 \\
J1831.1$-$6503* & $18^{\rm h}31^{\rm m}06\fs5552$ & $-65\degr03\arcmin57\farcs2498$ & 59835.7858 & 76 & 92 & 103 & 56 & 6.2 & 24.8 \\
J1845.8$-$2521* & $18^{\rm h}45^{\rm m}51\fs5552$ & $-25\degr21\arcmin30\farcs5983$ & 59978.1788 & 91 & 123 & 143 & 60 & 13.0 & 31.5 \\
J1906.0$-$1718 & $19^{\rm h}06^{\rm m}04\fs4604$ & $-17\degr18\arcmin55\farcs08$ & 59978.1931 & 81 & 129 & 153 & 60 & 9.7 & 28.3 \\
J1910.7$-$5320* & $19^{\rm h}10^{\rm m}47\fs7393$ & $-53\degr20\arcmin18\farcs6035$ & 59730.2738 & 75 & 143 & 163 & 60 & 5.8 & 24.3 \\
J1913.4$-$1526 & $19^{\rm h}13^{\rm m}24\fs3604$ & $-15\degr26\arcmin58\farcs5608$ & 59978.186 & 81 & 105 & 121 & 60 & 9.5 & 28.1 \\
J1916.8$-$3025 & $19^{\rm h}16^{\rm m}53\fs6865$ & $-30\degr25\arcmin28\farcs5622$ & -- & -- & -- & -- & -- & -- & -- \\
J1924.8$-$1035* & $19^{\rm h}24^{\rm m}49\fs27$ & $-10\degr35\arcmin27\farcs2415$ & 59978.2003 & 83 & 97 & 106 & 60 & 10.2 & 28.8 \\
J1947.6$-$1121 & $19^{\rm h}47^{\rm m}39\fs8657$ & $-11\degr21\arcmin54\farcs7202$ & -- & -- & -- & -- & -- & -- & -- \\
J2026.3+1431 & $20^{\rm h}26^{\rm m}23\fs7085$ & $+14\degr31\arcmin21\farcs0001$ & 59933.6116 & 68 & 96 & 115 & 60 & 5.2 & 23.7 \\
J2029.5$-$4237* & $20^{\rm h}29^{\rm m}30\fs3589$ & $-42\degr37\arcmin42\farcs6068$ & -- & -- & -- & -- & -- & -- & -- \\
J2043.9$-$4802 & $20^{\rm h}43^{\rm m}55\fs5615$ & $-48\degr02\arcmin23\farcs2755$ & 59835.6874 & 72 & 89 & 100 & 56 & 5.0 & 23.5 \\
J2112.5$-$3043 & $21^{\rm h}12^{\rm m}33\fs6035$ & $-30\degr43\arcmin45\farcs4784$ & 59835.7701 & 71 & 77 & 81 & 56 & 4.4 & 23.0 \\
J2121.8$-$3412 & $21^{\rm h}21^{\rm m}53\fs811$ & $-34\degr12\arcmin11\farcs8817$ & 59786.1333 & 65 & 110 & 124 & 60 & 4.1 & 22.6 \\
J2133.1$-$6432 & $21^{\rm h}33^{\rm m}10\fs8252$ & $-64\degr32\arcmin17\farcs8784$ & 59835.6802 & 71 & 87 & 98 & 56 & 4.2 & 22.8 \\
J2201.0$-$6928 & $22^{\rm h}01^{\rm m}01\fs6772$ & $-69\degr28\arcmin26\farcs0394$ & 59786.1406 & 65 & 87 & 101 & 60 & 4.2 & 22.7 \\
J2212.4+0708 & $22^{\rm h}12^{\rm m}25\fs9937$ & $+07\degr08\arcmin34\farcs0795$ & 59786.1182 & 64 & 83 & 95 & 60 & 3.8 & 22.3 \\
J2219.7$-$6837 & $22^{\rm h}19^{\rm m}47\fs2559$ & $-68\degr37\arcmin02\farcs287$ & 59786.1259 & 64 & 80 & 90 & 60 & 3.9 & 22.4 \\
J2241.4$-$8327 & $22^{\rm h}41^{\rm m}25\fs6348$ & $-83\degr27\arcmin40\farcs3088$ & 59835.7028 & 70 & 91 & 107 & 56 & 4.2 & 22.7 \\
J2355.5$-$6614* & $23^{\rm h}55^{\rm m}33\fs7939$ & $-66\degr14\arcmin04\farcs1895$ & 59835.71 & 69 & 92 & 109 & 56 & 3.9 & 22.4 \\
    \hline
  \end{tabular}
\end{table*}

\begin{table*}
	\caption{Initial information on slow pulsars. Note that the positions reported here correspond to approximate localisation within the incoherent beam due to the lack of coherent beam detection, except for PSR~J1303$-$4713, for which the position and uncertainties are from \texttt{SeeKAT}. The signal-to-noise ratios are measured from the discovery observations folded with \texttt{PulsarX}.}
    \label{T:slow_pulsars}
\begin{tabular}{lcccccc}
\hline
Name &
R.A. &
Decl. &
MJD &
Spin frequency &
Dispersion measure &
Signal-to-noise ratio\\
PSR &
(J2000) &
(J2000) &
&
$\nu$ (Hz) &
DM (pc cm$^{-3}$) &
S/N
\\
\hline
    J1207$-$45 & $12^{\rm h}07^{\rm m}(05)$ & $-45\degr36\arcmin(48)$ & 59563.0596209787 & 2.10004(3) & 23.76(20) & 15.2\\
    J1303$-$4713 & $13^{\rm h}03^{\rm m}31\fs5(5)$ & $-47\degr13\arcmin12\farcs8(4)$ & 59887.1308713678 & 0.39014(1) & 83.97(99) & 16.5\\
    J1816$-$64 & $18^{\rm h}16^{\rm m}(07)$ & $-64\degr05\arcmin(48)$ & 59786.0928676127 & 0.81840(1) & 50.91(54) & 14.3\\
    J1845$-$25 & $18^{\rm h}45^{\rm m}(04)$ & $-25\degr21\arcmin(48)$ & 59786.0724193998 & 1.72310(2) & 86.91(35) & 10.5\\
    J1924$-$10 & $19^{\rm h}24^{\rm m}(03)$ & $-10\degr35\arcmin(48)$ & 59978.1951990979 & 8.71501(2) & 60.48(5) & 16.4\\
    J2355$-$66 & $23^{\rm h}55^{\rm m}(08)$ & $-66\degr14\arcmin(48)$ & 59563.0599451504 & 8.23064(1) & 19.50(5) & 11.7\\
    \hline
    \end{tabular}
\end{table*}

\section{Orbital period variations}
\begin{figure*}
  \centering
  	\includegraphics[width=\textwidth]{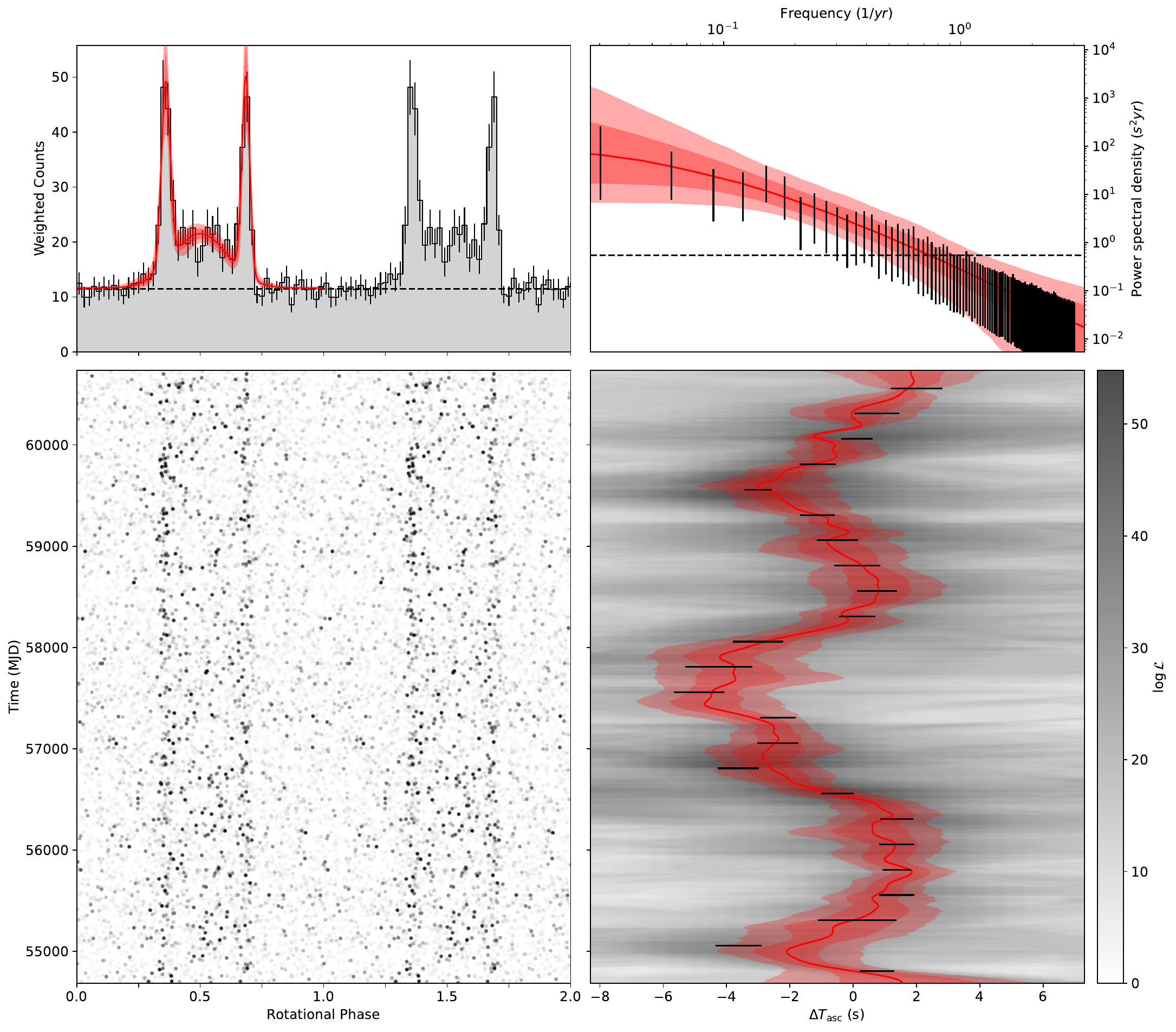}
        \caption{Gamma-ray pulsations and variations of the orbital phase over time for PSR~J1831$-$6503. The panels on the left show the weighted pulsar spin phases of each gamma-ray photon for the highest-likelihood timing solution (bottom-left panel) and the integrated pulse profile (top-left panel). The red curve and the shaded regions represent the highest-likelihood pulse-profile template and the $1\sigma$ and $2\sigma$ template confidence intervals estimated from the Monte-Carlo samples. The bottom-right panel shows the orbital phase variations as a function of time. The grey-scale image shows the log-likelihood for offsets from the pulsar's $T_{\rm asc}$ as measured in overlapping $800$-day windows, with $24$ evenly spaced gamma-ray ToAs (black bars) to guide the eye. The red curve and shaded regions represent the solution used to fold the gamma-ray data and the $1\sigma$ and $2\sigma$ confidence intervals on those deviations, obtained from Monte-Carlo timing analysis. The top-right panel shows the power spectral density of the orbital phase variations. The black bars show the measured $1\sigma$ confidence interval on the power spectral densities. The red curve and shaded regions illustrate the highest-likelihood solution and the $1\sigma$ and $2\sigma$ confidence intervals of the Mat\'{e}rn model. The dashed line estimates the white-noise floor.}
    \label{F:J1831_opv}
\end{figure*}

Figure \ref{F:J1831_opv} illustrates the gamma-ray pulsations and orbital phase variations for PSR~J1831$-$6503.

%%%%%%%%%%%%%%%%%%%%%%%%%%%%%%%%%%%%%%%%%%%%%%%%%%

% Don't change these lines
\bsp	% typesetting comment
\label{lastpage}
\end{document}